%% file: p.tex
\documentclass[10pt,letterpaper,twocolumn]{article}

\usepackage{usenix2020-09}
\usepackage[square,comma,numbers,sort&compress]{natbib}

\include{pkgs}

\DeclareMathAlphabet\mathbfcal{OMS}{cmsy}{b}{n}

\newcommand{\ACC}{\mbox{ACon$^2$}\xspace}

\newcommand{\MR}[1]{\textcolor{black}{#1}}
\newcommand{\MRM}[1]{\textcolor{black}{#1}}
\newcommand{\NMR}[1]{\textcolor{black}{#1}}

\usepackage{ifthen}
\newcommand\LONG{1}
\newcommand\LREF[2]{
  \ifthenelse{
    \equal{\LONG}{1}
  }
  {{\color{black}#1}\ignorespaces}
  {{\color{black}#2 of \cite{park2022acon}}\ignorespaces}
}

\input{cmds}
\input{cmds_math}

\begin{document}

\input{hdr}

\date{}
\maketitle

\sloppy

\input{abstract}
\input{intro}

\input{bg}
\input{problem}

\input{approach}

\input{eval}
\input{relwk}

\input{discussion}

\input{conclusion}

{
\input{ack}
}


\vspace{-1ex}
{
\bibliographystyle{plain}
\footnotesize
\setlength{\bibsep}{2pt}
\bibliography{ml,twsml}
}

\clearpage
\input{apdx}

\clearpage
\input{apdxext}


\end{document}

%% file: pkgs.tex
\usepackage{xurl}                
\usepackage{xcolor}             
\usepackage[]{hyperref}         
\hypersetup{                    
  colorlinks,
  linkcolor={green!80!black},
  citecolor={red!70!black},
  urlcolor={blue!70!black}
}

\usepackage{amsmath,amsopn,amssymb}
\usepackage{subfigure}
\usepackage{endnotes,xspace,graphicx,fancyvrb,multirow}
\usepackage{booktabs}
\usepackage{array,underscore,relsize}
\usepackage[T1]{fontenc}
\usepackage{times}
\usepackage{fancyhdr,lastpage}
\usepackage{enumitem}
\usepackage[labelfont=bf,font=small,skip=5pt]{caption}
\usepackage{makecell}

\usepackage{fp}
\usepackage{siunitx}

\usepackage{algorithm}
\usepackage[noend]{algpseudocode}

\usepackage{balance}

\usepackage{tikz}
\usetikzlibrary{shapes.geometric,shapes.arrows,decorations.pathmorphing}
\usetikzlibrary{matrix,chains,scopes,positioning,arrows,fit,patterns}

\usepackage{tablefootnote}

%% file: cmds.tex
\fvset{fontsize=\scriptsize,xleftmargin=8pt,numbers=left,numbersep=5pt}

\def\Snospace~{\S{}}

\newcommand{\x}{$\times$\xspace}

\newif\ifdraft\drafttrue
\newif\ifnotes\notestrue
\ifdraft\else\notesfalse\fi

\input{glyphtounicode}
\newcolumntype{R}[1]{>{\raggedleft\let\newline\\\arraybackslash\hspace{0pt}}p{#1}}

\newcommand{\includepdf}[1]{
  \includegraphics[width=\columnwidth]{#1}
}

\newcommand{\squishlist}{
\begin{itemize}[noitemsep,nolistsep]
  \setlength{\itemsep}{-0pt}
}
\newcommand{\squishend}{
  \end{itemize}
}

\usepackage{tikz}

\usepackage{xstring}
\newcommand{\PP}[1]{
\vspace{2px}
\noindent{\bf \IfEndWith{#1}{.}{#1}{#1.}}
}

\newcommand{\X}{{\footnotesize $\times$}\xspace}

\newcommand{\boxbeg}{
\vspace{2px}
\noindent\begin{tabular}{|l|}\hline
\begin{minipage}{3.2in}
\vspace{2px}
\noindent
}

\newcommand{\boxend}{
\vspace{2px}
\end{minipage}\\ \hline
\end{tabular}
\vspace{-10pt}
}

\newcommand{\ie}{\emph{i.e.,}~}
\newcommand{\eg}{\emph{e.g.,}~}

\usepackage{pifont}

\providecommand{\para}[1]{\vspace{2ex}\noindent\textbf{#1}}

%% file: cmds_math.tex
\usepackage{amsmath} 
\usepackage{bm} 
\usepackage{bbm} 
\usepackage{amsthm}
\usepackage{mathtools}

\newcommand{\vmid}{\;\middle|\;}
\newcommand{\ep}{\varepsilon}

\makeatletter
\@ifundefined{proposition}{%
    
}{}
\@ifundefined{definition}{%
    \newtheorem{definition}{Definition}%
}{}
\@ifundefined{lemma}{%
    \newtheorem{lemma}{Lemma}
}{}
\@ifundefined{corollary}{%
    \newtheorem{corollary}{Corollary}
}{}
\@ifundefined{theorem}{%
    \newtheorem{theorem}{Theorem}
}{}
\@ifundefined{corollary}{%
    
}{}
\@ifundefined{assumption}{%
    \newtheorem{assumption}{Assumption}
}{}
\@ifundefined{problem}{%
    
}{}
\@ifundefined{problem*}{%
    \newtheorem*{problem*}{Problem}
}{}
\@ifundefined{claim}{%
    
}{}
\@ifundefined{example}{%
    \newtheorem{example}{Example}
}{}
\@ifundefined{implication}{%
    
}{}
\makeatother

\newtheorem{innercustomgeneric}{\customgenericname}
\providecommand{\customgenericname}{}
\newcommand{\newcustomtheorem}[2]{%
  \newenvironment{#1}[1]
  {%
   \renewcommand\customgenericname{#2}%
   \renewcommand\theinnercustomgeneric{##1}%
   \innercustomgeneric
  }
  {\endinnercustomgeneric}
}
\newcustomtheorem{customclaim}{Claim}

\providecommand{\realnum}					{\mathbb{R}}

\providecommand{\naturalnum}				{\mathbb{N}}

\renewcommand{\(}						{\left(}
\renewcommand{\)}						{\right)}
\renewcommand{\[}						{\left[}
\renewcommand{\]}						{\right]}

\providecommand{\Prob}{\mathbbm{P}}

\providecommand{\Exp}{\mathbbm{E}}
\providecommand{\Expop}{\mathop{\Exp}}

\def\x{\mathbf{x}}
\def\y{\mathbf{y}}
\def\z{\mathbf{z}}

\def\P{\mathbf{P}}

\def\S{\mathbf{S}}

\def\X{\mathbf{X}}
\def\Y{\mathbf{Y}}

\def\yh{\hat{{y}}}

\def\Ch{\hat{{C}}}

\def\Es{\mathcal{{E}}}
\def\Fs{\mathcal{{F}}}

\def\Ks{\mathcal{{K}}}

\def\Ns{\mathcal{{N}}}

\def\Ps{\mathcal{{P}}}
\def\Qs{\mathcal{{Q}}}

\def\Ss{\mathcal{{S}}}

\def\Vs{\mathcal{{V}}}

\def\Xs{\mathcal{{X}}}
\def\Ys{\mathcal{{Y}}}

\def\Xbs{\mathbfcal{{X}}}
\def\Ybs{\mathbfcal{{Y}}}

%% file: hdr.tex
\title{\NMR{\ACC: Adaptive Conformal Consensus} for Provable Blockchain Oracles}

\ifdefined\DRAFT
 \pagestyle{fancyplain}
 \lhead{Rev.~\therev}
 \rhead{\thedate}
 \cfoot{\thepage\ of \pageref{LastPage}}
\fi


\author{
  \emph{Sangdon Park}$^{\dagger}$ \qquad
  \emph{Osbert Bastani}$^\ast$ \qquad
  \emph{Taesoo Kim}$^{\dagger}$ 
  \\\\
  \emph{$^\dagger$\emph{Georgia Institute of Technology}}\qquad
  \emph{$^\ast$\emph{University of Pennsylvania}}
}

%% file: abstract.tex
\begin{abstract}
  Blockchains with smart contracts are distributed ledger systems
  that achieve block-state consistency among distributed nodes by only allowing
  deterministic operations of smart contracts.
  However, the power of smart contracts is enabled by
  interacting with stochastic off-chain data, which in turn
  opens the possibility to undermine the block-state consistency.
  To address this issue, an oracle smart contract is used to provide
  a single consistent source of external data; but, simultaneously, this introduces
  a single point of failure, which is called the oracle problem.
  To address the oracle problem, we propose an adaptive conformal consensus (\ACC) algorithm
  that derives a consensus set of data from multiple oracle contracts
  via the recent advance in online uncertainty quantification learning.
  Interesting,
  the consensus set provides a desired correctness guarantee 
  under distribution shift and Byzantine adversaries.  
  We demonstrate the efficacy of the proposed algorithm on
  two price datasets and an Ethereum case study.
  In particular, the Solidity implementation of the proposed algorithm shows
  the potential practicality of the proposed algorithm, implying that
  online machine learning algorithms are applicable to address security issues in blockchains. 
\end{abstract}

%% file: intro.tex
\section{Introduction}
\label{s:intro}
\vspace{-1ex}

\begin{figure}[tb!]
  \centering
  \includegraphics[width=0.9\linewidth]{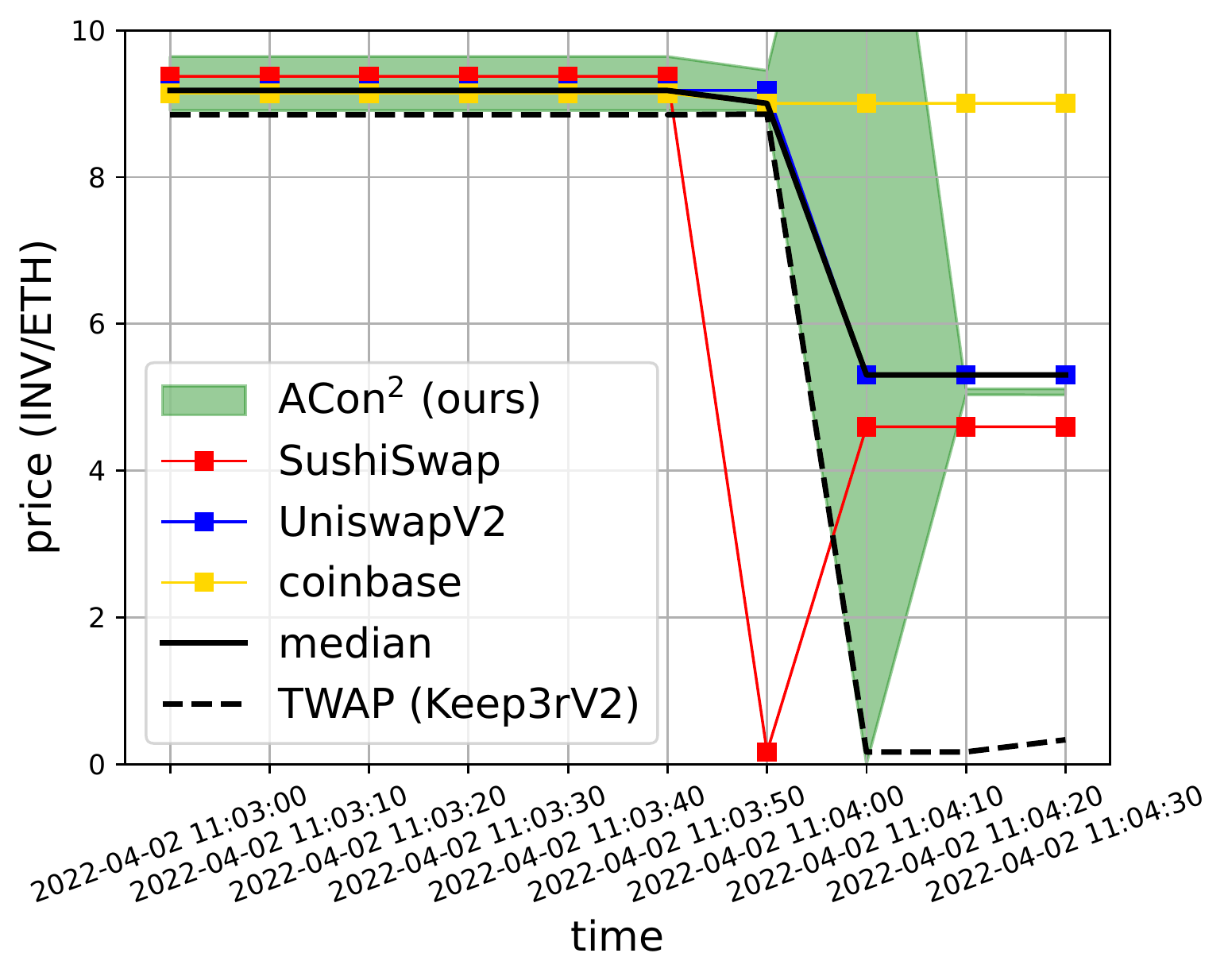}
  \caption{
    We propose to address prediction consensus via \ACC.
    For an application, we consider the recent price manipulation of an INV token price on SushiSwap.
    \ACC provides uncertainty on price prediction (in prediction intervals)
    along with a correctness guarantee under Byzantine adversaries (\eg price manipulators),
    resulting in providing uncertainty after the price manipulation at 2022-04-02 11:04:00
    for peculiarity (an empty interval but represented in a large interval for visualization).
    See Section \ref{sec:localshift} for details.
  }
  \label{fig:inveth}
\end{figure}


Blockchains are distributed ledger systems
in which a set of transactions forms a block, and
blocks are securely connected to form a chain via cryptography
to avoid record manipulation.
The concept of the ledger can be generalized to
executable programs, called \emph{smart contracts},
coined by Nick Szabo \cite{szabo1994smartcontracts}.
As smart contracts can be any program, they provide a great number of applications for blockchains. 
In particular,
a smart contract is used to provide collateralized lending services within a blockchain,
\eg lending USD via Ethereum.
To this end, the contract needs to interact with off-chain data, \eg an Ethereum price in US dollars (USD).
However, accessing and recording arbitrarily external data in a blockchain is prohibited
because of the deterministic property of distributed blockchains.
To maintain consistent block states across distributed nodes,
operations in blockchains need to be deterministic.
However, reading and writing stochastic data into blockchains break the consistency
among blockchains at each distributed node.
To avoid this inconsistency issue, a special smart contract, called an \emph{oracle smart contract},
is introduced as a single source of a data feed;
however, this in turn provides a single point of failure, called the \emph{oracle problem} \cite{ethereumoracles}.
In particular, malicious adversaries can feed invalid data into the oracle contracts
to achieve their goals (\eg price manipulation \cite{samczsun2019undercolattack,samczsun2020soyouwant,bzx2020,yvault2020,harvestflashloanattack2020,venus2021,enzyme2021,rekt2022inverse},
to lend Ethereum with a cheaper price in USD).

To address the oracle problem,
we may use traditional consensus solutions over diverse oracle contracts
(\eg consensus over Byzantine generals \cite{lamport1982byzantine},
consensus over abstract sensors \cite{marzullo1990tolerating}, or
robust statistics like median or truncated mean).
For example, the median prices from diverse price oracle contracts can be used for a consensus price. 
However, the main challenges to address the oracle problem include the following:
handling
\ding{182} the \emph{inevitable uncertainty} in external data for consensus
(\eg ETH/USD price is varying across markets),
\ding{183} \emph{distribution shift} along time (\eg ETH/USD price is varying across time),
\ding{184} the presence of adversaries to undermine consensus,
\ding{185} a \emph{correctness guarantee} on consensus in dealing with the previous challenges, and
\ding{186} whether the proposed algorithm is implementable in blockchains.
To our understanding, these challenges are not jointly considered in a single traditional consensus method.

In this paper,
we view the oracle problem as a \emph{prediction consensus learning} problem;
we consider each oracle smart contract as a predictor $\Ch_t$ at time $t$,
where it predicts
the uncertainty over labels $y$ based on the external data observations $\x_t$.
Given multiple predictors from multiple data sources,
we learn and predict the consensus over uncertain labels.
To address the prediction consensus learning,
we exploit the recent advance in
online machine learning for uncertainty quantification \cite{gibbs2021adaptive,bastani2022practical} based on conformal prediction \cite{vovk2005algorithmic}.

In particular,
we propose an adaptive conformal consensus (\ACC) algorithm that satisfies a correctness guarantee.
At time $t$, this algorithm returns a set-valued predictor $\Ch_t$, where
given an observation $\x_t$ from multiple sources,
we have a \emph{consensus set} $\Ch(\x_t)$ 
that likely contains the true consensus label even in the presence of distribution shift and adversaries. 
Here,
we model uncertainty via a set of labels, called a \emph{prediction set}; 
this set-of-labels notation is equivalent to having multiple votes on labels, \ie
if it is uncertain to choose one option,
it makes multiple uncertain choices instead of choosing one certain but wrong choice.
Based on this, we can handle uncertainty from data in Challenge \ding{182}.
For the $k$-th source at time $t$,
given the observation $\x_t$ along with a label $\y_{t, k}$,
the $k$-th base prediction set $\Ch_{t, k}$ is updated via any adaptive conformal prediction
(\eg \cite{bastani2022practical}).
As the adaptive conformal prediction learns a correct prediction set under distribution shift, 
this addresses Challenge \ding{183}.
For consensus, we consider that $K$ base prediction sets from $K$ sources are given,
and
adversaries can arbitrarily manipulate at most $\beta$ sources (thus $\beta$ base prediction sets) among $K$,
called \emph{$\beta$-Byzantine adversaries}. 
Given $K$ base prediction sets $\Ch_{t, k}(\x_t)$ for $k \in \{1, \dots, K\}$,
where $\beta$ of them are possibly manipulated,
we construct a consensus set $\Ch_t(\x_t)$,
which contains labels that are also contained in the $K-\beta$ base prediction sets;
by filtering out possibly wrong base prediction sets in this way,
the consensus set is not maliciously manipulated by the adversaries,
which addresses Challenge \ding{184}.
Finally, we provide the worst-case correctness guarantee on the consensus sets from our algorithm \ACC;
under any $\beta$-Byzantine adversaries and any distribution shift (along with mild assumptions),
the consensus sets from \ACC likely contain the true consensus labels at a desired miscoverage rate.
This is proven given the correctness guarantee of the base prediction sets,
thus addressing Challenge \ding{185}.

We demonstrate the efficacy of the proposed algorithm \ACC
via the evaluation over two datasets and one case study.
In particular, 
we use two datasets, obtained from the Ethereum blockchain: 
a USD/ETH price dataset, which manifests natural distribution shift, and
an INV/ETH price dataset, which embeds price manipulation attacks.
For the case study,
we implement our algorithm in Solidity to show its practicality on the Ethereum blockchain.
In short, we empirically show that
the consensus sets by \ACC achieve a desired miscoverage rate even under distribution shift and Byzantine adversaries.
Moreover, our Solidity implementation shows that
an online machine learning algorithm has \MR{opportunities} to be used in blockchains,
which also implies that it can enjoy the underlying security from blockchain-level consensus
(\eg proof-of-work); this \MR{partially} addresses Challenge \ding{186}
\footnote{\MR{Datasets and code, including Solidity, are publicly available at \\\url{https://github.com/sslab-gatech/ACon2}.}}.

%% file: bg.tex
\vspace{-1ex}
\section{Background}
\label{s:bg}
\vspace{-1ex}

Here,
we provide background on
blockchain oracles and online learning, in particular
adaptive conformal prediction.

\subsection{Blockchain Oracles}
\label{sec:blockchainoracles}
\vspace{-1ex}

A blockchain is a distributed ledger system that
consists of records, called \emph{blocks},
by securely connecting them in a \emph{chain} via cryptography.
The main use of the blockchain is
a distributed ledger for cryptocurrencies, like Bitcoin \cite{bitcoin} or Ethereum \cite{ethereum}.
The concept of the ledger is generalized to record
a computer program, called \emph{smart contracts},
practically realized in Ethereum \cite{ethereumwhitepaper};
the smart contracts are automatically executed on the blockchain
to provide additional functionalities beyond ledgers, like
decentralized finance (DeFi) or a non-fungible token (NFT).

\para{Blockchain oracles.}
One special type of smart contracts is an \emph{oracle smart contract}.
The blockchain is a distributed system, where each node of the system maintains
the exact identical information in a blockchain.
To this end, all smart contracts are \emph{deterministic}.
However, the blockchain needs to read information from the real world.
In particular, Ethereum can be exchanged based on agreed US dollars (USD) \cite{ethereumoracles},
but the value of Ethereum in USD depends on the off-chain \emph{stochastic} data.
Thus, by simply reading and feeding the stochastic data into on-chain
by executing a related smart contract at each node could potentially break the
consistency among blockchains at each node. 

To address this issue,
oracle smart contracts are used to provide a single point of feeding off-chain data;
as it is a single point of contact, all smart contracts that interact with it can
maintain the consistent blockchain state.
However, the oracle contract can be a single point of failure as well,
which is known as the \emph{oracle problem} \cite{ethereumoracles}.
To dive into the oracle problem, we first explain
one dominant application of smart contracts in DeFi \MRM{that is heavily related to
the oracle problem.} 

\para{Automated market maker.}
An automated market maker (AMM) is a smart contract that forms a market to swap tokens.
For example, Uniswap \cite{uniswap} is a smart contract protocol
that forms AMMs for various pairs of tokens, where
the price of tokens is decided by a constant product formula.
We take this as our concrete example in describing AMMs. 
In the constant product formula, the price of a pool of two tokens $A$ and $B$ is decided from
the equation $xy=k$ given a constant $k$,
where
$x$ and $y$ are the amounts of token $A$ and token $B$, respectively.
Letting $k=1$, the current price of token $A$ by token $B$ is $\frac{y}{x}$;
if \MRM{a trader sells} token $A$ by the amount of $x'$,
the amount of token $B$ that the trader will receive is $y' = y - \frac{xy}{x + x'}$ to maintain
the ratio of the amounts of two tokens to be $k=1$.
After this trade, the price of token $A$ by the token $B$ becomes
$\frac{y - y'}{x + x'}$.

In DeFi, the price formed by an AMM is mainly used for
an on-chain decentralized collateralized loaning service.
Specifically,
a user deposits assets (\eg ETH) to the lending service to borrow another asset (\eg USD)
proportional to the value of the deposited assets.
Taking an example from \cite{samczsun2020soyouwant},
suppose the collateralization ratio is $150\%$.
If the spot price (\ie the current price which can be sold immediately) of ETH is 400 USD,
the user can borrow $100,000$ USD
by the deposit of $375$ ETH, \ie
\begin{align}
  375~ (\text{ETH}) \times
  400 \( \frac{\text{USD}}{\text{ETH}} \) \times
  \frac{100}{150}
  = 100,000~ (\text{USD})
  \label{eq:colex}
\end{align}
To get the spot price of ETH, the lending service accesses a price oracle, possibly from an AMM.

\para{The oracle problem and oracle manipulation.}
The \emph{oracle problem} is a contradictory situation in which
a blockchain needs a single oracle smart contract that reads data from off-chain
to maintain consistency among distributed blockchains,
while the oracle contract can be a point of failure by \emph{manipulation}.
For example, 
the price formed by the AMM reflects the value of two tokens in the real world; thus,
the smart contract that provides a price is also the oracle contract.
But, considering how to decide a price in AMMs, it is susceptible to price manipulation.
In particular, an adversary can sell a huge number of token $A$ to an AMM; then
its price at this AMM is skewed compared to the price of other AMMs,
thus maliciously affecting other DeFi services that rely on the price from the manipulated AMM.

The price manipulation was executed
\cite{samczsun2019undercolattack,samczsun2020soyouwant,bzx2020,yvault2020,harvestflashloanattack2020,venus2021,enzyme2021,rekt2022inverse},
as recently as April 2, 2022 \cite{rekt2022inverse}, 
when we wrote this paper in October 2022. 
Here, we provide a simplified price manipulation attack from \cite{samczsun2020soyouwant};
see \cite{samczsun2019undercolattack} for a detailed analysis.
Suppose
the spot price of $1$ ETH is $400$ USD at Uniswap.
An adversary buys $5,000$ ETH for $2,000,000$ USD from Uniswap.
The spot price of $1$ ETH is now $1,733.33$ USD at Uniswap. 
The lending service fetches the spot price of ETH from Uniswap,
which is $1~ \text{ETH} = 1,733.33~ \text{USD}$.
The adversary can borrow $433,333.33$ USD by depositing $375$ ETH,
which is about four times higher than before the price manipulation in (\ref{eq:colex}).
Then, the adversary sells $5,000$ ETH for $2,000,000$ USD to return back to the original price.
This attack can be implemented to be atomic, so arbitrage is not possible.
Thus, the attacker enjoys the benefit by $333,333.33~\text{USD}$.

\para{Countermeasures.}
Practical countermeasures to defend against price manipulation have been proposed \cite{uniswaptwap,chainlink}.
Assuming a single price source is given,
we can consider the average price within a time frame, \ie
$
\frac{C_{t_1} - C_{t_0}}{t_1 - t_0},
$
where
$t_0$ is the previous timestamp,
$t_1$ is the current timestamp,
$C_{t_0}$ is the previous cumulative price, and
$C_{t_1}$ is the current cumulative price.
This is called Time-Weighted Average Price (TWAP) \cite{uniswaptwap}.
Here, the time frame $t_1 - t_0$ is a design parameter;
if it is large, the TWAP is hard to manipulate, as the manipulated price is exposed to arbitrage opportunities and also
the aggressively manipulated price is averaged out.
However, the choice of the large time frame sacrifices getting an up-to-date price.
If the time frame is set by a relatively small value, the price is heavily affected by the manipulation,
which is the main cause of the recent Inverse Finance incident \cite{rekt2022inverse}.
\MRM{Recent research shows that, even with a large time frame, manipulators collude with miners to reduce manipulation cost of TWAP \cite{mackinga2022twap}}.
Alternatively, 
when multiple price sources are given, we can consider price accumulation among multiple sources;
one traditional way is to consider robust statistics, \eg the median among prices. 
Chainlink \cite{chainlink} uses the median approach in sophisticated ways (\eg the median of medians).
\MR{However, price oracles for minor coins are not attractive for Chainlink node operators
  (\eg the Inverse Finance incident \cite{rekt2022inverse} is triggered by price manipulation on a minor coin, where Chainlink does not provide a price oracle). 
} 

Finally, the known practical solutions are considered to have limitations on guaranteeing security.
In particular, TWAP assumes setting the right time frame, and
\MRM{the median robust statistics do not provide the uncertainty of the median value}.
\MR{The quantile of multiple prices can be used as the measure of uncertainty but is easier to manipulate than the median, potentially introducing false alarms by manipulation for denial-of-service attacks.}
Instead, we rely on machine learning theories,
in particular online machine learning and conformal prediction,
to handle the oracle problem with
provable correctness guarantees by learning security-sensitive parameters.

\vspace{-1ex}
\subsection{Adaptive Conformal Prediction}
\label{sec:acp}
\vspace{-1ex}

Provable uncertainty quantification on prediction is essential to build trustworthy predictors,
where
conformal prediction \cite{vovk2005algorithmic} provides the provable uncertainty quantification is via
\emph{prediction sets} (\ie a set of predicted labels)
that comes with a correctness guarantee.
Conformal prediction originally assumes that
distributions on \MRM{training and test data} are not changing
(more precisely exchangibility),
but recent work \cite{gibbs2021adaptive,bastani2022practical} extends this to handle
distribution shift, making it applicable in more practical settings.

We describe a setup for adaptive conformal prediction,
an online machine learning variant of the conformal prediction.
Let
$\Xs$ be example space,
$\Ys$ be label space, and
$\Ps'$ be the set of \MRM{all distributions} over $\Xs \times \Ys$.
In conformal prediction,
we assume that a conformity score function $s_t: \Xs \times \Ys \to \realnum_{\ge 0}$ at time $t \in \{1, \dots, T\}$ for a time horizon $T$
is given,
which measures whether datum $(x, y) \in \Xs \times \Ys$ conforms to a score function $s_t$.
Then, a \emph{prediction set predictor} (or a \emph{conformal predictor}) $\Ch_t: \Xs \to 2^\Ys$ at time $t$ is defined by the score function $s_t$
and a scalar parameter $\tau_t \in \realnum_{\ge 0}$
as follows:
\begin{align}
  \Ch_t(x) \coloneqq \left\{ y \in \Ys \mid s_t(x, y) \ge \tau_t \right\}.
  \label{eq:psdef}
\end{align}
Here, we denote a set of all \MRM{conformal predictors by $\Fs'$},
\NMR{and the prediction set size represents uncertainty (\ie a larger set means more uncertain).
  Note that we consider that the empty set $\emptyset$ and the entire set $\Ys$ represent the largest uncertainty. 
}

The main goal of adaptive conformal prediction is to choose $\tau_t$ at time $t$ such that
a prediction set $\Ch_t$ likely contains the true label $y_t$.
To measure the goodness of prediction sets, we use the miscoverage
of the prediction set, defined as follows:
\begin{align}\label{eq:miscover}
  Miscover(\Ch_t, x, y) \coloneqq \mathbbm{1}\( y \notin \Ch_t(x)\).
\end{align}

\MR{
  Under distribution shift, choosing $\tau_t$ such that a prediction set
  covering a future label is challenging.
  In adaptive conformal prediction,
  whenever the prediction set does not cover a label,
  $\tau_t$ can be decreased to make the set size larger.
  Even under fast shift, $\tau_t$ is possibly zero such that
  the prediction set always covers a fast-shifting label to achieve
  a desired miscoverage rate. 
}
More precisely, 
we desire to find a learner $L$ that uses
all previous data and prediction sets, such that
the learner returns a distribution over \MRM{conformal predictors $\Fs'$} where
sampled \MRM{conformal predictors} achieve a desired miscoverage rate $\alpha$
on the worst-case data during time until $T$;
the mistakes of the learner are measured by a miscoverage value $\Vs'$ as follows:
\begin{multline}
  \Vs'(\Fs', T, \alpha, L) \coloneqq
  \max_{p_1 \in \Ps'}
  \Expop_{\substack{(X_1, Y_1) \sim p_1 \\ \Ch_1 \sim L(\cdot)}}
  \dots
  \max_{p_T \in \Ps'}
  \Expop_{\substack{(X_T, Y_T) \sim p_T \\ \Ch_T \sim L(\cdot)}}
  \\
  \left| \frac{1}{T} \sum_{t=1}^T Miscover(\Ch_t, X_t, Y_t)
  - \alpha \right|,
  \label{eq:baselearnervalue}
\end{multline}
where the $\max$ over distributions contributes to generating the worst-case data that lead the prediction sets to miscover the data. 
We say that a learner $L$ is $(\alpha, \ep)$-correct for $\Fs'$ and $T$ if
\begin{align}
  \Vs'(\Fs', T, \alpha, L) \le \ep. \label{eq:basecorrect}
\end{align}
The correctness of this base learner is used as a building block to address
the blockchain oracle problem.
\MR{
  In particular,
  the base learner is used to construct a base prediction set (with a correctness guarantee) on the output of an oracle smart contract,
  and multiple such base prediction sets are combined to derive consensus with quantified uncertainty on the outputs of oracle smart contracts, to address the oracle problem.
}



%% file: problem.tex
\vspace{-1ex}
\section{Prediction Consensus}
\label{s:prob}
\vspace{-1ex}

\begin{figure*}[t!]
  \centering
  \includegraphics[width=0.95\linewidth]{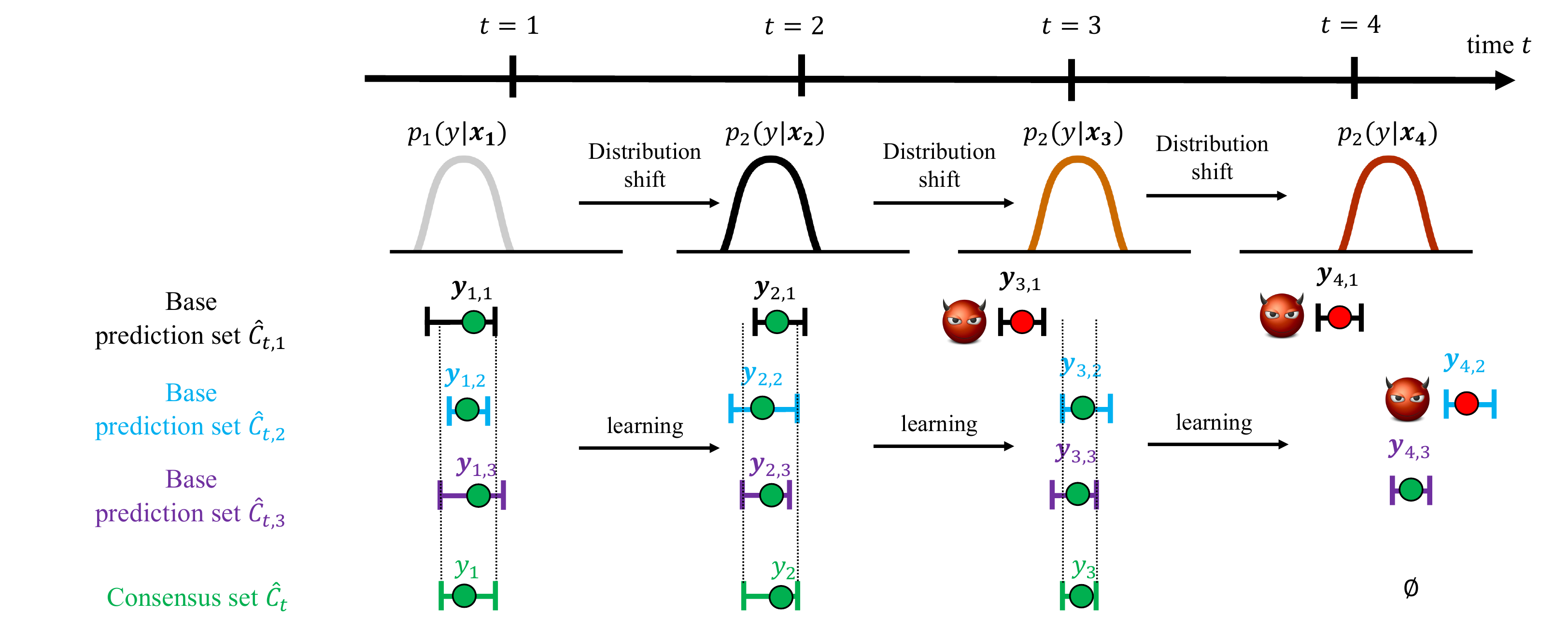}
  \caption{
    \MR{
      The summary of our approach for prediction consensus.
      Our goal is to learn consensus sets $\Ch_t$ that covers consensus labels $y_t$
      under distribution shift and Byzantine adversaries.
      We consider two steps to construct consensus sets:
      (1) construct base prediction sets $\Ch_{t, k}$ to handle distribution shift and
      (2) construct consensus sets from the base prediction sets to handle Byzantine adversaries.
      Here, under Assumption \ref{ass:mainassumption}, labels used to learn base prediction sets
      are considered to be consensus labels unless they are manipulated. Thus,
      consensus sets constructed using base prediction sets likely contain consensus labels.
      In particular, suppose $K=3$ and $k \in \{1, 2, 3\}$.
      At time $t=1$,
      each base prediction set $\Ch_{1,k}(\x_1)$ is learned to cover a label $y_{k,1}$.
      From these base prediction sets, we construct a consensus set $\Ch_1$ that contains labels,
      where each label is contained in the majority of base prediction sets,
      which includes an unknown consensus label $\y_1$.
      At time $t=2$,
      even under distribution shift,
      each base prediction set is learned 
      to cover a new label $\y_{k, 2}$ from a shifted distribution.
      At time $t=3$,
      one adversary arbitrarily manipulates the label $y_{3, 1}$ for the first prediction set
      (\ie $\beta=1$),
      so updating the base prediction set using the manipulated label $\y_{3, 1}$
      may not cover a future label.
      Even under this Byzantine adversary, the majority-based consensus set is constructed
      to cover an unknown consensus label $\y_3$.
      At time $t=4$,
      two adversaries manipulate the first and second prediction sets (\ie $\beta=2$).
      In this case, our consensus set does not conclude consensus among labels from base prediction sets;
      thus, it returns an empty set, which indicates the peculiarity of environments
      (\ie the violation of assumptions, like Assumption \ref{ass:mainassumption} or
      the majority base prediction sets are not manipulated).
      See Section \ref{s:approach} for details.
    }
  }
  \label{fig:approach}
\end{figure*}

We view the blockchain oracle problem as prediction consensus
under distribution shift and Byzantine adversaries, where
a consensus learner is derived based on multiple data sources.
Next, we consider setups on
the nature, an adversary, and a consensus learner,
\MR{followed by a problem statement with its connection to the oracle problem
in Section \ref{sec:formalprob}.}


\vspace{-1ex}
\subsection{Setup: Nature}

We have multiple sources, from which data are fed into a blockchain. 
In particular,
let
$K$ be the number of sources,
$\Xs_k$ be the example space of the $k$-th source,
and
$\Ys$ be the label space, shared by all sources.
Here, we consider $(\x_{t, k}, \y_{t, k}) \in \Xs_k \times \Ys$ as a datum from the $k$-th source at time $t \in \{1, \dots, T\}$, where \MRM{$T$ is a time horizon}.
Additionally,
let 
$\Xbs \coloneqq \Xs_1 \times \dots \times \Xs_K$ be the example space from all sources
and
$\Ybs \coloneqq \Ys^K$ be the label space from all sources,
where we consider 
$(\x_t, \y_t) \coloneqq (\x_{t, 1}, \dots, \x_{t, K}, \y_{t, 1}, \dots, \y_{t, K}) \in \Xbs \times \Ybs$
as a datum from all sources at time $t$.
Finally, let $y_t$ be a consensus label at time $t$.
Importantly, this consensus label is generally not observable
\MRM{(\eg the true price of ETH in USD is unknown)}.
\MR{
  We consider the following running example, which is used across this section.
  \begin{example}\label{ex:1}
  Consider three AMMs for ETH price sources in USD
  (\ie $\text{AMM}_1$, $\text{AMM}_2$, and $\text{AMM}_3$, where $K=3$).
  At time $t$, without ``arbitrage'' (which is explained later in this section),
  suppose the ETH price by $\text{AMM}_1$ is $1000$, where
  we denote all previous prices by $\x_{t, 1}$.
  Similarly,
  the ETH prices by $\text{AMM}_2$ and $\text{AMM}_3$ are $1001$ and $1200$, respectively.
  Here, without arbitrage,
  as transactions occur locally for each AMM,
  it is almost impossible to have the same price (\ie a consensus label)
  at the same time (\ie the consensus label is unknown).
  \end{example}
}

To address the prediction consensus problem,
we aggregate labels from the $K$ sources to estimate the consensus label.
However, as labels embed uncertainty, we need an uncertainty-aware consensus scheme.
To quantify the uncertainty,
we consider a set of distributions over $\Xbs \times \Ybs \times \Ys$, denoted by $\Ps$.
In particular, given a distribution $p_t \in \Ps$ at time $t$,
we have a datum $(\x_t, \y_t, y_t)$ sampled from $p_t$, where
we consider random variables as uppercase letters, \ie $(\X_t, \Y_t, Y_t) \sim p_t$.

As the consensus label $y_t$ is not generally observable, 
we connect it from the $K$ labels
based on the structural information from the following assumption on the consensus label and source labels;
we assume that the consensus label distribution is identical to
the label distribution of each source.
\begin{assumption}\label{ass:mainassumption}
  Assume $p_t(\y_{t, k} \mid \x_{t}) = p_t(y_t \mid \x_t)$ for any $t \in \{1, \dots, T\}$ and $k \in \{1, \dots, K\}$.
\end{assumption}
\MR{
  This assumption implies that labels obtained from each source are considered as consensus labels
  if the labels from each source are not manipulated by adversaries.
  In this case, learning consensus is not necessary.
  However, we consider a more practical setup where adversaries can manipulate labels from sources,
  where this assumption is used as a basis of the correctness of consensus, \ie\LREF{(\ref{eq:keylem:ass1})}{(17)}. 
}

A good example that justifies this assumption is the existence of \emph{arbitrage} in price markets, as follows.
\MR{
  \begin{example}\label{ex:2}
  \emph{Arbitrage} is process on buying one token from one market and selling the token to another market in order to take advantage of differing prices,
  and an \emph{arbitrageur} is an agent conducting arbitrage. 
  In our running example, the prices of ETH in $\text{AMM}_1$, $\text{AMM}_2$, and $\text{AMM}_3$ are
  $1000$, $1001$, and $1200$, respectively.
  An arbitrageur can benefit from buying ETH from $\text{AMM}_1$ (or $\text{AMM}_2$) with a low price
  and selling ETH to $\text{AMM}_3$ with a high price.
  This arbitrage continues until the arbitrageur cannot benefit from the price difference when
  the ETH price in $\text{AMM}_1$ (or $\text{AMM}_2$) increases
  and the ETH price in $\text{AMM}_3$ decreases
  due to the changing value of ETH for each market,
  thus resulting in the market prices reaching a similar price
  (\eg $\y_{t, 1}=1066$, $\y_{t, 2}=1070$, and $\y_{t, 3}=1068$).
  Here, in an ideal case
  (\eg there is no transaction fee),
  the two market prices reach to the same price, which is called the consensus price (\eg $y_t=1068$).
  \end{example} 
}

\MR{This example demonstrates that
arbitrage likely balances prices across markets,
which justifies Assumption \ref{ass:mainassumption},
\ie a consensus price $y_t$ tends to match to a local price $\y_{t, k}$.
}
Note that Assumption \ref{ass:mainassumption} can be handled or relaxed to cover \MR{more practical setups (\eg under a transient period when the assumption fails);
see Appendix\LREF{\ref{s:impl}}{C}~on the practical mitigation and }
see Section \ref{s:disc} on the discussion for a weaker assumption.

Additionally,
we assume the conditional independence of source labels $\y_{t, k}$ and a consensus label $y_t$
given $\x_t$.
\begin{assumption}\label{ass:condassumption}
  Assume $p_t(y_t \mid \x_t) \prod_{k=1}^K p_t(\y_{t, k} \mid \x_{t}) = p_t(\y_{t}, y_t \mid \x_{t})$
  for any $t \in \{1, \dots, T\}$ and $k \in \{1, \dots, K\}$.
\end{assumption}
\MR{
  Assumption \ref{ass:condassumption} implies
  $p_t(y_t \mid \y_t, \x_t) = p_t(y_t \mid \x_t)$
  and $p_t(\y_{t,k} \mid \y_{t, k+1}, \cdots,\y_{t, K}, \x_t) = p_t(\y_{t, k} \mid \x_t)$;
  the first equality means that a consensus label is independent of a label from each source given examples from all sources,
  and the second equality similarly means that
  a source label is independent of other source labels given examples from all sources.
}

\vspace{-1ex}
\subsection{Setup: Adversary}

We consider Byzantine adversaries. In particular,
let $e_t: \Xbs \to \Xbs$ be a \emph{Byzantine adversary} if it 
arbitrarily manipulates examples from arbitrarily chosen sources.
\MR{
  If it only manipulates \emph{at most} $\beta$ sources,
  we say that it is a \emph{$\beta$-Byzantine adversary}.
}
\begin{definition}[threat model]
  An adversary $e_t: \Xbs \to \Xbs$ at time $t$ is a $\beta$-Byzantine adversary if
  the adversary arbitrarily manipulates at most $\beta$ sources among $K$ sources.
\end{definition}
\MR{Here,
  we assume that only $\beta$, the upper bound of the number of manipulated sources, is known.
  However, we do not assume that
  we know sources chosen by the adversary
  and
  the number of actual adversaries to implement $e_t$.
  Knowing $\beta$ is seemingly strong, but
  our arguments are still valid by replacing $\beta$ with its estimate
  as long as $\beta$ is upper bounded by the estimation
  (see Section \ref{sec:consensussets} on the discussion for the choice of $\beta$).
  \begin{assumption}\label{ass:beta}
    $\beta$ is known.
  \end{assumption}
  \vspace{-4ex}
}

\MR{
  \begin{example}\label{ex:3}
    A $1$-Byzantine adversary can choose any one of three AMMs
  and change the ETH price by price manipulation at time $t$.
  If the adversary chooses $\text{AMM}_3$ for the manipulation by selling many ETH tokens,
  its ETH price will be decreased (\eg $\y_{t, 3} = 800$).
  Also, the adversary may not choose for manipulation even though it has the ability to do so.
  \end{example}
}

Note that the data distribution $p_t$ implicitly depends on $e_t$, \ie
$p_t(\x_t, \y_t, y_t) = p_t(\x_t) p_t(\y_t \mid e_t(\x_t)) p_t(y_t \mid \x_t)$,
where the adversary $e_t$ involves after $\x_t$ is drawn but does not affect the consensus label $y_t$.
We denote a set of all $\beta$-Byzantine adversaries by $\Es_\beta$.
In blockchains, price manipulation adversaries are considered $\beta$-Byzantine adversaries
in this paper. 

\subsection{Setup: Consensus Learner}\label{sec:prob:learner}
This paper considers a set-valued predictor to model prediction consensus.
In particular, let $\Ch_t: \Xbs \to 2^\Ys$ be a \emph{consensus set predictor},
where
it takes examples from $K$-sources at time $t$ to predict consensus labels as a set.
Here,
we denote the collection of consensus set predictors by $\Fs$.
Importantly, we consider the set-valued predictor instead of a point-estimator to explicitly model
the uncertainty of the prediction.

At time $t$, given all previous data $(\x_i, \y_i)$ and
consensus set predictors $\Ch_i$ for $1 \le i \le t-1$,
\ie $\z_{1:t-1} \coloneqq (\x_1, \dots, \x_{t-1}, \y_1, \dots, \y_{t-1}, \Ch_{1}, \dots, \Ch_{t-1})$,
we design a consensus learner $L: (\Xbs \times \Ybs \times \Fs)^* \to \Qs$ that returns
a distribution over \NMR{consensus set predictors} $\Fs$ for a future time step\footnote{$S^* \coloneqq \cup_{i=0}^\infty S^i$.}.
As we desire to design a learner $L$ that satisfies a correctness guarantee
even under distribution shift and Byzantine adversaries,
we consider the following correctness definition adopted from online machine learning \cite{sasha2012relax}.
In particular,
letting
$T$ be a time horizon,
\MRM{$Miscover$ be the miscoverage of a prediction set as in (\ref{eq:miscover})},
and
$\alpha$ be a desired miscoverage rate,
we denote the miscoverage value of the consensus learner $L$ under
distribution shift and $\beta$-Byzantine adversaries
by 
$\Vs(\Fs, T, \alpha, \beta, L)$, 
\begin{multline}\label{eq:mainvalue}
  \max_{\substack{p_1 \in \Ps \\ e_1 \in \Es_\beta}}
  \Exp
  \dots
  \max_{\substack{p_T \in \Ps \\ e_T \in \Es_\beta}}
  \Exp
  \frac{1}{T} \sum_{t=1}^T Miscover(\Ch_t, e_t(\X_t), Y_t)
  - \alpha.
\end{multline}
Here, the $t$-th expectation is taken over
$\X_{t} \sim p_t(\x)$,
$\Y_{t} \sim p_t(\y\mid e_t(\X_{t}))$,
$Y_t \sim p_t(y \mid \X_t)$,
and
$\Ch_t \sim L(\z_{1:t-1})$.
Intuitively,
the $\max$ over distributions $\Ps$ models the worst-case distribution shift,
and
the $\max$ over $\Es_\beta$ models the worst-case Byzantine adversaries.
Moreover, we consider that the consensus label is not affected by the adversaries, as represented in
$Y_t \sim p_t(y \mid \X_t)$, while
the source label can be affected by the adversaries $e_t$, as in $\Y_{t} \sim p_t(\y\mid e_t(\X_{t}))$.
Note that the learner $L$ cannot observe a consensus label $Y_t$, but
$Y_t$ is only used to evaluate the learner via the value $\Vs$.

Considering the value of the learner as a correctness criterion,
we aim to design the learner that is correct under distribution shift and Byzantine adversaries.
\begin{definition}[correctness]
  \label{def:correct}
  A consensus learner $L: (\Xbs \times \Ybs \times \Fs)^* \to \Qs$ is $(\alpha, \beta, \ep)$-correct for $\Fs$ and $T$ if we have
  \begin{align*}
    \Vs(\Fs, T, \alpha, \beta, L) \le \ep.
  \end{align*}
\end{definition}
\MR{
\begin{example}\label{ex:4}
  The ETH prices of $\text{AMM}_1$, $\text{AMM}_2$, and $\text{AMM}_3$ change over time (\ie distribution shift). Even worse, the $1$-Byzantine adversary manipulates prices across time (\ie local shift by Byzantine adversaries).
  Under these conditions and a desired miscoverage rate $\alpha=0.1$,
  our goal is to find price intervals that include consensus prices at least $100(1 - \alpha)\%$ (=$90\%$) of the time (assuming $\epsilon = 0$).
\end{example}
}

This correctness definition does not consider the size of the consensus set;
if a learner can return prediction sets that output the entire label set,
this learner is always correct, but its uncertainty measured by the set size is not informative.
So, we also consider minimizing the prediction set size $S(\Ch_t(e_t(\x_t)))$ at time $t$
based on some application-specific size metric $S: 2^\Ys \to \realnum_{\ge 0}$.
Note that the miscoverage value $\Vs'$ in (\ref{eq:baselearnervalue}) accounts for the size by
considering the absolute value of the difference of the miscoverage rate and a desired miscoverage;
this may require a scalar parameterization of a prediction set as in (\ref{eq:psdef}),
but we consider a general setup to cover various adaptive conformal predictors.

\vspace{-1ex}
\subsection{Problem}\label{sec:formalprob}
\vspace{-1ex}
In this paper, we view the blockchain oracle problem as a prediction consensus problem.
In particular, \MR{under previously mentioned setups},
for any given $\Fs$, $T$, $\ep$, $\alpha$, and $\beta$,
we find an $(\ep, \alpha, \beta)$-correct consensus learner while
minimizing the size of prediction sets across time.
The main challenges include
\ding{182},
\ding{183},
\ding{184},
\ding{185}, and 
\ding{186}.
\MR{
  In connection to the price oracle problem,
  the Nature setup models the behavior of price markets and
  the Adversary setup models the behavior of price manipulators.
  Each price oracle is viewed as a base prediction set predictor,
  which returns a price prediction with uncertainty (represented in a set),
  and the consensus learner exploits multiple price prediction sets to derive
  consensus over prices under distribution shift in price markets and Byzantine adversaries for price manipulation.
  See Figure \ref{fig:approach} for a summary on our approach in addressing the prediction consensus problem. 
}




%% file: approach.tex
\section{Adaptive Conformal Consensus}
\label{s:approach}
\vspace{-1ex}

We propose an \emph{adaptive conformal consensus} (\ACC) approach
for prediction consensus under distribution shift and Byzantine adversaries.
Intuitively, our approach aggregates votes on labels from
base prediction sets to form a consensus set, a set of labels if they are voted from at least $K - \beta$ base prediction sets. 
We provide the correctness guarantee on the consensus set, \ie
the consensus set probably contains the true consensus label
even under distribution shift and $\beta$-Byzantine adversaries,
where
this guarantee relies on the correctness guarantee of the base prediction sets.

\subsection{Consensus Sets} \label{sec:consensussets}
We define a consensus set predictor,
a set-valued function to handle uncertainty to address Challenge \ding{182}.
In particular, given base prediction sets $\Ch_{t, k}$ for $k \in \{1, \dots, K\}$
at time $t$, the consensus set $\Ch_t$ contains a label \MRM{if the label is included in at least} $K - \beta$ base prediction sets,
as follows
\begin{align}
  \Ch_{t}(\x_t) \coloneqq \left\{ y \in \Ys \vmid \sum_{k=1}^K \mathbbm{1}\( y \in \Ch_{t, k}(\x_{t}) \) \ge K - \beta \right\}.
  \label{eq:consensusset}
\end{align}

Here, we suppose that the base prediction sets $\Ch_{t, k}$ that satisfy correctness guarantees
are given at time $t$,
where
we introduce a way to construct the base prediction sets in Section \ref{sec:baseps}.
Moreover, we can view that the consensus set is based on a special conformity score
$\sum_{k=1}^K \mathbbm{1}\( y \in \Ch_{t, k}(\x_t) \)$, thus also denoting it by conformal consensus sets.
\MRM{Note that the uncertainty by consensus sets accounts for two sources of uncertainty: 
  the uncertainty of data-generating Nature via the base prediction sets and
  the uncertainty of Byzantine adversaries via the consensus set. 
}

\begin{figure}[tb!]
  \centering
  \newcommand\Rad{1.0cm}
  \subfigure[Non-adversarial]{
    \label{fig:consnonadv}
    \footnotesize
    \centering
    \scalebox{0.9}{
    \begin{tikzpicture} 
      \begin{scope}
        \clip (1,0) circle (\Rad);
        \clip (0.5,-1) circle (0.75*\Rad);
        \fill[gray!25](1,0) circle(\Rad);
      \end{scope}
      \begin{scope}
        \clip (1,0) circle (\Rad);
        \clip (1.5,-0.8) circle (0.7*\Rad);
        \fill[gray!25](1,0) circle(\Rad);
      \end{scope}
      \begin{scope}
        \clip (0.5,-1) circle (0.75*\Rad);
        \clip (1.5,-0.8) circle (0.7*\Rad);
        \fill[gray!25](0.5,-1) circle(0.75*\Rad);
        \fill[pattern=north east lines, pattern color=black](1,0) circle(\Rad);
      \end{scope}
      
      \node [draw, circle, black, minimum size=2*\Rad, align=center, label={above: \makecell{$\Ch_{t, 1}(\x_t)$}}] at (1.0,0)   {};
      \node [draw, circle, black, minimum size=1.5*\Rad, align=center, label={left: \makecell{$\Ch_{t, 2}(\x_t)$}}] at (0.5,-1)   {};
      \node [draw, circle, black, minimum size=1.4*\Rad, align=center, label={below: \makecell{$\Ch_{t, 3}(\x_t)$}}] at (1.5,-0.8) {};
      
    \end{tikzpicture}
    }
  }
  ~~~
  \subfigure[Adversarial]{
    \label{fig:consadv}
    \small
    \centering
    \footnotesize
    \scalebox{0.9}{
    \begin{tikzpicture}
      \begin{scope}
        \clip (1,0) circle (\Rad);
        \clip (0.5,-1) circle (0.75*\Rad);
        \fill[gray!25](1,0) circle(\Rad);
        \fill[pattern=north east lines, pattern color=black](1,0) circle(\Rad);        
      \end{scope}
      \begin{scope}
        \clip (1,0) circle (\Rad);
        \clip (2.0,-1.6) circle (0.7*\Rad);
        \fill[gray!25](1,0) circle(\Rad);
      \end{scope}
      \begin{scope}
        \clip (0.5,-1) circle (0.75*\Rad);
        \clip (2.0,-1.6) circle (0.7*\Rad);
        \fill[gray!25](0.5,-1) circle(0.75*\Rad);
      \end{scope}
      
      \node [draw, circle, black, minimum size=2*\Rad, align=center, label={above: \makecell{$\Ch_{t, 1}(\x_t)$}}] at (1.0,0)   {};
      \node [draw, circle, black, minimum size=1.5*\Rad, align=center, label={left: \makecell{$\Ch_{t, 2}(\x_t)$}}] at (0.5,-1)   {};
      \node [draw, circle, red, minimum size=1.0*\Rad, align=center, label={below: \makecell{$\Ch_{t, 3}(e_t(\x_t))$}}] at (1.8,-1.4) {};      
    \end{tikzpicture}
    }
  }
  \caption{
    Consensus sets at time $t$, where $K=3$ and $\beta=1$.
    The consensus set is a set of labels voted by $K-\beta$ prediction sets as highlighted in gray.
    Without adversarial manipulation as in Figure \ref{fig:consnonadv},
    a consensus label is likely contained in the consensus set
    due to the correctness of base prediction sets;
    under manipulation as in Figure \ref{fig:consadv},
    the consensus label is still likely contained in the consensus set
    as the majority of base prediction sets are still probably correct.
    See Section \ref{sec:consensussets} for details.
  }
\label{fig:approachsummary}
\end{figure}
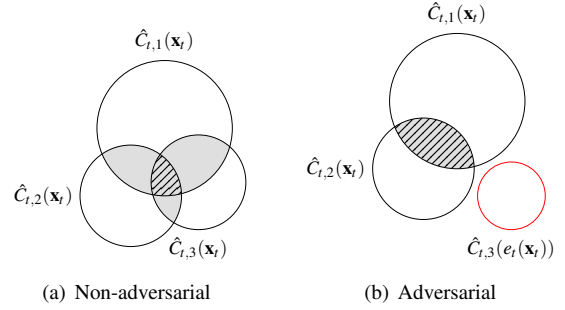

\para{Intuition on correctness from a special case.}
\MR{
To provide intuition on the correctness of a consensus set (\ref{eq:consensusset}),
suppose a special case where $T \to \infty$ and a Byzantine adversary $e$ is fixed but unknown.
In particular, Figure \ref{fig:approachsummary} illustrates base prediction sets (in circles)
along with consensus sets (in gray areas), where $K=3$ and $\beta = 1$.
Consider Figure \ref{fig:consnonadv} where the adversary does not manipulate any base prediction sets.
Under Assumption \ref{ass:mainassumption}, if each base prediction set contains the consensus label with probability $1 - \alpha_k$ for $k \in \{1, 2, 3\}$, where $\alpha_k$ is usually tiny,
the striped area contains the consensus label with probability at least $1 - \sum_{k=1}^3\alpha_k$
due to a union bound.
This means that the consensus set in the gray area also contains the consensus label with probability at least $1 - \sum_{k=1}^3\alpha_k$.
This intuition still holds when an adversary manipulates a prediction set as in Figure \ref{fig:consadv}.
As before, let each base prediction set contains the consensus label with probability $1 - \alpha_k$, but
the third base prediction set (in red) is manipulated.
Due to the union bound, the striped area contain
the consensus label with probability at least $1 - \sum_{k=1}^2\alpha_k$;
thus, the consensus set in gray also contains the consensus label with the same probability.
Here, we assume that we know the third base prediction set is manipulated, but
this is unknown in general so we cannot know the correctness probability $1 - \sum_{k=1}^2\alpha_k$.
Thus, we consider a more conservative probability $1 - \sum_{k=1}^3 \alpha_k$
as the probability of the consensus set for including the consensus label,
since this does not require the knowledge of manipulated prediction sets.
See Appendix\LREF{\ref{apdx:specialcaseanalysis}}{D.1}~for a rigorous proof on this special case
and 
Theorem \ref{thm:consensuscorrectness} for the correctness guarantee in a general case.
}

\MR{
Note that \cite{cherubin2019majority} constructs a majority vote prediction set, similar to (\ref{eq:consensusset}), under a non-adversarial setup (\ie $\beta=0$) and stronger assumptions on data than ours;
it assumes the exchangeability on samples from each source (along with other ensemble approaches \cite{balasubramanian2015conformal,toccaceli2017combination,linusson2020efficient}),
which implies that a data distribution is not changing,
but we consider a general assumption (\ie a distribution is changing over time).
}

\para{Use case in DeFi.}
\MR{
  In DeFi, a consensus set is an interval over prices, instead of a point price value.
  But, the price interval is still compatible with existing DeFi applications, like loaning services.
  In particular, a loaning service can use a conservative price (\ie the smallest price in the interval) as a price reference, while using the interval size as a measure of uncertainty. 
}


\begin{algorithm}[th!]
  \caption{Adaptive Conformal Consensus (\ACC)}
  \label{alg:acc}
  \small
  \begin{algorithmic}[1]

    \For{$t = 1, \dots, T$}
    \State Observe $\x_t$
    \State Construct a consensus set $\Ch_t(\x_t)$ via (\ref{eq:consensusset})
    \For{$k = 1, \dots, K$}
    \State Observe $\y_{t, k}$
    \State $\Ch_{t+1, k} \gets \text{update}(\Ch_{t, k}, \x_t, \y_{t, k})$ \label{alg:acc:update}
    \EndFor
    \EndFor
  \end{algorithmic}
\end{algorithm}

\para{Algorithm.}
To construct a consensus set, we propose a meta algorithm, which uses
base prediction sets constructed from $K$ sources.
\MRM{In particular, for each time $t$,
the algorithm observes $K$ base prediction sets of $\x_t$ from the $K$ sources
to construct a consensus set via (\ref{eq:consensusset})}.
Once each base prediction set observes a label $\y_{t, k}$,
it is updated via a base prediction set algorithm.
The proposed meta algorithm is described in Algorithm \ref{alg:acc}, which we denote by
\emph{Adaptive Conformal Consensus} (\ACC).
An update function $\text{update}(\cdot)$ in Line \ref{alg:acc:update}
consists of two update functions, \ie
the update on $\Ch_{t, k}$ by an adaptive conformal prediction, denoted by $\text{update}_{\text{ACP}}(\cdot)$,
and the update on the score function of $\Ch_{t, k}$ by a conventional online learning algorithm, denoted by $\text{update}_{\text{score}}(\cdot)$.
Here, we consider the following update, but the order of update is not critical:
$\text{update}(\Ch, \x, y) = \text{update}_{\text{score}}(\text{update}_{\text{ACP}}(\Ch, \x, y), \x, y)$.
Note that enumerating over $\Ys$ to construct the consensus set is not trivial
depending on the score function if $\Ys$ is continuous;
see Appendix\LREF{\ref{s:impl}}{C}~for details.

\para{Theory.}
The consensus set constructed by Algorithm \ref{alg:acc} is correct
under Byzantine adversaries and distribution shift (along with other assumptions in Section \ref{s:prob}).
In particular, to measure the correctness, 
we first consider the miscoverage value of each base learner $L_k$ for the $k$-th source.
Specifically,
at time $t \in \{1, \dots, T\}$,
the $k$-th source can be manipulated by a $\beta$-Byzantine adversary $e_t$; thus,
an example $\X_t$ and the corresponding label $\Y_{t, k}$ are manipulated.
The learner $L_k: (\Xbs \times \Ys \times \Fs_k )^* \to \Qs_k$ uses
observed normal or manipulated labeled examples $(\x_1', \y_{1, k}), \dots, (\x_{t-1}', \y_{t-1, k})$
along with the previous conformal predictors $\Ch_{1, k}, \dots, \Ch_{t-1, k}$
to find a distribution over conformal predictors such that the predictor $\Ch_{t, k}$ drawn from this distribution achieves a desired miscoverage rate $\alpha_k$.
Similar to the miscoverage value of the consensus learner in (\ref{eq:mainvalue}),
we define the miscoverage value $\Vs_k(\Fs_k, T, \alpha_k, \beta, L_k)$ of each base learner as follows:
\begin{multline*}
  \max_{\substack{p_1 \in \Ps \\ e_1 \in \Es_\beta}}
  \Exp
  \dots
  \max_{\substack{p_T \in \Ps \\ e_T \in \Es_\beta}}
  \Exp
  \frac{1}{T} \sum_{t=1}^T Miscover(\Ch_{t, k}, e_t(\X_t), \Y_{t, k})
  - \alpha_k,
\end{multline*}
where the $t$-th expectation is taken over
$\X_{t} \sim p_t(\x)\!$,
$\Y_{t, k} \sim p_t(y \mid e_t(\X_{t}))$,
and
$\Ch_{t, k} \sim L_k(\z_{1:t-1})$.
Suppose that the miscoverage value of the base learner is bounded by $\ep_{T, k}$, \ie
\begin{align*}
  \Vs_k(\Fs_k, T, \alpha_k, \beta, L_k) \le \ep_{T, k}.
\end{align*}
Then, we prove that the miscoverage value of a consensus set constructed by Algorithm \ref{alg:acc} is
bounded
even under distribution shift and $\beta$-Byzantine adversaries, suggesting that
the consensus set eventually achieves a desired miscoverage rate;
see Appendix\LREF{\ref{apdx:proof:them:consensuscorrectness}}{D.5}~for a proof.
\vspace{-1ex}
\begin{theorem}
  \label{thm:consensuscorrectness}
  Under Assumption \ref{ass:mainassumption}, \ref{ass:condassumption}, and \ref{ass:beta},
  a consensus learner $L$ satisfies
  \vspace{-1ex}
  \begin{align*}
    \Vs\(\Fs, T, \sum_{k=1}^K\alpha_k, \beta, L\) \le \sum_{k=1}^K \ep_{T, k}
  \end{align*}
  if
  a base learner $L_{k}$ for any $k \in \{1, \dots, K\}$ satisfies
  \begin{align*}
    \Vs_k(\Fs_k, T, \alpha_k, \beta, L_k) \le \ep_{T, k}.
  \end{align*}
\end{theorem}
Intuitively, the miscoverage value of the proposed consensus learner is bounded by
the sum of the miscoverage value bounds of its base learners.
This means that if the value of each base learner is bounded by a decreasing function,
the consensus learner does as well,
implying it eventually converges to a desired miscoverage $\sum_{k=1}^K \alpha_k$,
under distribution shift and $\beta$-Byzantine adversaries.
This theorem proves that the proposed consensus learner is
$\left(\sum_{k=1}^K \alpha_k, \beta, \sum_{k=1}^K \ep_{T, k} \right)$-correct for $\Fs$ and $T$;
thus, this addresses Challenge \ding{183}, \ding{184}, and \ding{185}.

Theorem \ref{thm:consensuscorrectness} requires that
the bound of each base learner's miscoverage value $\Vs_k$ are known.
Here, we connect this value to known miscoverage values $\Vs'$
in (\ref{eq:baselearnervalue}) of adaptive conformal prediction.
In particular, 
if the miscoverage value of adaptive conformal prediction is bounded,
our miscoverage value of a base learner under Byzantine adversaries is also bounded
using the same algorithm; see Appendix\LREF{\ref{apdx:proof:lem:valueconnection}}{D.6}~for a proof.

\begin{lemma}\label{lem:valueconnection}
  We have
  $\Vs_k(\Fs_k, T, \alpha_k, \beta, L_k) \le \Vs'(\Fs_k, T, \alpha_k, L_k)$.
\end{lemma}

\para{On the choice of $\beta$.}
Importantly, Theorem \ref{thm:consensuscorrectness} holds for any $\beta$,
suggesting that
the correctness does not depend on the parameter of the Byzantine adversary.
\MR{However, $\beta$ is unknown in practice
  though we need this in Algorithm \ref{alg:acc} and Theorem \ref{thm:consensuscorrectness}.
  To this end, we replace $\beta$ by its estimate $\hat{\beta}$, and
  as long as $\beta \le \hat\beta$,
  the correctness by the theorem holds with $\hat{\beta}$ since
  a $\hat\beta$-Byzantine adversary is a $\beta$-Byzantine adversary.
  We can have $\hat\beta = K$ as $\beta \le \hat\beta$ always holds;
  however, this produces trivially large prediction sets due to (\ref{eq:consensusset}).
  Alternatively,
  we can have $\hat\beta = 0$, but it always violates $\beta \le \hat\beta$ if $\beta \ge 1$.
  Thus, we use $\hat\beta = \lfloor \frac{K}{2} \rfloor$
  as this produces a reasonably small prediction set,
  while satisfying the correctness guarantee under a reasonable assumption
  (\ie the majority of sources are not manipulated). 
}
\MR{Note that $\lfloor \frac{K}{3} \rfloor$ condition for the correctness guarantee by
Theorem 1 in \cite{lamport1982byzantine} is considered in a different setup, \ie
consensus among all participants without uncertainty,
but we consider the consensus by one participant under uncertainty.}

\vspace{-1ex}
\para{On the choice of $\alpha$ and $K$.}
\MR{
  We provide an implication of Theorem \ref{thm:consensuscorrectness} on varying $K$ given $\alpha$.
  The desired miscoverage rate $\alpha$ is a user-specified parameter
  (\eg $\alpha=0.01$ means that the generated consensus sets do not cover consensus labels at most $1\%$ of the time).
  The number of sources $K$ is decided by possible data sources (\eg the number of markets where we can exchange ETH and USD).
  Given $\alpha$, $K$ can be increased to enhance the correctness of consensus sets from strong Byzantine adversaries.
  In this case, $\alpha_k$ needs to be decreased to achieve the desired miscoverate rate $\alpha$.
  In particular, Theorem \ref{thm:consensuscorrectness} implies that the miscoverate rate of consensus sets is bounded by $\sum_{k=1}^K \alpha_k$. But, we want $\sum_{k=1}^K \alpha_k = \alpha$, and
  in this paper, we distribute $\alpha$ ``budget'' equally to each source, \ie $\alpha_k = \alpha / K$.
  This implies that, given $\alpha$, if $K$ increases,
  $\alpha_k$ needs to be reduced to satisfy the equality.
}


\vspace{-1ex}
\para{Implications.}
We highlight the implication of Byzantine robustness.
In particular,
at time $t$, if less than $K$ observations are made (\eg by the failure of some base prediction sets),
we can consider the missing observation as the result of the Byzantine adversaries;
thus, the correctness guarantee in Theorem \ref{thm:consensuscorrectness} still holds.
Interestingly, in the context of blockchains,
if all sources of base prediction sets are from on-chain information
(\eg base prediction sets from AMMs),
we can enjoy the lower-level consensus mechanism (\eg proof-of-work) to have
reliable base prediction sets.

\vspace{-1ex}
\subsection{Base Prediction Sets} \label{sec:baseps}
\vspace{-1ex}
The proposed adaptive conformal consensus is used with
\emph{any} $(\alpha, \ep)$-correct adaptive conformal prediction for constructing base prediction sets.
The following includes possible options for blockchain applications,
where computational efficiency is one of critical measures.

\subsubsection{A Score Function for Regression}\label{sec:scorefunc}
The conformal prediction is generally used with any application-dependent score function.
Considering that our main application in blockchains is price regression,
we propose to use the Kalman filter \cite{kalman1961new} as a score function.
In particular, 
the Kalman filter is used to estimate states based on the sequence of observations.
Here, we use it to estimate a price based on the previous price data.
In the Kalman filter,
we consider a price datum $\y_{t, k}$ as an observation from the $k$-th source at time $t$,
from which we estimate the price state.
To this end, we have the identity matrices as an observation model and state-transition model.
Moreover,
let $w_{t, k}^2$ be the variance of zero-mean Gaussian state noise, and
$v_{t, k}^2$ be the variance of zero-mean Gaussian observation noise.

\para{Prediction.}
At time $t$ before observing the price $\y_{t, k}$,
the Kalman filter predicts the distribution over observations
given the previous data $\y_{1, k}, \dots, \y_{t-1, k}$ as follows:
\begin{align*}
  \bar{s}_{t, k}(\x_t, \y_{t, k}) \coloneqq \Ns( \y_{t, k}; \mu_{t-1, k}, \sigma_{t-1, k}^2 + w_{t-1, k}^2 + v_{t-1, k}^2 ),
\end{align*}
where
\MRM{$\Ns(y; \mu, \sigma^2)$
denotes the Gaussian probability density function (PDF)} at $y$ with the mean of $\mu$ and the variance of $\sigma^2$,
and
$\mu_{t-1, k}$ and $\sigma_{t-1, k}^2$ are estimated from $\y_{1, k}, \dots, \y_{t-1, k}$
via the Kalman prediction.
This prediction over observations is used to define the score function $s_{t, k}$.
In particular,
let $\x_{t} \coloneqq (\y_{1, k}, \dots, \y_{t-1, k})$ be
an example from the $k$-th source, a list of price data up to time $t-1$.
Then, we define the score function $s_{t, k}$ by the \MR{scaled} Gaussian distribution over observations, which
measures how likely $\y_{t, k}$ will be observed at time $t$, \ie
\begin{align}
  s_{t, k} (\x_t, \y_{t, k}) \coloneqq
  \bar{s}_{t, k}(\x_t, \y_{t,k}) / \MR{(2s^{\text{max}}_{t, k})}, \label{eq:kfnormscore}
\end{align}
where
\MR{$s^{\text{max}}_{t, k}$ is the maximum of $\bar{s}_{t, k}$
  \ie $\bar{s}_{t, k}(\x_t, \mu_{t-1, k})$.
Here, we consider the scaled Gaussian as this normalization reduces the effort in tuning hyperparameters; see Appendix\LREF{\ref{s:impl}}{C}~for details.
}
Note that we assume that the $k$-th source only uses $\x_{t, k}$,
but this can be used with the entire $\x_t$ from all sources.

\para{Update.}
After observing $\y_{t, k}$, the score function $s_{t, k}$ is updated in two ways:
noise update and Kalman update.
For the noise update, we consider a gradient descent method;
see Appendix\LREF{\ref{s:impl}}{C}~for details.
For the state update, via the Kalman update,
the state Gaussian PDF $\Ns ( \y_{t, k}; \mu_{t, k}, \sigma_{t, k}^2 )$ is
updated as follows:
\begin{align*}
  \text{update}_{\text{Kalman}}(\Ch, \x_t, \y_{t, k}) :~~
  \mu_{t, k} &\gets \mu_{t-1, k} - K_t \( \y_{t, k} - \mu_{t-1, k} \) \\
  \sigma_{t, k}^2 &\gets (1 - K_t) ( \sigma_{t-1, k}^2 + w_{t, k}^2 ),
\end{align*}
where $K_t = (\sigma_{t-1, k}^2 + w_{t, k}^2) / (\sigma_{t-1, k}^2 + w_{t, k}^2 + v_{t, k}^2)$.
Based on the two update functions, we define the update for the score function as follows:
$\text{update}_{\text{score}}(\Ch, \x, y) = \text{update}_{\text{Kalman}}(\text{update}_{\text{noise}}(\Ch, \x, y), \x, y)$.

\para{Pros and cons.}
Sequential data can be modeled via non-linear filtering approaches
(\eg extended Kalman filtering or particle filtering) or
recurrent neural networks.
Compared to these, the Kalman filter is computationally light,
preferred in blockchains.
The downside of the Kalman filter is its strong Gaussian assumption.
\MR{
However, this assumption does not affect the correctness of base prediction sets as
the correctness holds for any score function,
which is a property of conformal prediction 
and also our choice of an adaptive conformal predictor in Lemma \ref{lem:mvpbound}.
Note that the assumption on a score function (\eg the Gaussian assumption)
affects the size of prediction sets
like any conformal prediction method by the definition of a prediction set (\ref{eq:psdef}).
}

\subsubsection{Adaptive Conformal Prediction}

Given
a score function $s_{t, k}$
and
a labeled example $(\x_{t}, \y_{t, k})$
at time $t$,
the $k$-th base prediction set is updated via 
adaptive conformal prediction.
Here, we use multi-valid conformal prediction (MVP) \cite{bastani2022practical},
where we consider a special variation of MVP (\ie MVP with a single group).

\para{Prediction.}
MVP considers that the prediction set parameter $\tau_{t, k}$ is roughly quantized via binning.
Let
$m$ be the number of bins,
$\tau_{\text{max}}$ is the maximum value of $\tau_{t, k}$, 
$B_i = \big[ \tau_{\text{max}}\frac{i-1}{m}, \tau_{\text{max}}\frac{i}{m} \big)$ be the $i$-th bin,
and
$B_m = \[ \tau_{\text{max}}\frac{m-1}{m}, \tau_{\text{max}} \]$ be the last bin.
We consider a prediction set parameterized by $\tau_{t, k}$, which falls in the one of these bins, \ie
$\Ch_{t, k}(\x_t) \coloneqq \left\{ y \in \Ys \mid s_{t, k}(\x_t, y) \ge \tau_{t, k} \right\}$.
Moreover, we consider that $\Ch_{t, k}$ representation includes
internal states $n_{t, k} \in \realnum^m$ and $v_{t, k} \in \realnum^m$,
initialized zero and 
updated as the algorithm observes labeled examples $(\x_t, \y_{t, k})$.
In prediction, we construct $\Ch_{t, k}(\x_t)$ as an input for the consensus set.
Here, constructing $\Ch_{t, k}(\x_t)$ over $\Ys$ is not trivial if $\Ys$ is continuous;
see Appendix\LREF{\ref{s:impl}}{C}~for details.

\para{Update.}
Algorithm\LREF{\ref{alg:smvp}}{2}~updates
$n_{t, k}$ and $v_{t, k}$, and
compute $\tau_{t, k}$ to update $\Ch_{t, k}$ given learning rate $\eta$ \MRM{on the change of the scalar parameter $\tau_{t, k}$}.
We use $\text{update}_{\text{MVP}}$ for $\text{update}_\text{ACP}$.

\para{Correctness.}
The MVP learner by Algorithm\LREF{\ref{alg:smvp}}{2}~is $(\ep, \alpha)$-correct for a set of quantized thresholds and $T$.
In particular, 
the correctness bound of the MVP in \cite{bastani2022practical} is proved for a multi-valid guarantee.
Here, we explicitly connect that the MVP bound is used to bound the miscoverage value of the MVP learner as follows (see Appendix\LREF{\ref{apdx:proof:lem:mvpbound}}{D.7}~for a proof):
\begin{lemma}
  \label{lem:mvpbound}
  Letting
  $S_i = \{ t \in \{1, \dots, T\} \mid \tau_{t, k} \in B_i \}$ and
  $f(n) \coloneqq \sqrt{(n+1) \log_2^2 (n+2)}$,
  the MVP learner $L_{\text{MVP}}$ satisfies the following \MR{for any score function $s_t$}:
  \begin{align*}
    \Vs'(\Fs, T, \alpha, L_{\text{MVP}}) \le
    \sum_{i=1}^m
    \frac{f(|S_i|)}{|S_i|} \sqrt{13.6 m \ln m}
  \end{align*}
  if a distribution over scores $s_t(X_t, Y_t)$ for any $t \in \{1, \dots, T\}$ is smooth enough
  \footnote{In \cite{bastani2022practical}
  the smoothness is parameterized by $\rho$ and
  we assume $\rho \to 0$.
  }
  and
  $\sqrt{\frac{\ln m}{6.8 m}} \le \eta \le \sqrt{\frac{\ln m}{6.6 m}}$.  
\end{lemma}

%% file: eval.tex
\section{Evaluation}
\label{s:eval}

\begin{figure*}[th!]
  \centering
  \subfigure[A single data source ($K=1$)]{
    \centering
    \includegraphics[width=0.3\linewidth]{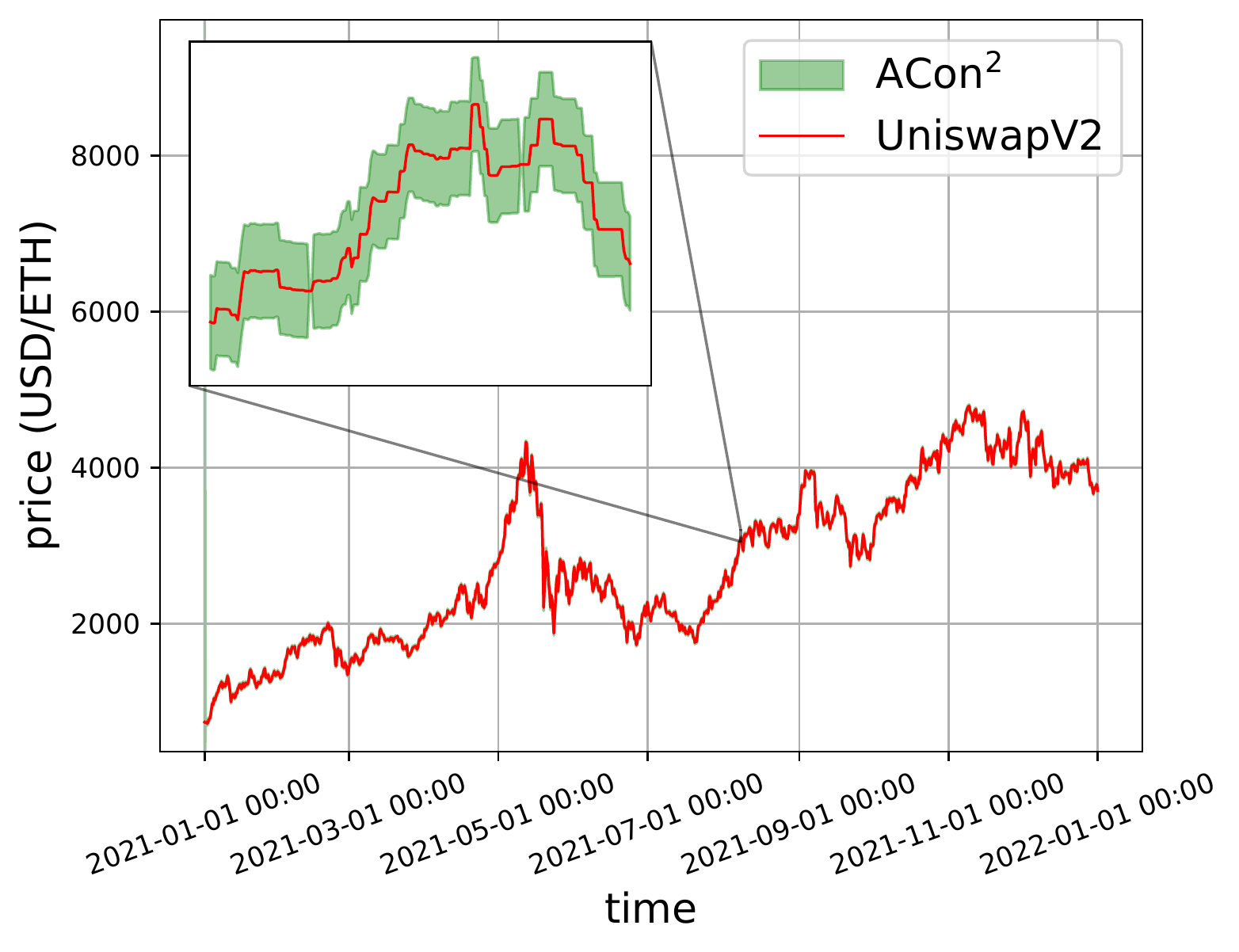}
  }
  \subfigure[Two data sources ($K=2$)]{
    \centering
    \includegraphics[width=0.3\linewidth]{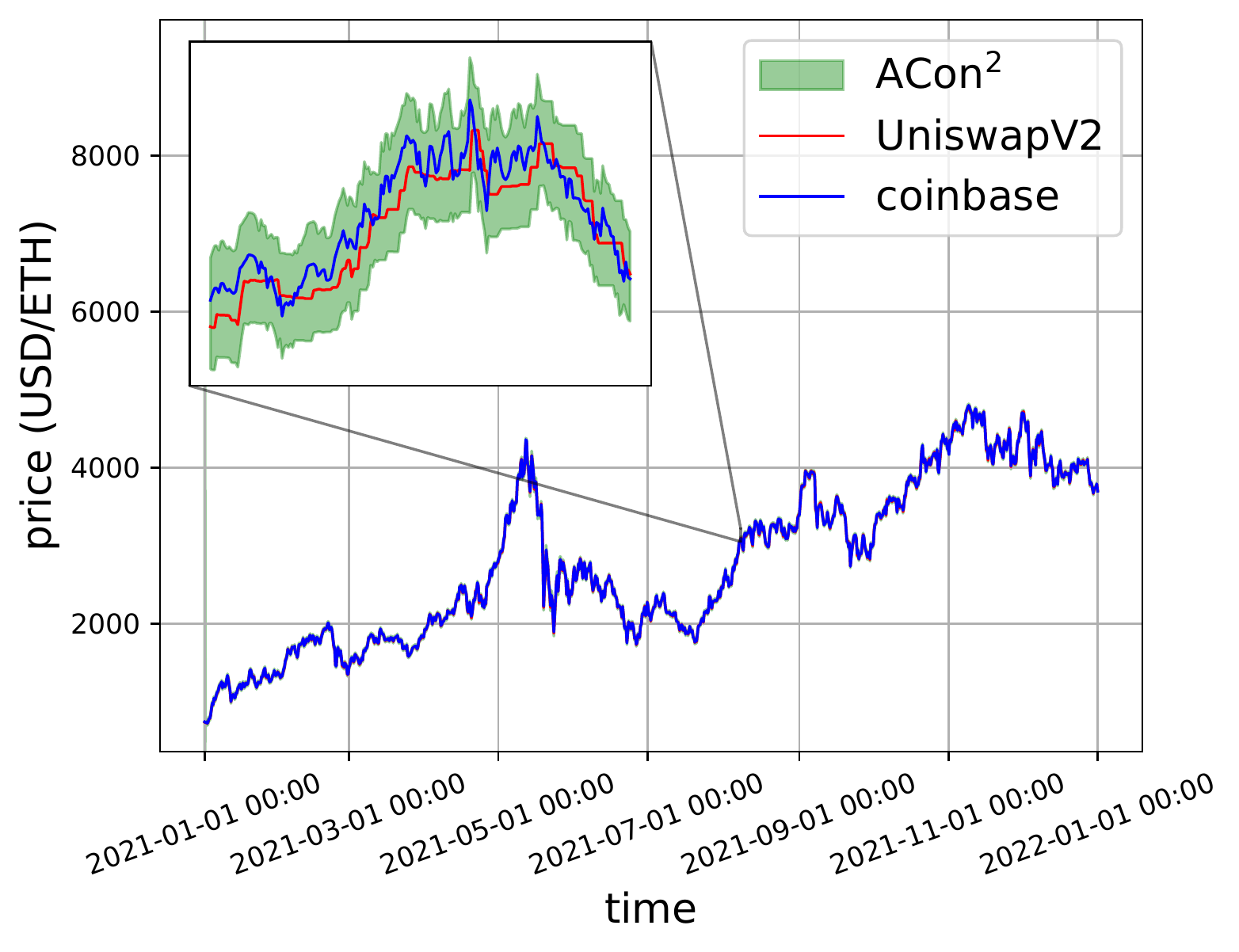}
  }
  \subfigure[Three data sources ($K=3$)]{
    \centering
    \includegraphics[width=0.3\linewidth]{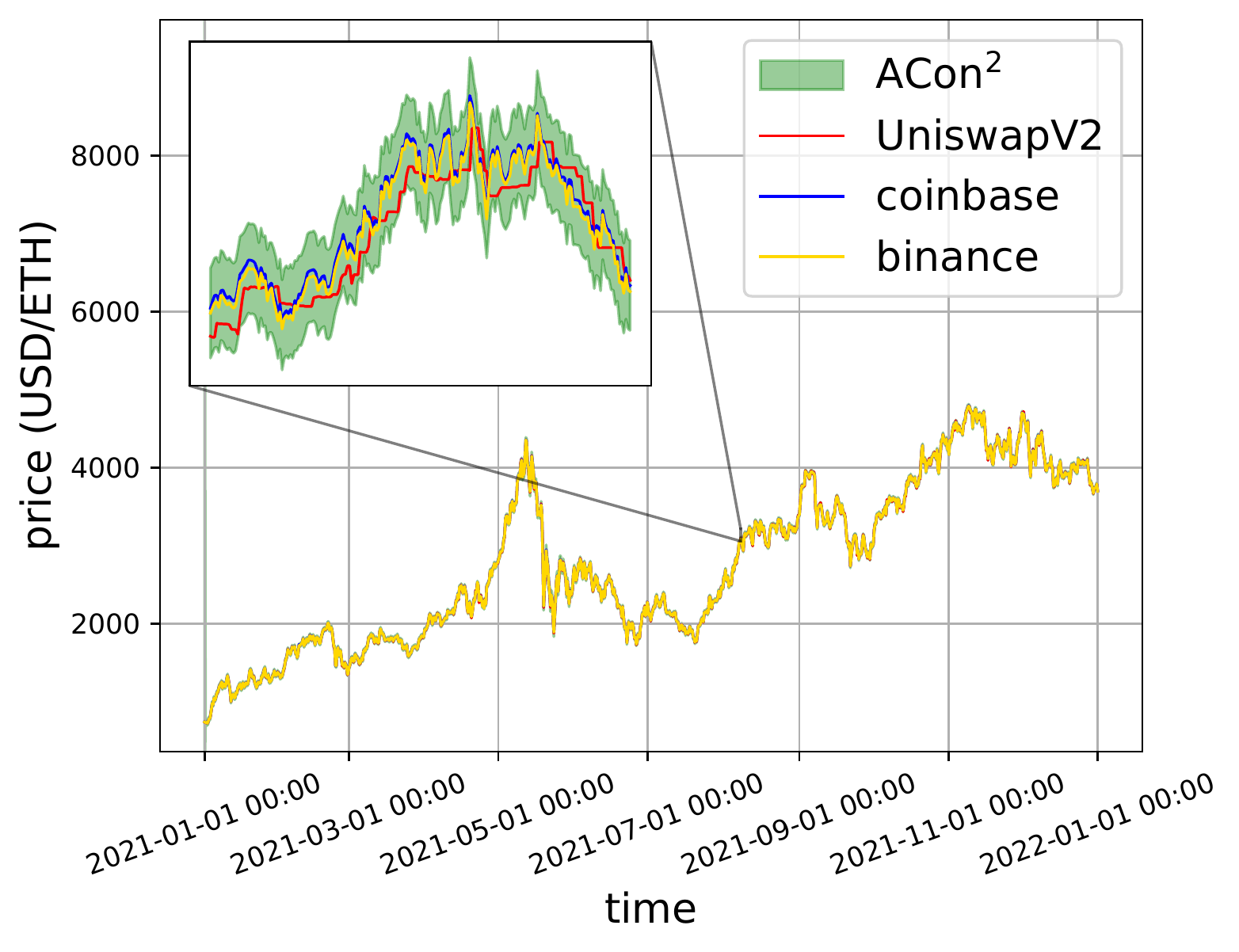}
    \label{fig:global:overall:K3}
  }
  \caption{
    Prediction consensus results on USD/ETH data.
    The \ACC \MR{(represented in a green area)} of each figure uses a different number of sources $K$ \MR{(where source data are represented in colored solid lines)}, as denoted in captions.
    For any number of sources with $\beta = \lfloor \frac{K}{2} \rfloor$,
    the \ACC finds intervals that closely follow the price of each source,
    demonstrating that the correctness of the consensus sets by \ACC under distribution shift,
    while maintaining relatively small consensus set size
    (Challenge \ding{183} and \ding{185}).
    See Figure \ref{fig:global:miscoverage} and \ref{fig:global:size} for additional details.
  }
  \label{fig:global:overall}
\end{figure*}

We evaluate the efficacy of the proposed \ACC based on
two price datasets,
which manifest global distribution shift (for Challenge \ding{182}, \ding{183} and \ding{185}) and
local shift due to Byzantine adversaries (for Challenge \ding{182}, \ding{184}, and \ding{185}),
and
one case study over the Ethereum blockchain (for Challenge \ding{186}).
Note that we mainly focus on price oracles as it is the current major issue of the oracle problem,
but the proposed approach can be applicable in more general setups.
See Appendix\LREF{\ref{s:impl}}{C}~for implementation details.


\para{Metric.}
For evaluation metrics of approaches,
we use a miscoverage rate of consensus sets and
a size distribution of consensus sets.
In particular,
the miscoverage rate is $\frac{1}{T}\sum_{t=1}^T \text{Miscover}(\Ch_t, \x_t, y_t)$,
as in (\ref{eq:mainvalue}),
given a sequence of $(\x_t, y_t)$ for $1 \le t \le T$ at a time horizon $T$.
Here, the consensus label $y_t$ is unknown, so
for price data,
we choose the median of $\y_{t, k}$ for $1 \le k \le K$
as pseudo-consensus labels only for evaluation purposes, and
\NMR{we call a miscoverage rate a \emph{pseudo-miscoverage} rate, to highlight this.}
The size of a consensus set is measured by a metric $S: 2^\Ys \to \realnum_{\ge 0}$;
for the price data, as the consensus set is an interval, we measure the length of the interval.

\para{Baselines.}
We consider three baselines;
a \emph{median baseline} is the median value of values from multiple sources,
and
a \emph{$\text{TWAP}_{\text{Keep3rV2}}$ baseline} is the
time-weighted average price (TWAP) implemented in Keep3rV2 \cite{keep3rv2}.
These two baselines are not directly comparable to our approach, as
it does not quantify uncertainty.
But, they are considered to be effective solutions for price manipulation,
so we use them in local shift experiments to show their efficacy in price manipulation.

Additionally, we consider a baseline that considers uncertainty.
In particular, for the base prediction set,
we use one standard deviation prediction set $\sigma$-BPS,
which returns all labels deviated from the mean by $\sigma$, \ie
$\Ch_{t, k}(\x_t) \coloneqq [\mu - \sigma, \mu + \sigma]$, where
$\mu$ and $\sigma$ are the mean and standard deviation of a score $s_{t, k}(\x_t, \y_{t, k})$.
This base prediction set is used with \ACC for the baseline, denoted by $\sigma$-\ACC.

\subsection{Global Shift: USD/ETH Price Data}

\begin{figure*}[th!]
  \centering
  \subfigure[A single data source ($K=1$)]{
    \centering
    \includegraphics[width=0.3\linewidth]{{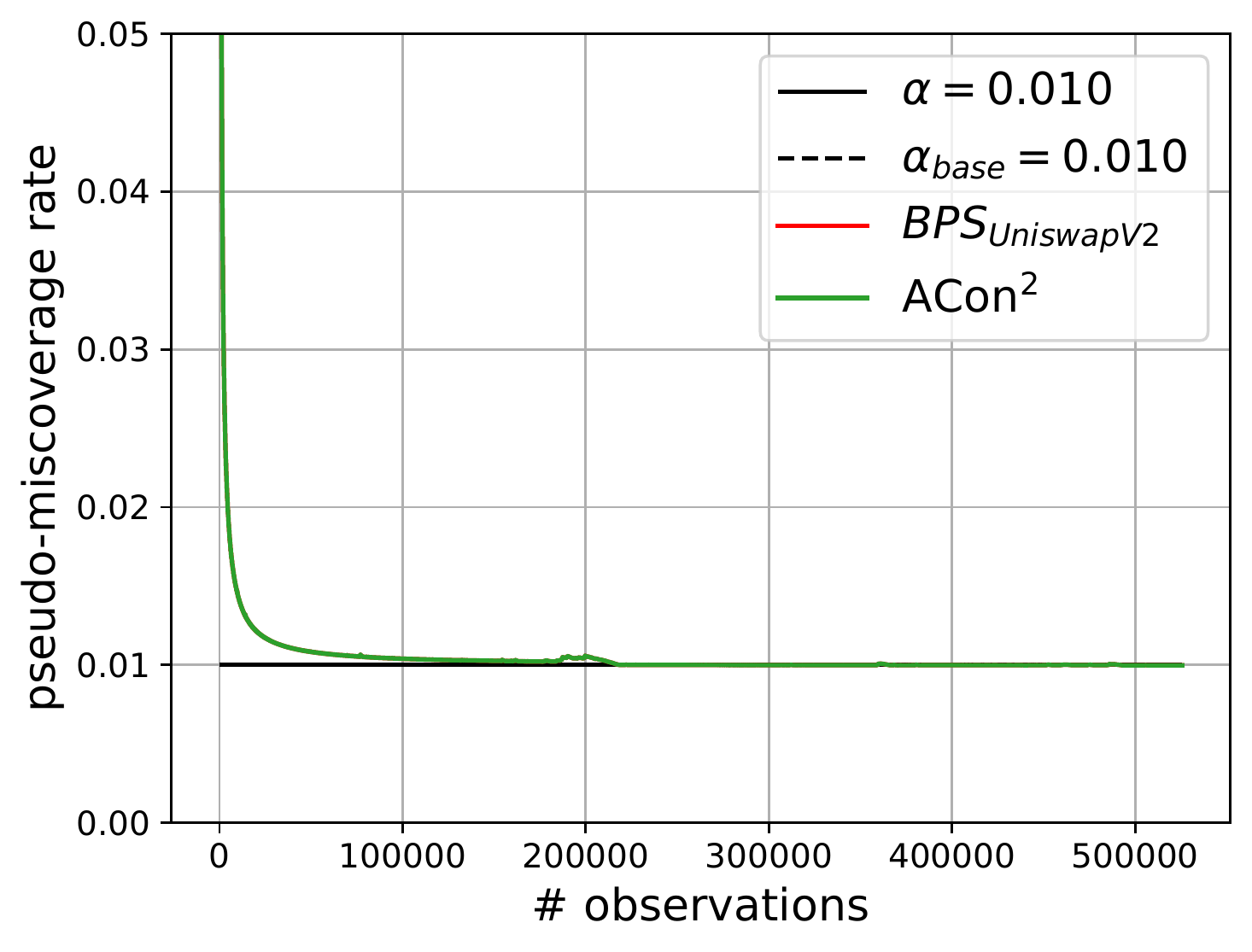}}
    \label{fig:global:miscoverate:K1}
  }
  \subfigure[Two data sources ($K=2$)]{
    \centering
    \includegraphics[width=0.3\linewidth]{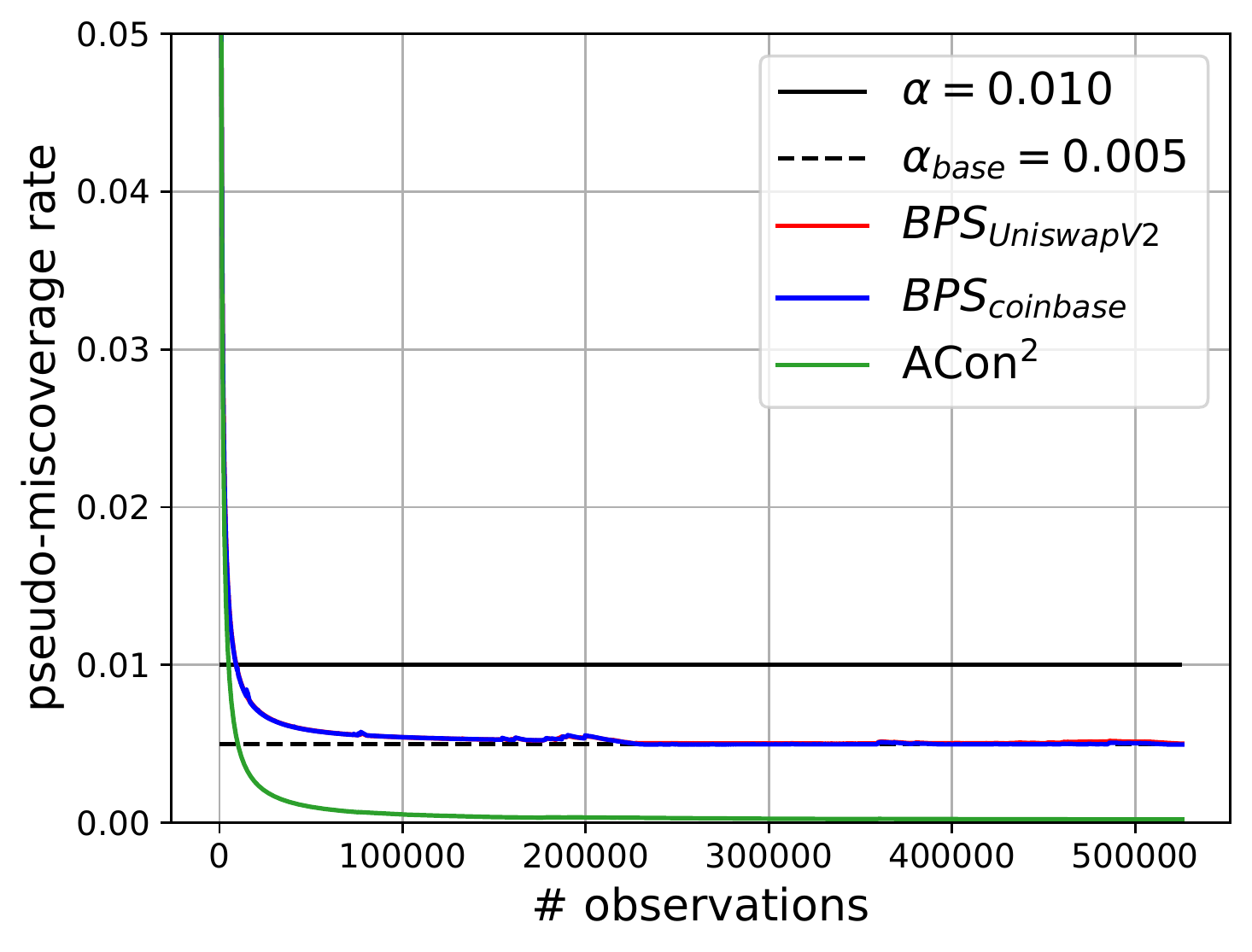}
  }
  \subfigure[Three data sources ($K=3$)]{
    \centering
    \includegraphics[width=0.3\linewidth]{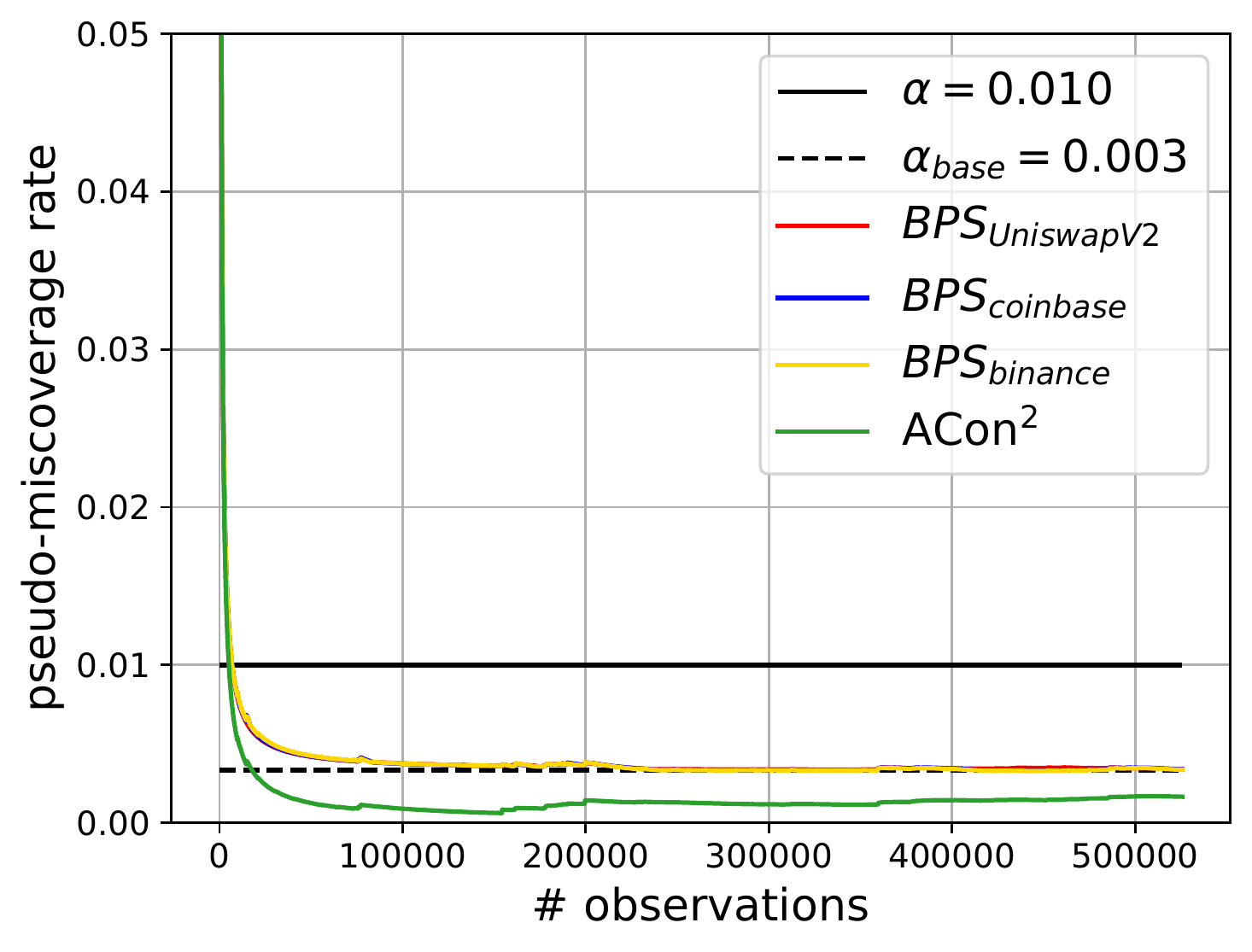}
  }
  \caption{
    Pseudo-miscoverage rate on USD/ETH data.
    \MR{Desired miscoverage rate $\alpha$ of \ACC is depicted as a black solid line, where
    empirical pseudo-miscoverate rates in green need to be around or below of the $\alpha$ line for
    justifying the correctness of \ACC (Definition \ref{def:correct}).
    The empirical miscoverate rates of base prediction sets (BPS) require to be close to
    desired miscoverate rates $\alpha_{\text{base}}$ to be correct as in (\ref{eq:basecorrect}).
    }
    For each $K$, the \ACC satisfies a desired miscoverage rate $\alpha$.
    In particular,
    for all $K$, we set a desired miscoverage rate by $\alpha=0.01$.
    Moreover, based on Theorem \ref{thm:consensuscorrectness},
    we set the desired miscoverage rate of each base prediction set
    by $\alpha_{\text{base}} = \frac{\alpha}{K}$.
    As can be seen, the empirical miscoverage rate of each base prediction set 
    converges to $\alpha_{\text{base}}$, so
    the empirical pseudo-miscoverage rate of \ACC is below of $\alpha$, while
    while affected by the abrupt price change. 
    This demonstrates the approximate correctness of consensus sets by \ACC under distribution shift.
    Note that when $K=1$, \ACC and a BPS are identical,
    so empirical miscoverate rates are overlapped in Figure \ref{fig:global:miscoverate:K1}.
  }
  \label{fig:global:miscoverage}
\end{figure*}

We evaluate our approach \ACC in price data that contains global shift.
In particular, we use USD/ETH price data from 2021-1-1 to 2021-12-31 for every 60 seconds,
obtained from
\NMR{UniswapV2 via \texttt{Web3.py}},
Coinbase via Coinbase Pro API,
and
Binance via Binance API.
The results are summarized in
Figure \ref{fig:global:overall}, 
Figure \ref{fig:global:miscoverage},
and
Figure \ref{fig:global:size}.

Figure \ref{fig:global:overall} shows
the consensus sets of \ACC in green for $K \in \{1, 2, 3\}$,
which closely follows the abrupt change of USD/ETH prices,
while their interval size is sufficiently small to cover price data from different sources.
Note that the initial part of Figure \ref{fig:global:overall:K3} shows that
\ACC returns empty sets, meaning ``I don't know'' (IDK), which is
simply because base prediction sets are still learning to find proper score functions $s_{t, k}$ and
scalar parameters $\tau_{t, k}$.


Figure \ref{fig:global:miscoverage} and \ref{fig:global:size} demonstrate more details
on the correctness guarantee and the efficacy of the consensus sets in their size. 
In particular,
Figure \ref{fig:global:miscoverage} illustrates the
pseudo-miscoverage rate at each time step.
Up to the time $200$K, the base prediction sets and consensus sets by \ACC closely
satisfy the desired miscoverage rate.
After this time frame, due to the abrupt increase of the USD/ETH price,
the pseudo-miscoverage rates slightly increase, while recovering as time passes.
This is mainly because locally increased miscoverage errors
introduced by base prediction sets due to the abrupt shift.
Figure \ref{fig:global:size} shows the distribution on the size of consensus sets.
As can be seen, the average set size is about 40;
considering that the price is not perfectly synchronized due to transaction fee,
the set size is reasonable.
Here, the consensus sets when $K=2$ tend to produce larger set size, and
this is mainly because of the choice of $\beta = \lfloor \frac{K}{2} \rfloor = 1$,
where the consensus set includes all labels from two base prediction sets. 
In short, from these observation,
we empirically justify that
\ACC closely achieves a desired miscoverage rate, while maintaining a small set size under global distribution shift,
addressing Challenge \ding{183} and \ding{185}.
Finally, see the $\sigma$-BPS baseline results in Figure \ref{fig:sigmabps}, which shows
the necessity of careful parameter adaptation for the correctness guarantee and reasonable prediction set size of base prediction sets.

\subsection{Local Shift: INV/ETH Price Data}
\label{sec:localshift}

\begin{table*}
  \centering
  \footnotesize
  \renewcommand{\arraystretch}{0.7}
  \begin{tabular}{c||cccccc}
    \toprule
    time &
    \makecell{2022-04-02 \\ 11:03:40} &
    \makecell{2022-04-02 \\ 11:03:50} &
    \makecell{2022-04-02 \\ 11:04:00} &
    \makecell{2022-04-02 \\ 11:04:10} &
    \makecell{2022-04-02 \\ 11:04:20} & 
    \makecell{2022-04-02 \\ 11:04:30}
    \\
    \midrule
    SushiSwap &
    9.3734 & 9.3734 & {\color{red}0.1667} & 4.5976 & 4.5976 & 4.5976
    \\
    UniswapV2 &
    9.1802 & 9.1802 & 9.1802 & 5.3045 & 5.3045 & 5.3045
    \\
    Coinbase &
    9.1418 & 9.1418 & 9.0041 & 9.0041 & 9.0041 & 9.0041
    \\
    \midrule
    median &
    9.1802 & 9.1802 & 9.0041 & 5.3045 & 5.3045 & 5.3045
    \\
    {\footnotesize\makecell{TWAP of SushiSwap by Keep3rV2 \cite{keep3rv2}}} &
    8.8497 & 8.8497 & 8.8575 & {\color{red}0.1667} & {\color{red}0.1667} & {\color{red}0.3317}
    \\
    \midrule
    $\text{BPS}_{\text{SushiSwap}}$ &
    [9.10,9.65] & [9.10,9.65] & {\color{red}[0.10,0.65]} & {\color{blue}[3.98,5.02]} & [4.08,5.11] & [4.08,5.11]
    \\
    $\text{BPS}_{\text{UniswapV2}}$ &
    [8.91,9.45] & [8.91,9.45] & [8.91,9.45] & {\color{blue}[5.13,5.68]} & [5.03,5.58] & [5.03,5.58]
    \\
    $\text{BPS}_{\text{Coinbase}}$ &
    [8.64,9.64] & [8.64,9.64] & [8.51,9.51] & {\color{blue}[8.50,9.51]} & [8.50,9.51] & [8.50,9.51]
    \\
    \ACC ($K=3$) &
    [8.91,9.64] & [8.91,9.64] & [\textbf{8.91},\textbf{9.45}] & $\emptyset$ & [5.03,5.11] & [5.03,5.11]
    \\
    \midrule
    $\text{BPS}_{\text{SushiSwap}}$ &
    [9.10,9.65] & [9.10,9.65] & {\color{red}[0.10,0.65]} & [4.23,4.78] & [4.32,4.87] & [4.08,5.11]
    \\
    $\text{BPS}_{\text{UniswapV2}}$ &
    [8.91,9.45] & [8.91,9.45] & [8.91,9.45] & [5.13,5.68] & [5.03,5.58] & [5.03,5.58]
    \\
    \ACC ($K=2$) &
    [8.91,9.65] & [8.91,9.65] & \underline{[0.10,9.45]} & [4.23,5.68] & [4.32,5.58] & [4.08,5.58]
    \\
    \midrule
    $\text{BPS}_{\text{SushiSwap}}$ &
    [9.10,9.65] & [9.10,9.65] & {\color{red}[0.10,0.65]} & [4.23,4.78] & [4.32,4.87] & [4.32,4.87]
    \\
    \ACC ($K=1$) &
    [9.10,9.65] & [9.10,9.65] & {\color{red}[0.10,0.65]} & [4.23,4.78] & [4.32,4.87] & [4.32,4.87]
    \\
    \bottomrule
  \end{tabular}
  \caption{
    Prediction consensus results on INV/ETH price manipulation data.
    The price of SushiSwap in red is manipulated (\ie {\color{red}$0.1667$}).
    The updated consensus set by \ACC with $K=3$ is not affected by the manipulation, while
    the base prediction set for SushiSwap, $\text{BPS}_{\text{SushiSwap}}$, follows the manipulated price (\ie ${\color{red}[0.10, 0.65]}$).
    At the next time step, prices are spread due to arbitrage in blue, resulting in failing to make consensus;
    thus, \ACC returns the interval for ``no consensus'' (\ie $\emptyset$), which signals the abnormality of the market while the median does not provide that.
    This suggests the importance of quantifying uncertainty (Challenge \ding{182}).
    In short, intervals by \ACC do not meaningfully manipulated by adversaries (Challenge \ding{184}),
    thus providing benefits to downstream applications through quantified uncertainty.
  }
  \label{tab:inveth}
\end{table*}

To show the efficacy of our approach under local shift by Byzantine adversaries,
we adopt the recent incident on Inverse Fiance \cite{rekt2022inverse},
which occurs on April 2nd in 2022 and about $\$15.6$M was stolen.
This incident began from
the price manipulation on the INV-ETH pair of SushiSwap.
Interesting, Inverse Finance uses the TWAP oracle by the Keeper \cite{keep3rv2},
but the TWAP oracle was heavily manipulated due to the short time window of the Keeper's TWAP.
We collect the associated price data from Ethereum blockchain between
2022-04-01 and 2022-04-02.
In particular, we collect the spot price of Sushiswap and UniswapV2 from all blocks in the time frame. 
Also, we read the Keeper TWAP oracle data.
For the additional price source, we use Coinbase. 

Table \ref{tab:inveth} shows the details on the behavior of
our approach, baselines, and price data around the price manipulation; see Figure \ref{fig:inveth} for its visualization.
The trend of consensus sets along with pseudo-miscoverage rate curves in Figure \ref{fig:local:miscoverage} and size distributions in Figure \ref{fig:local:size}
is similar to the USD/ETH data.
Here, we focus on the interpretation of Table \ref{tab:inveth}.

Table \ref{tab:inveth} includes the price data of INV/ETH from three sources
(\ie SushiSwap, UniswapV2, and Coinbase),
baseline results (\ie the median of three prices and the TWAP of SushiSwap by Keep3rV2),
and
consensus sets by \ACC with different options on $K \in \{1, 2, 3\}$.
Here, the price manipulation occurs around 11:04 on April 2nd, 2022 by swapping
500 ETH for 1.7K INV to the Sushiswap pool, so the
spot price of ETH by INV was significantly decreased.
Here, the consensus set by \ACC ($K=3$) is not affected by the price manipulation as shown in bold,
but in the next time step, it can detect the failure of consensus among three markets,
highlighted in blue,
producing the IDK interval, $\emptyset$; this demonstrates the positive effect in modeling uncertainty (Challenge \ding{182}).
This consensus failure is due to the manipulation followed by the slow arbitrage,
and
the IDK interval is a still valid indicator of anomaly
as IDK intervals imply that our belief (\eg Assumption \ref{ass:mainassumption} or \ref{ass:beta}) is violated.
\MR{Note that Assumption \ref{ass:mainassumption} may not hold in this transient period.
  But, the price manipulation rarely happens in practice so has negligible effects on the correctness guarantee,
  as empirically shown in Figure \ref{fig:inveht:miscoverage:K3}.
}

Compared to this, the consensus sets by \ACC with $K=2$ is affected by the price manipulation, as shown in the underline,
but it still shows the abnormal signal via the larger interval size, also showing the usefulness of uncertainty (Challenge \ding{182}).
However, when $K=1$, the consensus set by \ACC closely follows the manipulated price, which make sense
as \ACC with $K=1$ assumes that there is no adversaries (\ie $\hat\beta = 0$) and
consider local shift by manipulation as global distribution shift, to which \ACC needs to be adapted.
For the median and TWAP (by Keeper) baselines,
they do not provide uncertainty, so it is unclear to capture the unusual event in the markets.
Moreover, the TWAP tends to maintain the manipulated price for a while, suggesting that
choosing a good time window is crucial.
Note that for the INV/ETH pair, we could not identify a Chainlink oracle at the time of the incidence,
partially because the minor coin, like INV, is not attractive to Chainlink node operators.
This implies that our approach could provide an oracle for minor tokens
if Challenge \ding{186} is addressed.




\vspace{-1ex}
\subsection{Case Study: Ethereum Blockchain}
\label{sec:casestudy}
\vspace{-1ex}


Beyond evaluating our approach on real data,
we demonstrate that our approach can \MRM{be implemented} in blockchains.
The application running over blockchains requires to have computational limitations, so
using machine learning techniques on blockchains is counter-intuitive in the first place.
Here, we implement our approach in \texttt{Solidity} and measure its gas usage on the Ethereum network,
\MR{as the gas usage summarizes computational and memory cost of smart contracts \cite{ethereumyellow}.}

\begin{table}[tbh!]
  \centering
  \footnotesize
  \renewcommand{\arraystretch}{0.5}
  \begin{tabular}{c|ccc}
    \toprule
    & \makecell{Swap \\ $(G_{\text{Swap}})$} &
    \makecell{Swap+BPS \\ $(G_{\text{Swap}} + G_{\text{BPS}})$} &
    \makecell{\ACC \\ $(G_{\texttt{ACon}^2})$} 
    \\
    \midrule
    gas used & \makecell{$112,904.23$ \\ {\footnotesize$\pm 5,115.41$}} & \makecell{\MR{$726,219.60$}
    \\ {\footnotesize$\pm 31,582.52$}} & \makecell{\MR{$105,518.16$} \\ {\footnotesize $\pm 257.39$}}
    \\
    \bottomrule
  \end{tabular}
  \caption{
    \MR{Gas usage averaged over 500 transactions.
    A token swap operation (\ie \texttt{swap$(\cdot)$} in \texttt{UniswapV2Pair.sol} takes $0.1$M gas units,
    while the swap followed by a base prediction set (BPS) update takes $0.7$M gas units.
    \ACC related to form a consensus set requires $0.1$M gas units.
    This is the basis for holistic cost analysis in Table \ref{tab:eth:gas:e2e}.
    }
  }
  \label{tab:eth:gas}
\end{table}


\begin{table}[tbh!]
  \centering
  \footnotesize
  \renewcommand{\arraystretch}{0.5}
  \setlength{\tabcolsep}{5pt}
  \begin{tabular}{c|cccc}
    \toprule
    & baseline & $K=1$ & $K=2$ & $K=3$ \\
    \midrule
    \makecell{gas used} &
    \footnotesize{112,904.23} 
    &
    \footnotesize{831,737.76} 
    &
    \footnotesize{1,557,957.36} 
    &
    \footnotesize{2,284,176.96} 
    \\
    \bottomrule
  \end{tabular}
  \caption{
    \MR{
    The average gas usage of the entire oracle consensus system in varying $K$.
    Each used gas is computed by $K(G_{\text{Swap}} + G_{\text{BPS}}) + G_{\text{\texttt{Acon$^2$}}}$,
    where $G_{\text{Swap}}$, $G_{\text{BPS}})$, and $G_{\text{\texttt{Acon$^2$}}}$ are gas used for each component as denoted in Table \ref{tab:eth:gas}.
    The baseline gas usage is $G_{\text{Swap}}$.
    Based on the gas price at 2022-2-18,
    the baseline requires $\$4.78$, and
    the proposed consensus system requires $\$35.24$, $\$66.02$, and $\$96.80$ for $K=1, 2, 3$, respectively, for a transaction fee to construct a consensus set,
    where the total cost can be distributed across $K+1$ users (\ie $K$-traders and the consensus system user). 
    This demonstrates that \ACC and BPS have opportunities to be implemented in blockchains (Challenge \ding{186}), while showing that efficient implementation (in particular for the base prediction sets) is necessary.
    }
  }
  \label{tab:eth:gas:e2e}
\end{table}

In particular, Table \ref{tab:eth:gas} shows the gas usage averaged over 500 transactions.
``Swap'' means the original swap operation, \ie UniswapV2 \texttt{swap}$(\cdot)$, which uses
112,904 \MR{($=G_{\text{Swap}}$)} gas on average. The base prediction set (BPS) is attached right at the end of the
swap function, which increases the gas usage by about 7 times (\ie \MR{$G_{\text{BPS}}$ in Table \ref{tab:eth:gas}}).
This is mainly due to the iteration in the MVP algorithm,
which is generally unavoidable in machine learning algorithms.
In constructing a consensus set by (\ref{eq:consensusset}) only requires read operations
(\ie reading from intervals of base prediction sets, written in blockchains after the swap operation),
thus it does not require gas for transaction in reading.
\MR{However, we consider a case where this consensus set is used in downstream smart contracts for writing, which requires 105,518 gas units \MR{($=G_{\texttt{ACon}^2}$).}}

\MR{
  Table \ref{tab:eth:gas:e2e} shows the gas usage of the entire consensus system in varying $K$,
  where for each consensus set,
  $K(G_{\text{Swap}} + G_{\text{BPS}}) + G_{\text{\texttt{Acon$^2$}}}$ gas units are required, linear in $K$.
  Importantly, the total cost for using the entire consensus system can be distributed across
  $K+1$ system users.
  In particular for $K=3$ in Table \ref{tab:eth:gas:e2e}, one trader pays the cost of $G_{\text{Swap}} + G_{\text{BPS}}$ (\ie $\$30.8$) and the consensus set user lay the cost of $G_{\text{\texttt{Acon$^2$}}}$ (\ie$\$4.5$).
}
This case study in the Ethereum blockchain demonstrates that
our algorithm based on online machine learning has \MR{opportunities to be implemented in operation-parsimonious blockchains, partially addressing Challenge \ding{186}}.
Finally, Figure\LREF{\ref{fig:ethnetwork}}{10}~shows the consensus sets on a local Ethereum blockchain
with three AMMs, a trader, an arbitrageur, and an adversary.
As before, the consensus sets cover the price data, while robust to price manipulation,
shown by a spike in Figure\LREF{\ref{fig:eth:ps}}{10(a)}.
\MR{See Figure\LREF{\ref{fig:ethnetwork:variousKalpha}}{11}~for pseudo-miscoverate rates in varying
  $K$ and $\alpha$, showing the generality of \ACC. 
}

%% file: relwk.tex
\vspace{-1ex}
\section{Related work}
\label{s:relwk}

This section include related work on
blockchain oracles, consensus problems, and conformal prediction.

\subsection{Blockchain Oracles}
The blockchain oracle problem has been considered around the birth of the first smart-contract-enabled Ethereum blockchain \cite{ethereum};
methods to address this issue also have been proposed
\cite{provable,zhang2016town,truthcoin,de2017witnet,peterson2015augur,delphi,adler2018astraea,hess2017aeternity,chainlink}.
Here, considering that the oracle smart contract is an external data feeder,
we view existing methods in two categories:
whether data is authenticated, assuming their validity (\emph{data authentication})
and
whether data is valid (\emph{data validation}). 

Methods providing data authentication mainly propose protocols that deliver un-tampered data
(\eg via TLSNotary proof \cite{provable} or via Intel Software Guard Extensions \cite{zhang2016town}).
The blockchains generally do not have functionality to proactively establish secure channels toward off-chain, so
external services need to push data into the blockchain in reliable ways.

Along with data authentication approaches,
we also need to validate whether data is correct mainly via a voting mechanism or Schelling-point scheme.
For the voting mechanism, a (weighted) voting scheme on data $\yh(x) \in \Ys$ on possible data choices $\Ys$ is mainly used.
In particular,
Truthcoin \cite{truthcoin} and Witnet \cite{de2017witnet}
construct a weighted voting matrix on data choices, from which they extract common vote patterns via singular value decomposition
and incentivizes or penalizes inlier or outlier voters, respectively. 
Augur \cite{peterson2015augur} and Astraea \cite{adler2018astraea} use
a reputation token or stakes in general for weighted voting on data choices $\Ys$ to achieve consensus. 
Aeternity \cite{hess2017aeternity} reuses blockchain consensus mechanisms (\eg proof-of-work) for the consensus over data choices.
The Schelling-point schemes consider the fact that people tend to choose the same solution without communication, which could be a basis for consensus via robust statistics. In particular, 
Oracle Security Module by MakerDAO \cite{osm} and Chainlink \cite{chainlink} exploit a single or multi-layer median over possible data values (mainly prices), produced by node operators.


Compared to the aforementioned approaches to achieve data validity,
we account for uncertainty of real data via
online machine learning, providing the correctness guarantee on consensus.
Data authentication could be achievable
if the proposed algorithm is fully implemented within blockchains as demonstrated in Section \ref{sec:casestudy}.

\subsection{Consensus Problems}

\begin{table*}[t!]
  \centering
  \footnotesize
  \begin{tabular}{c||c|c|c}
    \toprule
    problem &
    \makecell{predictive consensus} &
    \makecell{true consensus} &
    \makecell{correctness on predictive consensus}
    \\
    \hline
    \hline
    \makecell{
      Byzantine generals 
      \cite{lamport1982byzantine}
    } &
    \makecell{
      $\yh(x) \in \Ys$ 
      (point prediction)
    } &
    \makecell{$y$ (discrete)} &
    \makecell{
      exactly correct, \ie
      {\footnotesize
        $\forall x, y,~ y = \yh(x)$
      }
    }
    \\
    \hline
    \makecell{
      approximate agreement 
      \cite{dolev1986reaching}
    } &
    \makecell{
      $\yh(x) \in \Ys$ 
      (point prediction)
    }
    &
    \makecell{$y$ (continuous)} &
    \makecell{
      approximately correct, \ie
      {\footnotesize
        \makecell{
          $\forall x, y, ~| y - \yh(x)| \le \epsilon$
        }
      }
    }
    \\
    \hline
    \makecell{
      triple modular redundancy
      \cite{von1956probabilistic,lyons1962use}
    } &
    \makecell{
      $\yh(x) \in \Ys$ 
      (point prediction)
    }
    &
    \makecell{$y$ (discrete)} &
    \makecell{
      approximately correct, \ie
      {\footnotesize
        $\Prob_{x, y}\{y = \yh(x)\} \ge 1 - \epsilon$
      }
    }
    \\
    \hline
    \makecell{
      abstract sensing 
      \cite{marzullo1990tolerating}
    } &
    \makecell{
      $\Ch(x) \subseteq \Ys$ 
      (set prediction)
    }
    &
    \makecell{$y$ (continuous)} &
    \makecell{
      exactly correct, \ie
      {\footnotesize
        $\forall x, y,~ y \in \Ch(x)$
      }
    }
    \\
    \hline
    \hline
    \makecell{
      prediction consensus 
      (ours)
    } &
    \makecell{
      $\Ch(x) \subseteq \Ys$
      (set prediction)
    }
    &
    \makecell{$y$ (discrete or continuous)} &
    \makecell{
      $(\alpha, \beta, \ep)$-correct, \ie 
      {\footnotesize
        $\Vs(\Fs, T, \alpha, \beta, L) \le \ep$
      }
    }
    \\
    \bottomrule
  \end{tabular}
  \caption{
    Related consensus problems (simplified in a prediction setup for comparison).
    Each problem relies on different setups and correctness definitions.
    Prediction consensus, considered in this paper, explicitly models a learner $L$ that returns predictive consensus $\Ch$,
    updated via online machine learning to handle distribution shift.
  }
  \label{tab:relworkcomparison}
\end{table*}

The consensus problem is traditionally considered in distributed systems.
In distributed systems, we need a protocol that enables consensus on values 
to maintain state consistency among system nodes
even in the existence of faulty nodes; see \cite{fischer1983consensus} for a survey.
In this paper, we consider each system node as a machine learned predictor.
Thus, the goal is to achieve consensus on predicted labels, and
we reinterpret and simplify setups in
\cite{lamport1982byzantine,dolev1986reaching,von1956probabilistic,lyons1962use,marzullo1990tolerating}
for comparison.
The following includes a short introduction along with comparison.
See Table \ref{tab:relworkcomparison} for a comparison summary.

In Byzantine generals \cite{lamport1982byzantine},
each prediction node (\ie a lieutenant in generals metaphor)
makes an observation $x$ on the environment (\ie a lieutenant receives messages from other generals) and
predicts a discrete label $\yh(x) \in \Ys$ (\eg whether ``attack'' or not).
Then, label predictions from all prediction nodes are aggregated to predict the true  consensus label $y$
(\ie an original order from a commander).
Here, achieving the consensus is non-trivial as a subset of prediction nodes can be Byzantine adversaries. 
This problem mainly considers applications on computer systems, so a discrete, point prediction is considered;
given the true consensus label $y$, it requires to achieve the exactly same prediction $\yh(x)$
(\ie $y = \yh(x)$ for all $x$ and $y$). This correctness guarantee is achievable in deterministic systems,
but if $x$ and $y$ are stochastic due to uncertainty or $y$ is continuous, it is not possible.
The setup for approximate agreement \cite{dolev1986reaching}
assumes uncertainty on $x$ and $y$ along with continuous labels $y$.
In particular, the setup considers that the predicted label is approximately correct, \ie
$| y - \yh(x) | \le \ep$ for all $x$ and $y$, and some $\ep \in \realnum_{\ge 0}$.

In electronic engineering,
a similar consensus concept in designing circuits is considered
as triple modular redundancy \cite{von1956probabilistic,lyons1962use}.
This mainly computes
the discrete output of redundant circuits from the same input $x$ and
the majority of the outputs to produce a single output $\yh(x)$.
This setup considers that the predicted consensus label is statistically correct,
\ie $\Prob_{x, y}\{ y = \yh(x) \} \ge 1 - \ep$ for some $\ep \in [0, 1]$.

In system control,
the concept of abstract sensing \cite{marzullo1990tolerating} is used to build
fault-tolerant sensor fusion, where
we consider the outputs from multiple sensors given the observation $x$ are intervals and
denote the intersection over intervals from multiple sensors by a consensus interval $\Ch(x)$.
This setup assumes that the interval should include a consensus label $y$ such that
the consensus interval eventually includes the consensus label as well, \ie $y \in \Ch(x)$ for all $x$ and $y$.
But, constructing an interval that must include the consensus label is impractical.

The aforementioned setups consider that distributions over the prediction $\yh(x)$ and the consensus label $y$ are stationary,
so they  do not explicitly consider an online machine learning setup, where the distributions shift along time.

\vspace{-1ex}
\subsection{Conformal Prediction}
Conformal prediction is a method that quantifies the uncertainty of any predictors \cite{vovk2005algorithmic}. It constructs a prediction set that possibly contains the true label. 
Here, we consider conformal prediction under various assumptions on data distributions.

\para{No shift.}
The original conformal prediction assumes exchangeability on data
(where a special case of exchangeability is the independent and identical distribution (i.i.d.) assumption on data).
This mainly assumes that in prediction, the conformal predictor observes the data
which are drawn from the same distribution as in training.
Under this assumption,
a marginal coverage guarantee \cite{vovk2005algorithmic,angelopoulos2020uncertainty} and
a conditional coverage guarantee (\ie probably approximately correct (PAC) guarantee) \cite{vovk2013conditional,Park2020PAC,bates2021distribution} can be achievable
for the correctness of constructed prediction sets.
However, this strong assumption on no distribution shift can be broken in practice. 

\para{Covariate or label shift.}
To address the no shift assumption,
the conformal prediction approach can be extended to handle
covariate shift \cite{shimodaira2000improving},
where the covariate distribution $p(x)$ can be shifted in prediction, while
the labeling distribution $p(y|x)$ is unchanged, 
or label shift \cite{lipton2018detecting},
where the label distribution $p(y)$ can be shifted in prediction, while
the conditional covariate distribution $p(x | y)$ is not changing. 
In particular, weighted conformal prediction can achieve a desired marginal coverage rate
given true importance weights \cite{tibshirani2019conformal};
similarly, PAC prediction sets, also called training-conditional inductive conformal prediction,
achieves a desired PAC guarantee under a smoothness assumption on distributions \cite{park2022pac}. 
The conformal prediction can achieve the marginal coverage guarantee under label shift by exploiting
true importance weights \cite{podkopaev2021distribution}.

\para{Meta learning.}
Covariate and label shift setups involve two distributions, while
in meta learning multiple distributions for each data sources are considered,
assuming a distribution of each data source is drawn from the same distribution. 
Under this setup, conformal prediction tailored to meta learning can attain
a desired coverage guarantee \cite{fisch2021fewshot},
a conditional guarantee \cite{fisch2021fewshot},
and
a fully-conditional guarantee \cite{park2022pacmeta}.

\para{Adaptive conformal prediction.}
The previous conformal prediction methods consider the non-sequential data, assuming
shift among a discrete set of distributions.
In sequential data, the data shift is continuous,
where the online learning of conformal prediction sets is considered. 
In particular,
adaptive conformal inference \cite{gibbs2021adaptive} updates a desired marginal coverage of conformal prediction for every time step to have a prediction set that satisfies a desired marginal coverage rate on any data from an arbitrarily shifted distribution. 
Multi-valid prediction \cite{bastani2022practical} constructs a prediction set at each time that achieves
threshold-calibrated validity as well as a desired marginal coverage.
Adaptive conformal prediction methods consider a single data source,
where the data distribution can be arbitrarily shifted (also by adversaries).
In this paper, we consider the consensus over base conformal prediction sets based on any adaptive conformal prediction methods for multiple data sources.

\para{Ensemble conformal prediction.}
\MR{
Under no shift assumption (more precisely the exchangeable assumption),
ensemble approaches in combining conformal predictors have been explored.
Existing work \cite{balasubramanian2015conformal,toccaceli2017combination,cherubin2019majority,linusson2020efficient} mainly assumes batch learning setups with the exchangeability assumption on samples from each source,
which implies that a data distribution is not changing.
In particular, 
\cite{balasubramanian2015conformal,toccaceli2017combination} consider combining p-values from data sources with the exchangeable assumption and
\cite{linusson2020efficient} combines non-conformity scores, constructed using data from different data sources with the exchangeable assumption.
The most closely related work is \cite{cherubin2019majority}; this approach
computes consensus sets, similar to ours, over conformal prediction sets, which are constructed under
the exchangeable assumption. 
Contrast to these, we consider an online learning setup where data distributions of sources are shifted over time and data from some sources are maliciously manipulated. 
}


%% file: discussion.tex
\vspace{-1ex}
\section{Discussion}
\label{s:disc}
\vspace{-1ex}

\para{Consensus assumption.}
Assumption \ref{ass:mainassumption}
considers a situation where the consensus label distribution is identical to
the distribution over source labels.
This is likely to be true due to the arbitrageur in price markets
but may not hold in general.
\MR{
For example, even in the existence of the arbitrageurs,
delay in arbitrage places market prices non-synchronized in a transient period
if arbitrageurs cannot notice the arbitrage opportunities immediately.
}
One way to relax the assumption is considering a
distributionally robust setup,
\ie
given $\rho \in \realnum_{\ge 0}$,
assume $D_f \( p_t(y_t \mid \x_t) \| p_t(\y_{t, k} \mid \x_{t, k}) \) \le \rho$ for any $t \in \{1, \dots, T\}$ and $k \in \{1, \dots, K\}$,
where $D_f$ is $f$-divergence.
This assumption can be used instead of\LREF{(\ref{eq:keylem:labeldistassumption}) in
proving Lemma \ref{lem:local2global}}{(15) in proving Lemma 4}, while finding $p_t(y_t | \x_t)$ is non-trivial.


\para{Efficient base prediction set update.}
As shown in Table \ref{tab:eth:gas},
the base prediction set update requires more gas.
Considering that adaptive conformal prediction is a recently developing area \cite{gibbs2021adaptive,bastani2022practical},
we believe that our choice of the base prediction set algorithm is a best possible option, while
developing computational efficient algorithms would be interesting. 

\para{The choice of adversary models.}
\MR{
We assume Byzantine adversaries in this paper,
but ``rational'' adversaries
driven by monetary benefits
could be considered with additional assumptions.
In particular,
we say that an adversary is \emph{rational} if
the adversary has a limited budget to manipulate examples.
In price manipulation, it is reasonable to assume that the adversary
has a limited number of tokens, so the manipulation is bounded,
\ie $\Es_{\text{rational}}= \{ e: \Xs \to \Xs \mid \forall x\in \Xs, \| e(x) - x \| \le \delta \}$, where
$\|\cdot\|$ is any norm and $\delta$ is the budget of the rational adversary.
This assumption makes the consensus procedure simpler by considering
the majority voting among the worst-case, deterministic  base prediction set, \ie
the consensus set is constructed via (\ref{eq:consensusset}) with
the worst case base prediction sets
$\hat{C}_{t,k}(\x_t) = \{ \yh_{t, k}(e_t(\x_t)) \mid e_t \in \Es_{\text{rational}} \}$,
where $\yh(e_t(\x_t))$ is a price by the $k$-th market at time $t$.
Here, we do not need to learn a score function and a prediction set for each data source.
However, the major limitation is that estimating $\delta$ is not easy and wrong estimation potentially undermines the correctness of the consensus set.
Our Byzantine adversary model relies on less assumption than the rational adversary model as
the Byzantine adversary is the rational adversary when $\delta \to \infty$.
}

%% file: conclusion.tex
\vspace{-1ex}
\section{Conclusion}
\label{s:conclusion}
\vspace{-1ex}

This paper proposes an adaptive conformal consensus (\ACC) to address
the oracle problem in blockchains to achieve consensus over multiple oracles.
In particular,
under some assumptions,
the proposed approach addresses five challenges
(\ding{182} handling uncertainty,
\ding{183} consensus under distribution shift,
\ding{184} consensus under Byzantine adversaries,
and
\ding{185} correctness guarantee on the consensus
)
and partially addresses one challenge
(\ding{186} practicality in blockchains),
which are
theoretically and empirically justified.

%% file: ack.tex
\vspace{-1ex}
\section{Acknowledgment}
\label{s:ack}
\vspace{-1ex}

We thank the anonymous reviewers,
our shepherd,
Jungwon Lim,
Yonghwi Kwon,
and
Seulbae Kim
for their helpful feedback.
This research was supported, in part, by the NSF award
CNS-1749711 and CCF-1910769;
ONR under grant N00014-23-1-2095;
DARPA V-SPELL N66001-21-C-4024;
and gifts from Facebook, Mozilla, Intel, VMware, and Google.

%% file: apdx.tex
\appendix
\onecolumn


\section{Additional Results}
\label{apdx:additionalresults}


\begin{figure*}[htb!]
  \centering
  \subfigure[USD/ETH data]{
    \centering
    \includegraphics[width=0.25\linewidth]{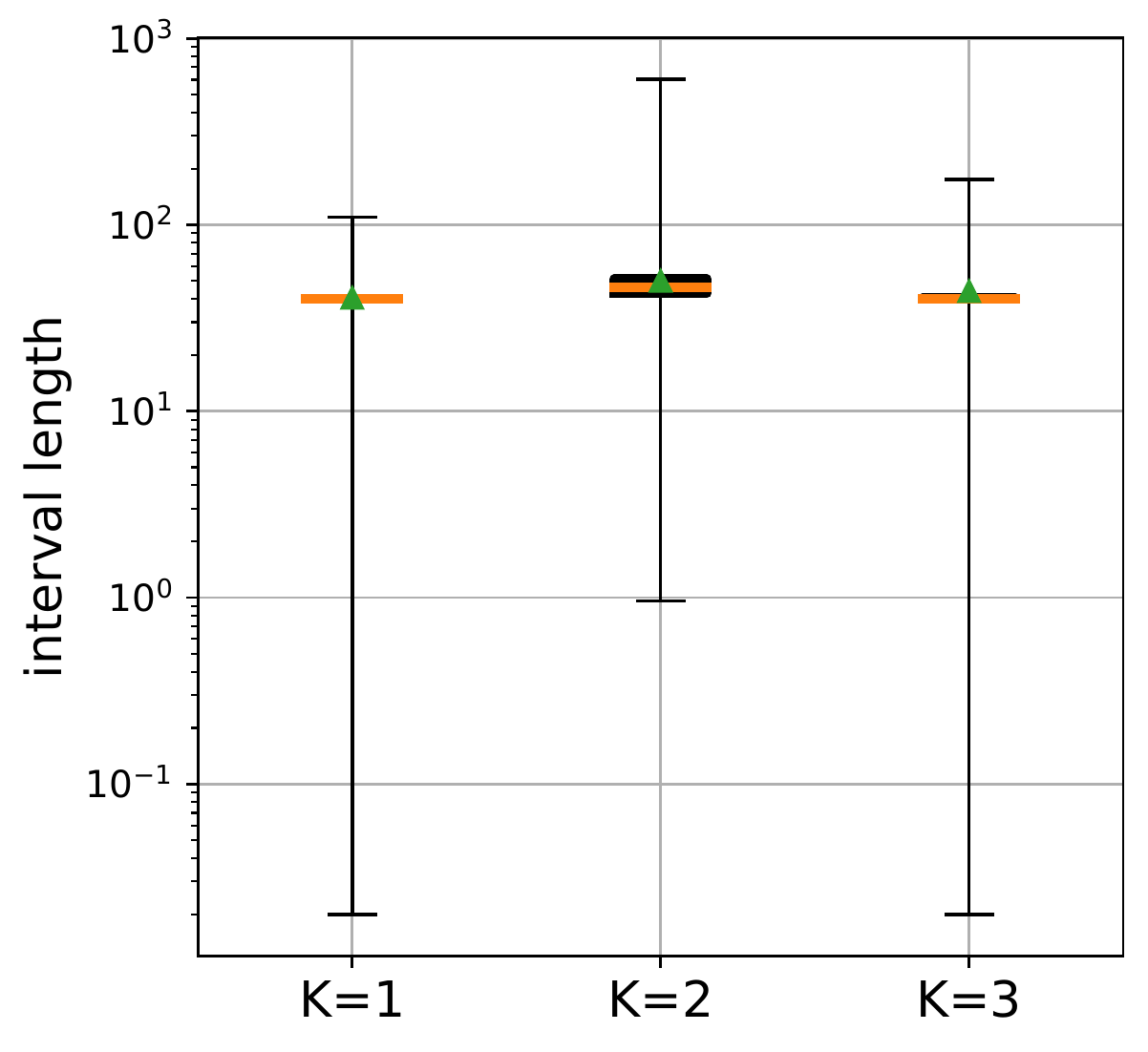}
    \label{fig:global:size}
  }
  \subfigure[INV/ETH data]{
    \centering
    \includegraphics[width=0.25\linewidth]{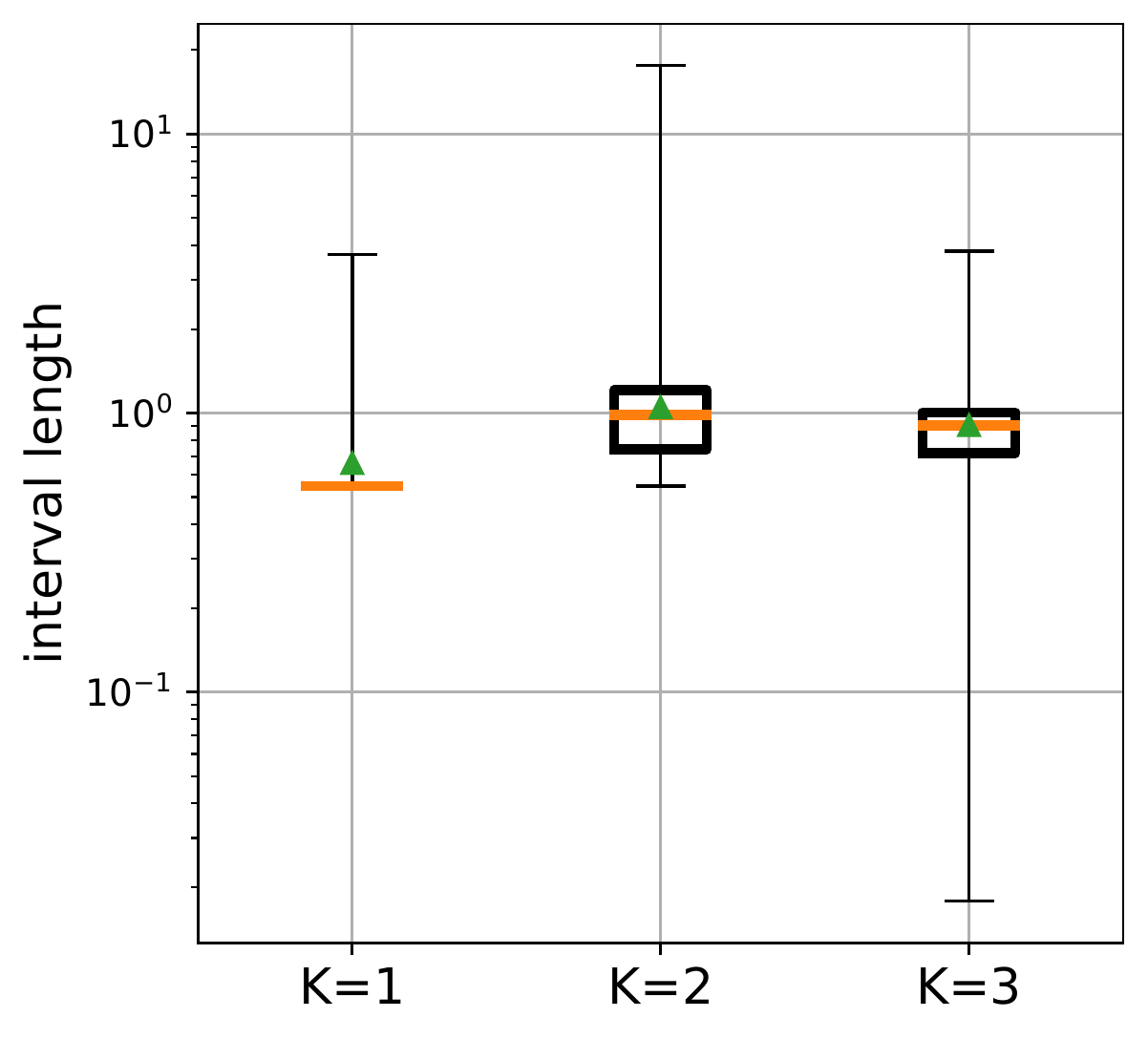}
    \label{fig:local:size}
  }
  \caption{
    Size distributions of consensus sets.
    \MR{A box plot contains $50\%$ of size samples, where an orange bar show a median value,
    and a whisker plot shows the minimum and maximum size values.}
    For USD/ETH data in Figure \ref{fig:global:size},
    the consensus sets for $K=1$ and $K=3$ are relatively smaller than the consensus sets of $K=2$,
    but
    the consensus sets by $K=1$ are prone to manipulation.
    Here, we do not count the IDK intervals as
    only $0.01\%$ consensus sets for $K=3$ are the IDK intervals.
    For INV/ETH data in Figure \ref{fig:local:size},
    the consensus sets are mostly less than one, which looks reasonable, considering the price scale.
    The consensus sets for $K=1$ and $K=3$ tends to be smaller than the consensus sets for $K=2$, but
    the sets with $K=1$ are prone to
    adversarial manipulation.
    Here, we do not count for the IDK intervals as
    only $0.3\%$ of the consensus sets for $K=3$ are the IDK intervals.
  }
\end{figure*}


\begin{figure*}[tbh!]
  \centering
  \subfigure[A single data source ($K=1, \beta=1, \hat\beta=0$)]{
    \centering
    \includegraphics[width=0.32\linewidth]{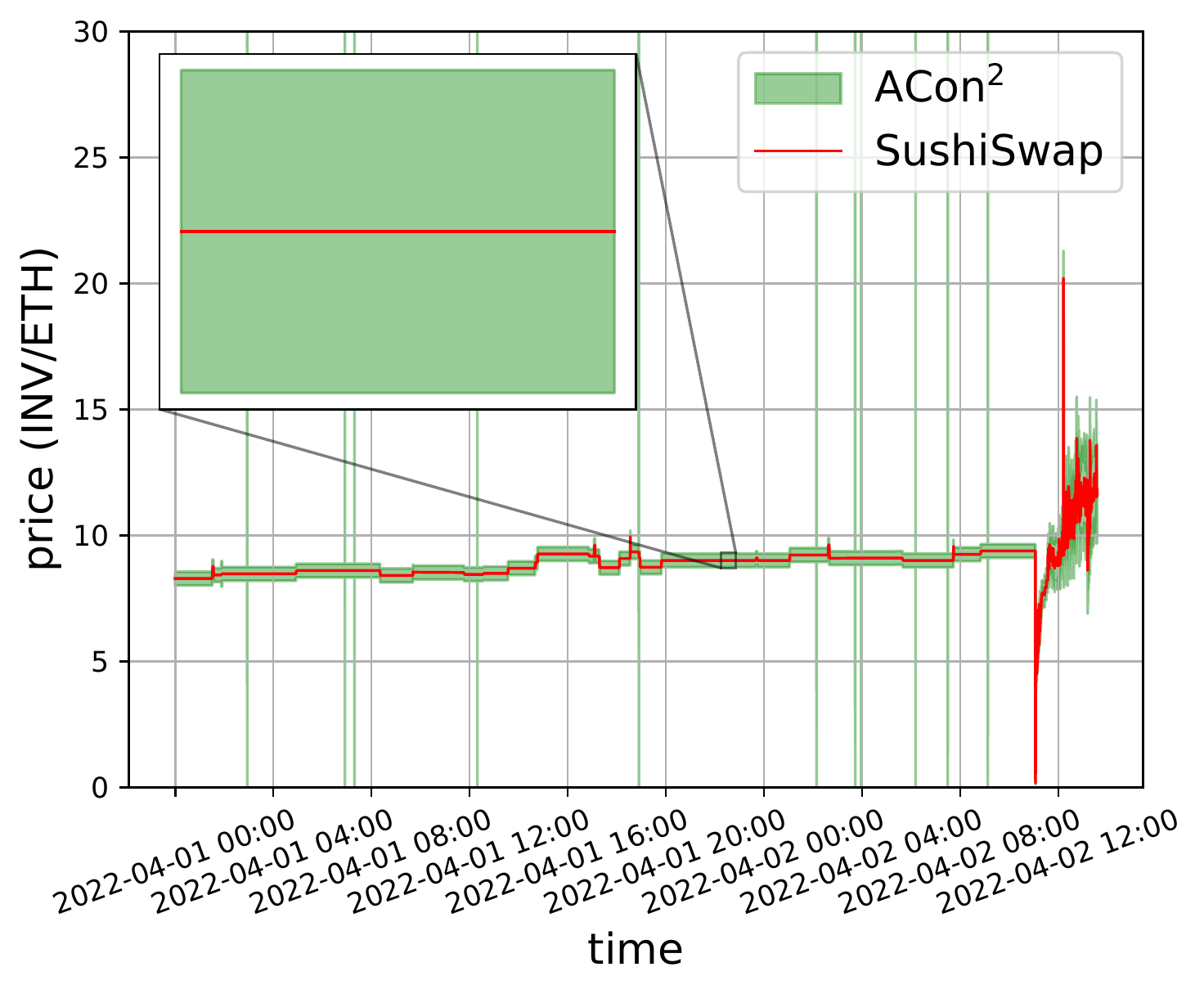}
  }
  \subfigure[Two data sources ($K=2, \beta=1, \hat\beta=1$)]{
    \centering
    \includegraphics[width=0.32\linewidth]{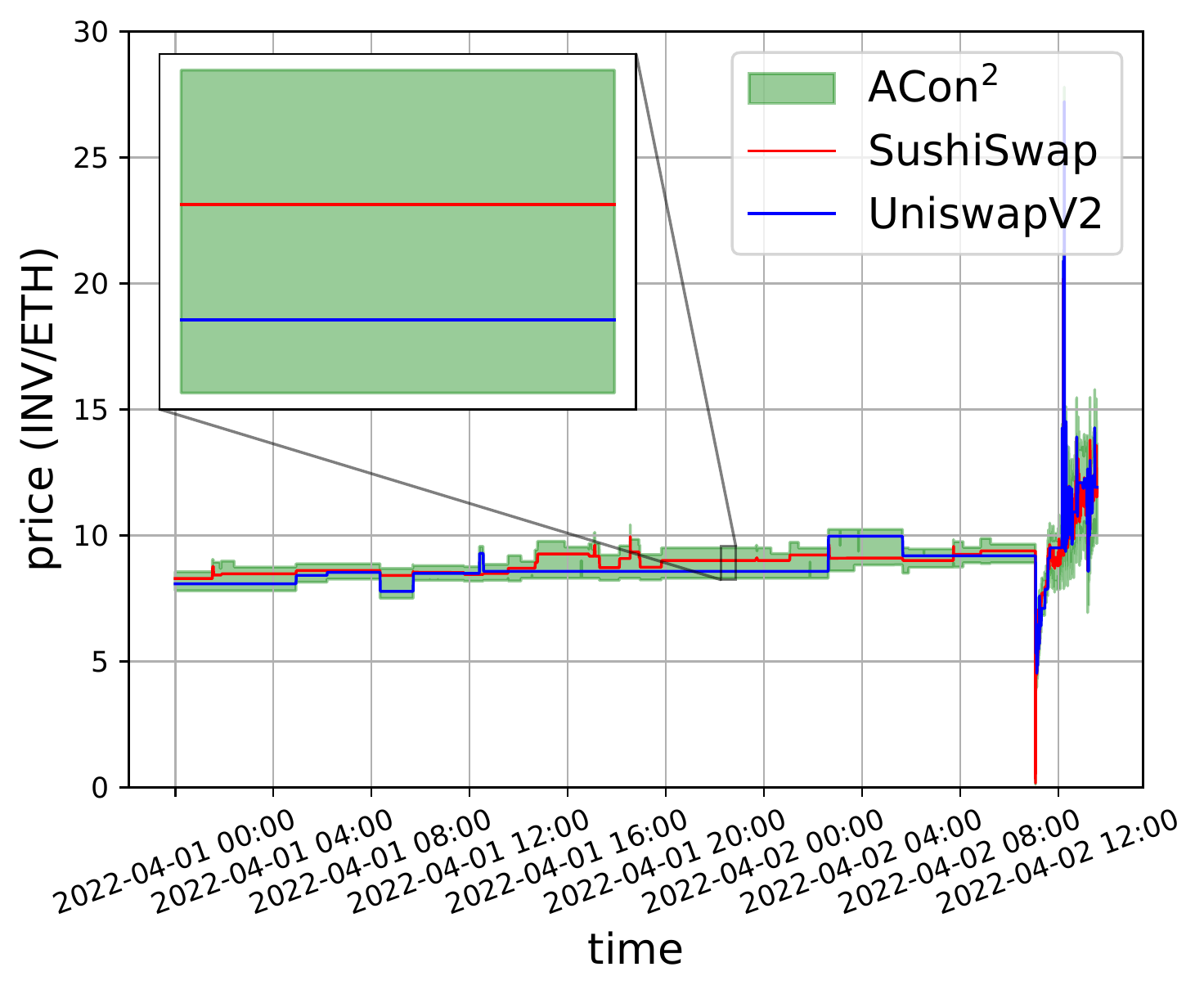}
  }
  \subfigure[Three data sources ($K=3, \beta=1, \hat\beta=1$)]{
    \centering
    \includegraphics[width=0.32\linewidth]{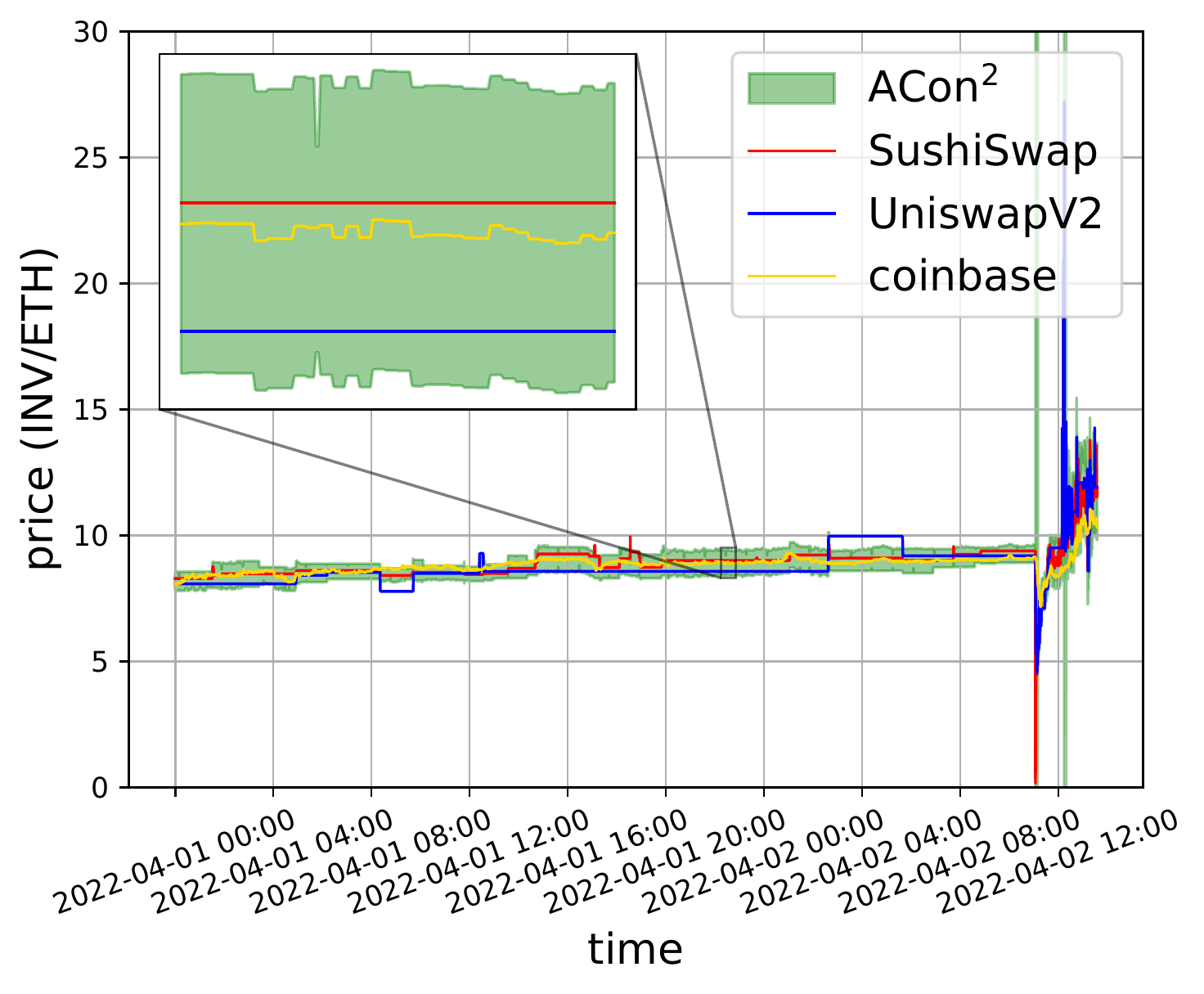}
  }
  \caption{
    Prediction consensus results on INV/ETH price manipulation data.
    For $K \in \{1, 2, 3\}$, the consensus sets by \ACC closely follow
    the price data, but when price manipulation occurs (around 2022-04-02 11:04),
    the behaviors of consensus sets differ.
    In particular, when $K=3$, the consensus sets are not meaningfully manipulated by
    the Byzantine adversaries, while the consensus sets for $K=1$ and $K=2$ are affected by the
    adversaries.
    \MR{After the manipulation,
      there is a transient period where huge manipulation followed by slow arbitrage introduces
      the violation of Assumption \ref{ass:mainassumption}.
      In this case, the consensus set does not make consensus, returning an empty set
      (where we plot $\Ys$ for a visualization purpose). 
    }
    See Table \ref{tab:inveth} for the detailed analysis on the consensus sets under manipulation.
  }
\end{figure*}

\begin{figure*}[thb!]
  \centering
  \subfigure[A single data source ($K=1, \beta=1$, $\hat\beta=0$)]{
    \centering
    \includegraphics[width=0.32\linewidth]{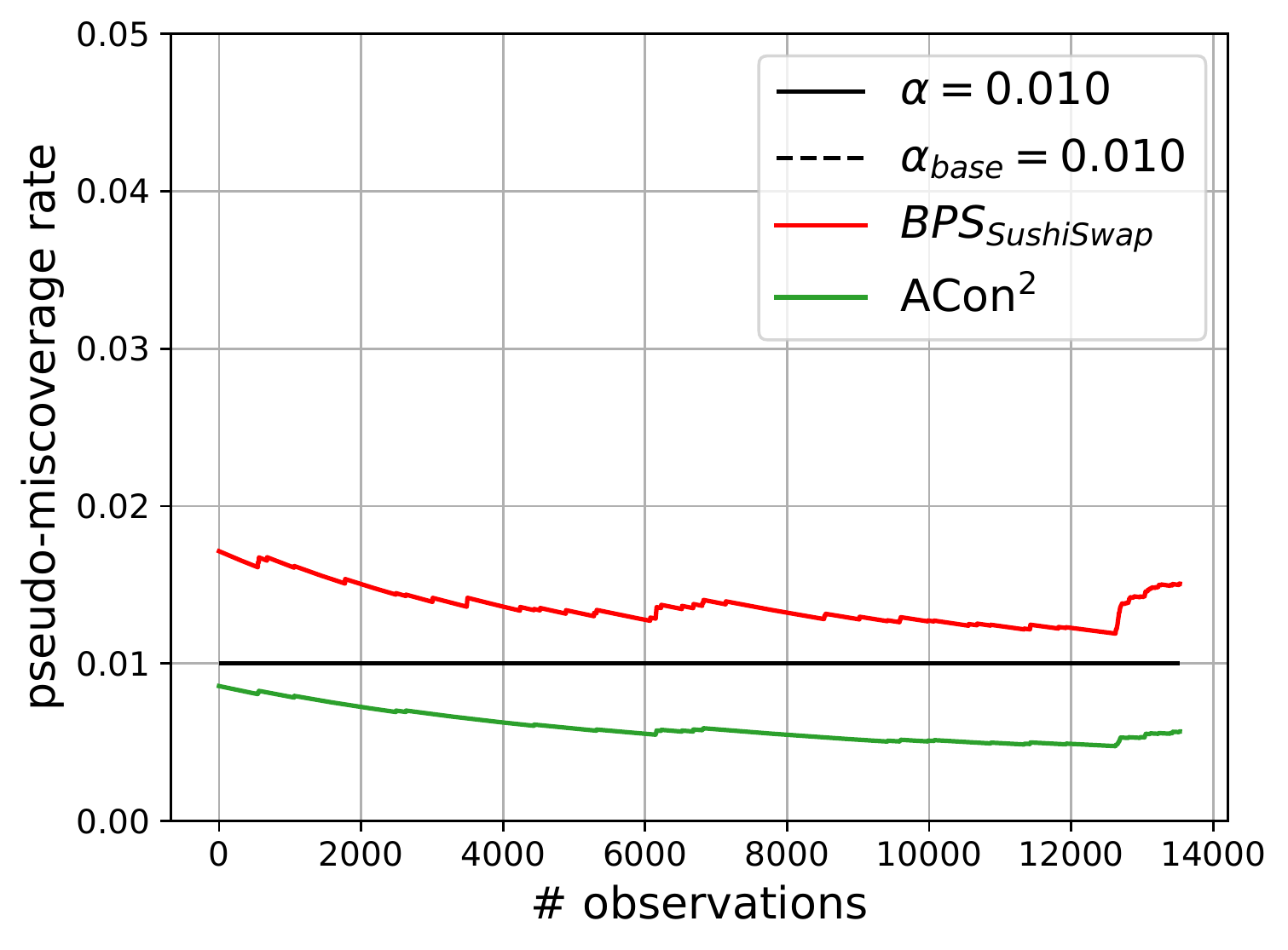}
    \label{fig:inveth:mis:K1}
  }
  \subfigure[Two data sources ($K=2, \beta=1, \hat\beta=1$)]{
    \centering
    \includegraphics[width=0.32\linewidth]{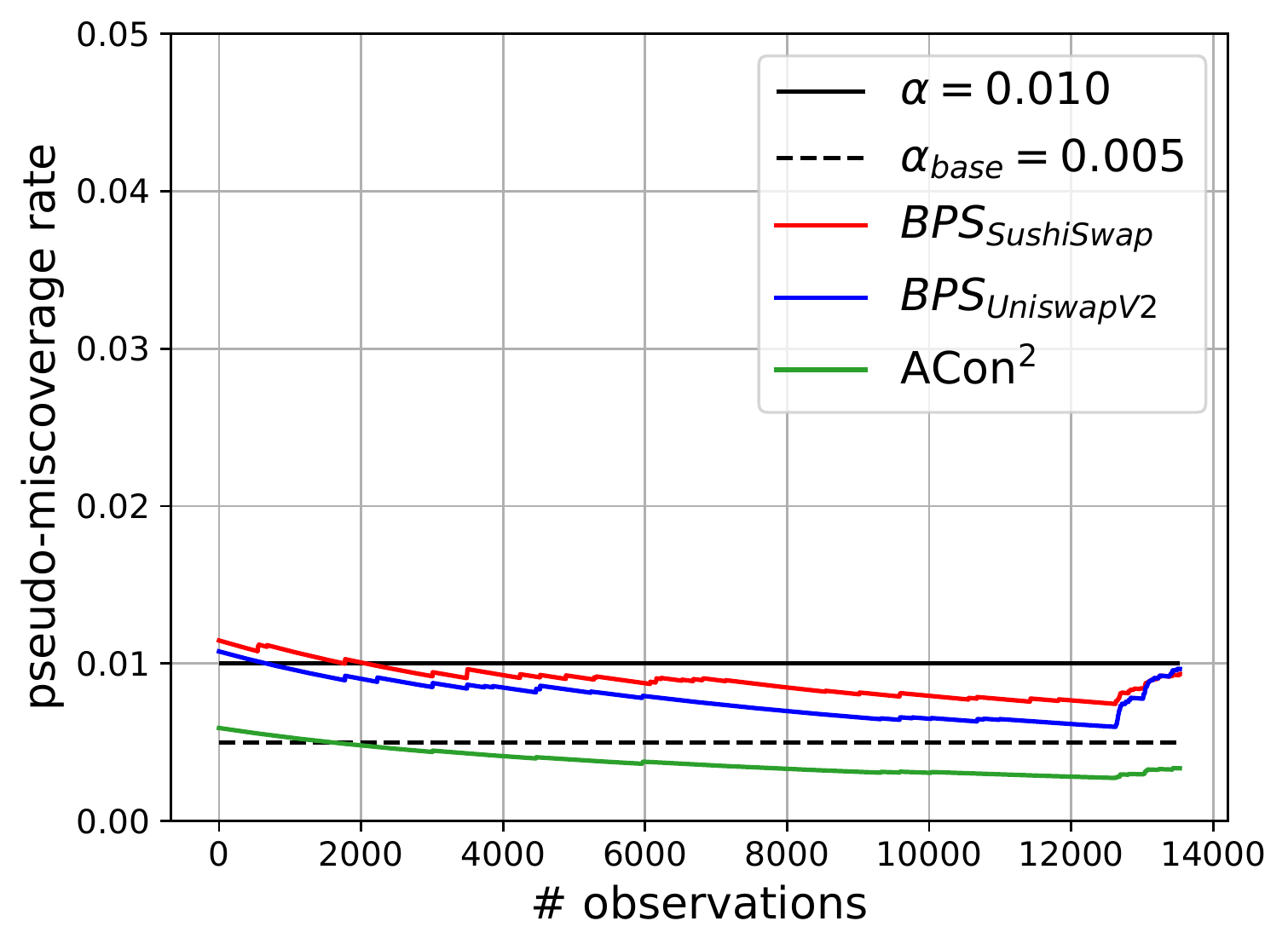}
  }
  \subfigure[Three data sources ($K=3, \beta=1,
    \hat\beta=1$)]{
    \centering
    \includegraphics[width=0.32\linewidth]{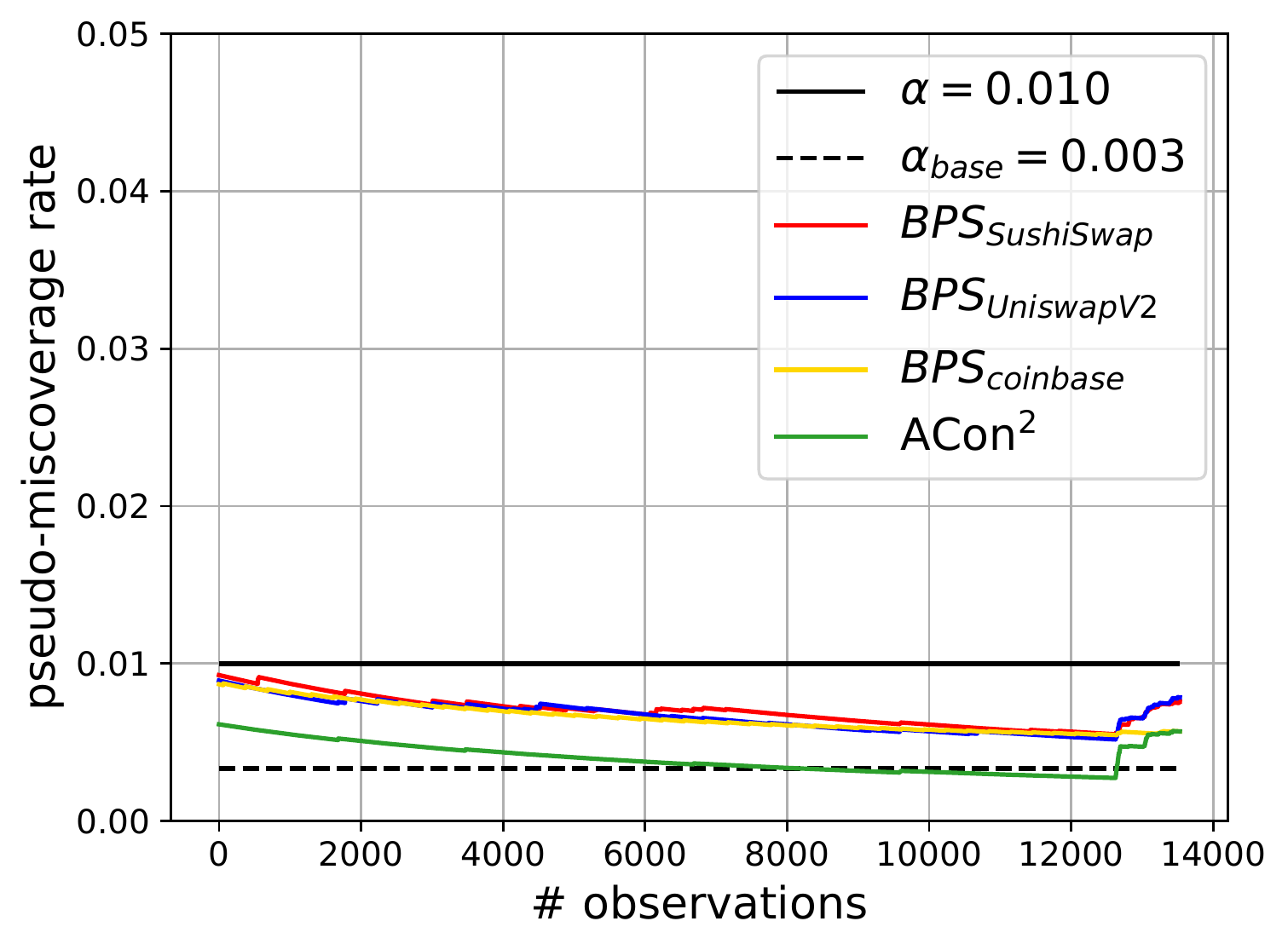}
    \label{fig:inveht:miscoverage:K3}
  }
  \caption{Miscoverage rates on INV/ETH data.
  The desired miscoverage rate $\alpha$ of \ACC is depicted as a black solid line, where
    an empirical pseudo-miscoverate rate in green needs to be around or below of the $\alpha$ line for
    justifying the correctness of \ACC (Definition \ref{def:correct}).
    The empirical miscoverate rates of base prediction sets (BPS) require to be close to
    desired miscoverate rates $\alpha_{\text{base}}$ to be correct as in (\ref{eq:basecorrect}).
    The miscoverage rates of base prediction sets and consensus sets closely follow
    the desired coverage rates (though they require more samples to be fully converged as Figure \ref{fig:global:miscoverage}).
    The pseudo-miscoverage rate of \ACC is below of a desired coverage rate $\alpha$,
    empirically supporting that it satisfies a desired correctness guarantee.
    Note that
    due to inactive arbitrage of this market,
    we consider a practical extension of a consensus set in \protect\LREF{(\ref{eq:consensussetnu})}{(9)}~with non-zero $\nu$.
    As $\nu$ is non-zero, the pseudo-miscoverage rate of \ACC in Figure \ref{fig:inveth:mis:K1} is more conservative than that of BPS, different to Figure \ref{fig:global:miscoverate:K1}.
    Also, in Figure \ref{fig:inveht:miscoverage:K3}, the empirical pseudo-miscoverage rate of the consensus sets (in green) is
    slightly increased around price manipulation,
    which are related to the behavior of base prediction sets and the potential violation of Assumption \ref{ass:mainassumption}.
    In particular, the empirical miscoverate rates of base prediction sets are affected by the price manipulation, which in turn affect consensus sets.
      Also, the violation of Assumption \ref{ass:mainassumption} is observed around the manipulation
    due to huge price manipulation followed by slow arbitrage.
    Considering that price manipulation is rare in practice,
    the empirical pseudo-miscoverage rate of \ACC would be reduced afterward. 
  }
  \label{fig:local:miscoverage}
\end{figure*}



\begin{figure*}[th!]
  \centering
  \subfigure[Consensus sets]{
    \centering
    \includegraphics[width=0.32\linewidth]{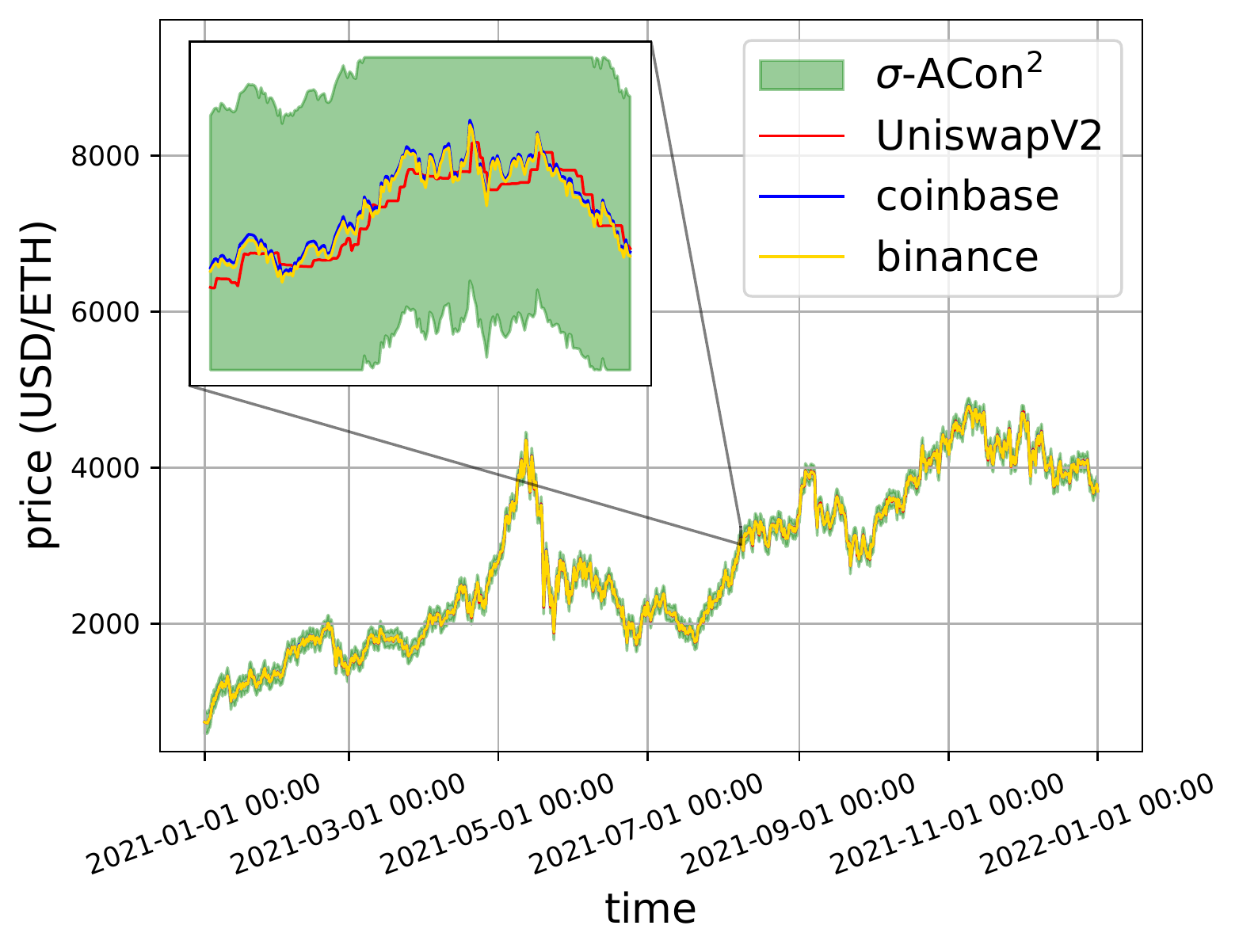}
    \label{fig:sigmabps:set}
  }
  \subfigure[Miscoverage rate]{
    \centering
    \includegraphics[width=0.32\linewidth]{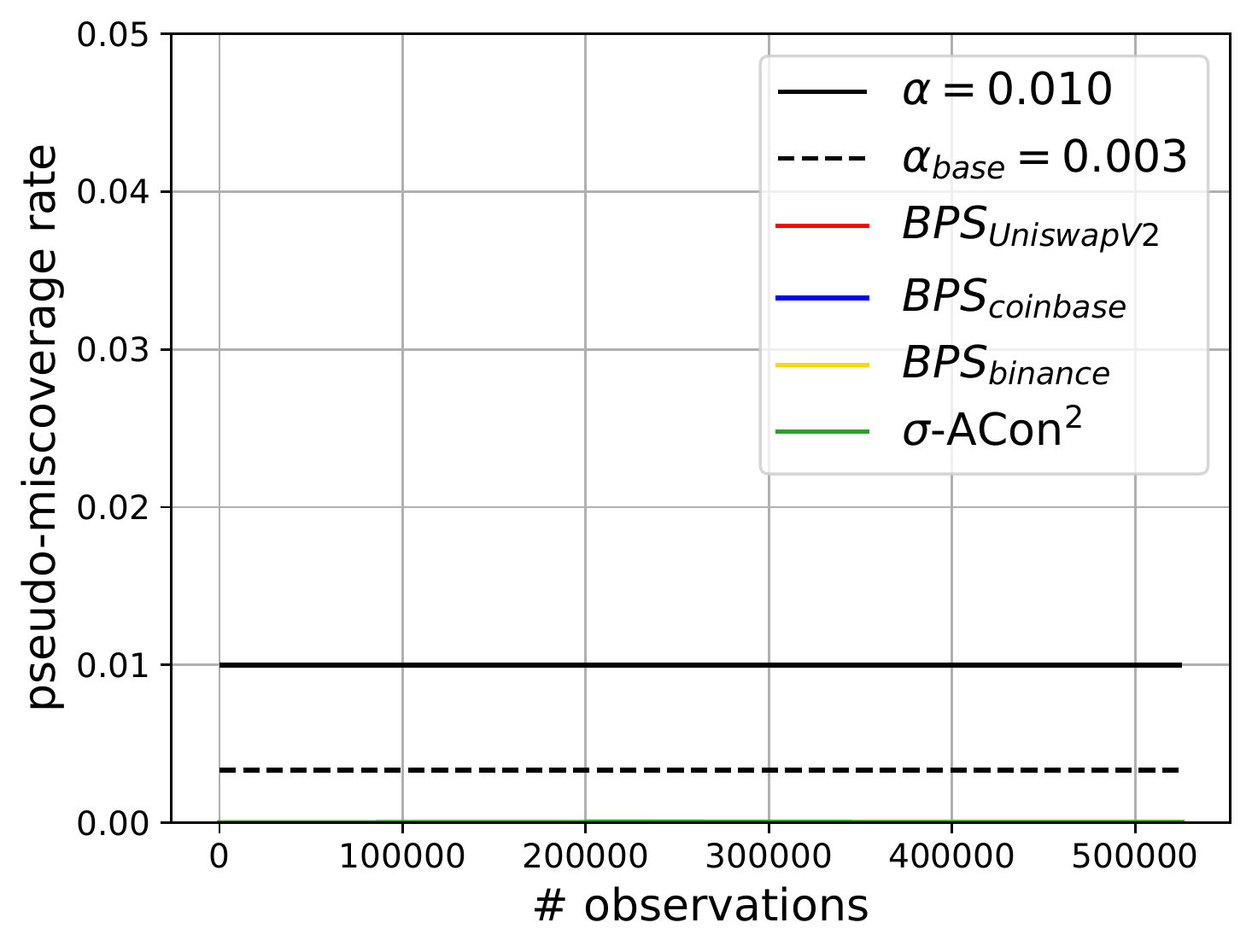}
    \label{fig:sigmabps:mc}
  }  
  \caption{
    $\sigma$-BPS results on USD/ETH data.
    We apply $\sigma$-BPS along with \ACC, denoted by $\sigma$-\ACC.
    Figure \ref{fig:sigmabps:set} demonstrates that the baseline produces large intervals,
    meaning that they do not provide useful uncertainty for downstream applications. 
    The empirical pseudo-miscoverage rate in Figure \ref{fig:sigmabps:mc} is too conservative
    (\ie almost zero).
    This suggests that learning threshold in adaptive conformal prediction is necessary. 
  }
  \label{fig:sigmabps}
\end{figure*}

%% file: apdxext.tex
\twocolumn

\begin{figure*}[t!]
  \centering
  \subfigure[Consensus sets]{
    \centering
    \includegraphics[width=0.32\linewidth]{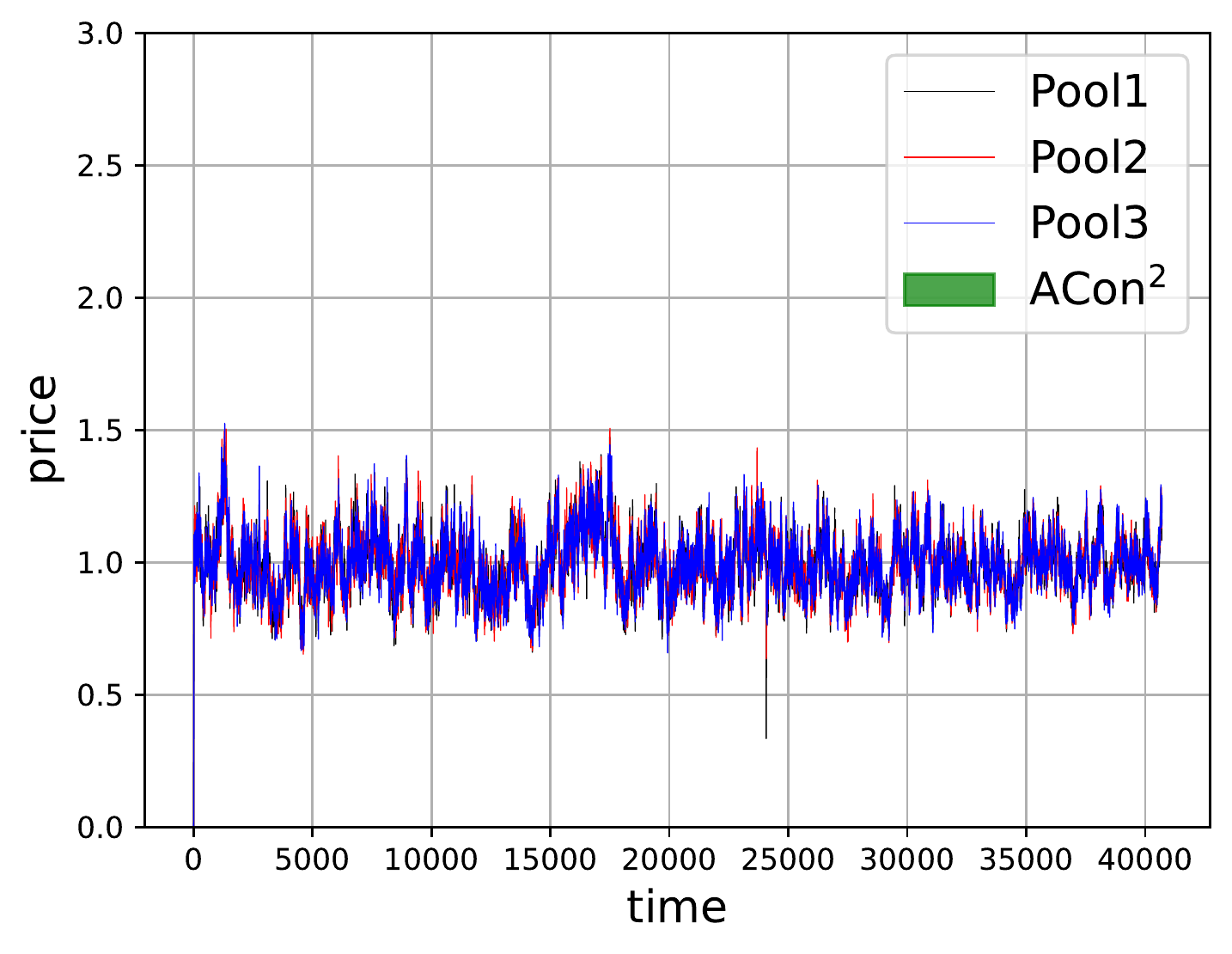}
    \label{fig:eth:ps}
  }
  \subfigure[Miscoverage rate]{
    \centering
    \includegraphics[width=0.32\linewidth]{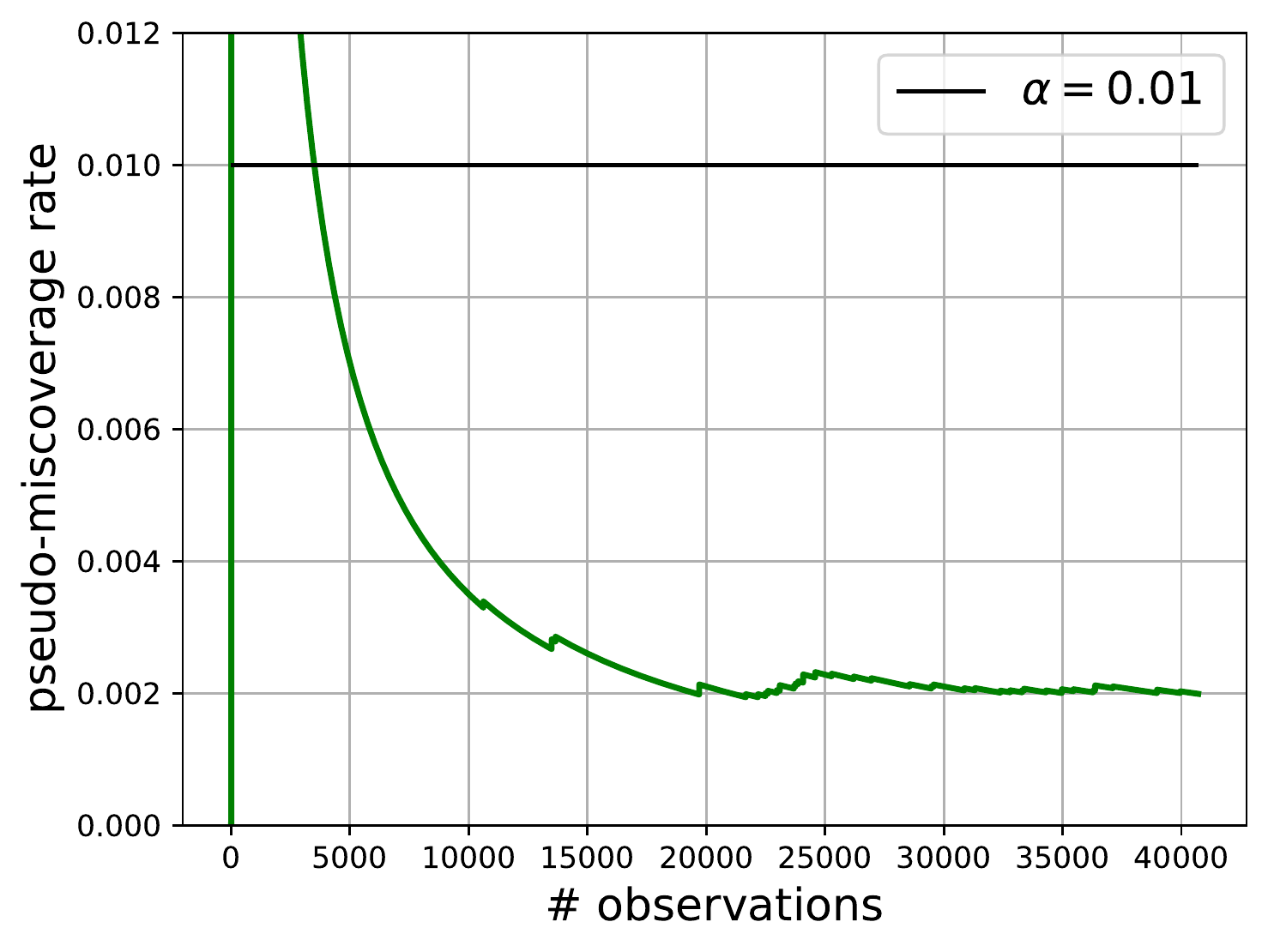}
    \label{fig:eth:miscover}
  }

  \caption{
    \ACC results on local Ethereum network data.
    As mentioned in Section \ref{s:impl},
    we establish a simulated environment on the Ethereum network for evaluating our approach.
    This environment consists of three AMMs, one trader, one arbitrageur, and
    one adversary.
    Figure \ref{fig:eth:ps} shows the consensus sets over time. 
    As can be seen, the prices from three AMMs are synchronized due to the arbitrageur,
    where the consensus sets cover the prices from three AMMs.
    Around 24000, an adversary manipulates the price of Pool3, while the consensus set
    is not affected by this. 
    Figure \ref{fig:eth:miscover} shows the empirical pseudo-miscoverage rate.
    It is below of the desired miscoverage rate $\alpha=0.01$, as specified.
    At around 24000, the miscoverage rate increased due to the price manipulation;
    this is mainly because
    the miscoverage rate increases from base prediction sets, which affects the miscoverage of consensus sets, and
    there can be 
    a transient period when Assumption \ref{ass:mainassumption} is violated
    from huge price manipulation followed by slow arbitrage. 
  }
  \label{fig:ethnetwork}
\end{figure*}

\begin{figure*}[t!]
  \centering
  \subfigure[$K=3$]{
    \centering
    \includegraphics[width=0.32\linewidth]{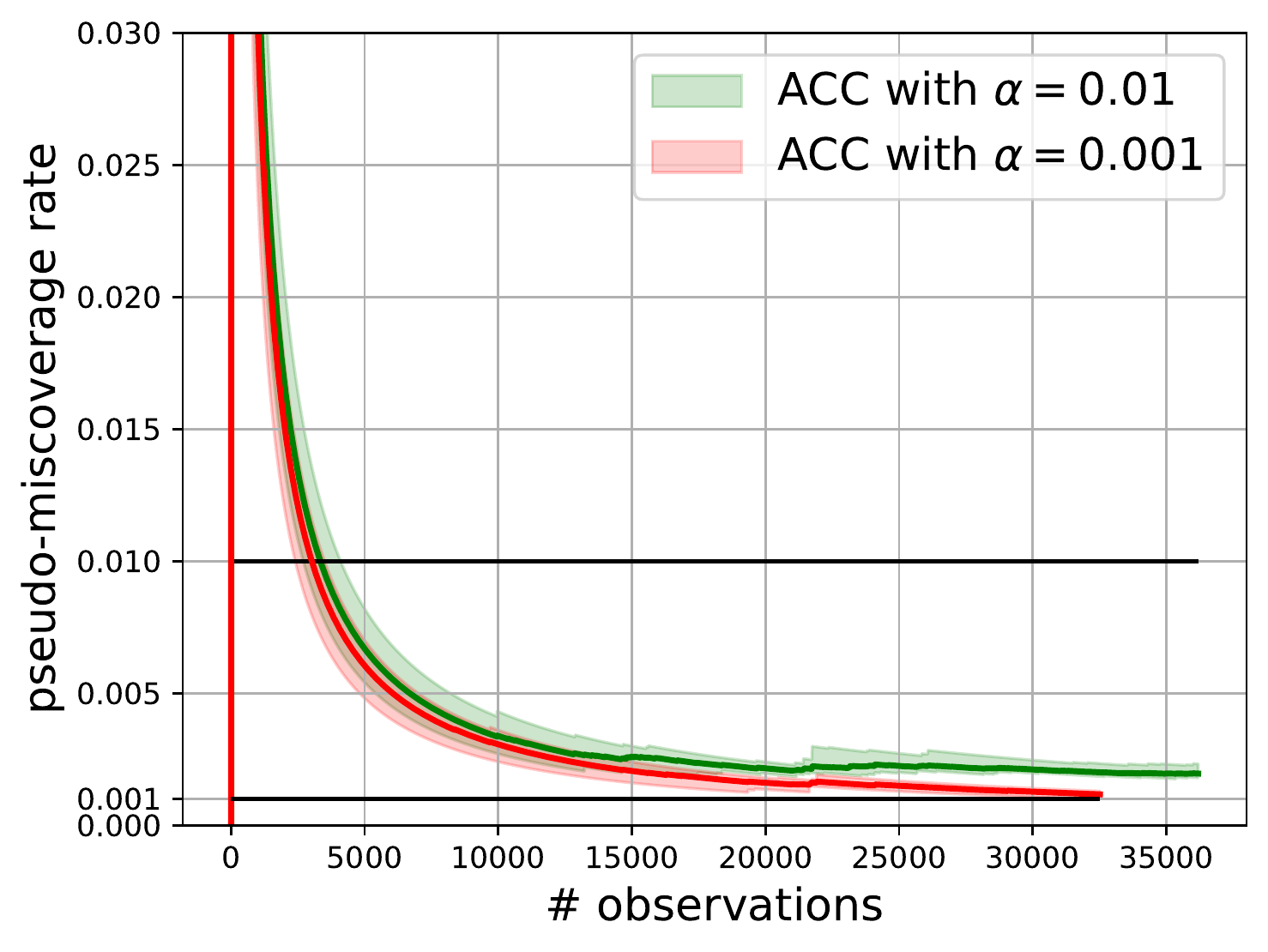}
    \label{fig:eth:K3}
  }
  \subfigure[$K=4$]{
    \centering
    \includegraphics[width=0.32\linewidth]{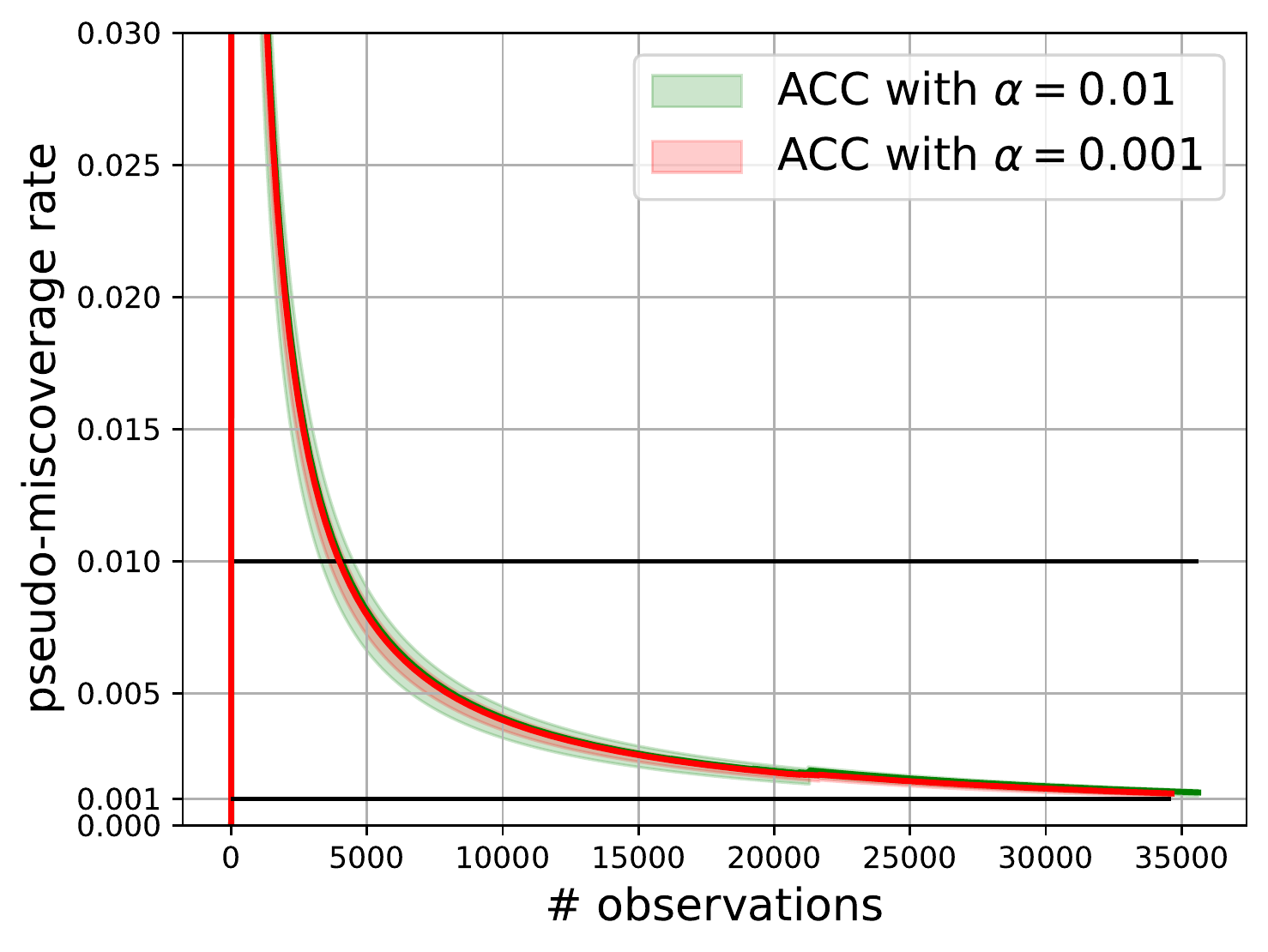}
    \label{fig:eth:K4}
  }
  \subfigure[$K=5$]{
    \centering
    \includegraphics[width=0.32\linewidth]{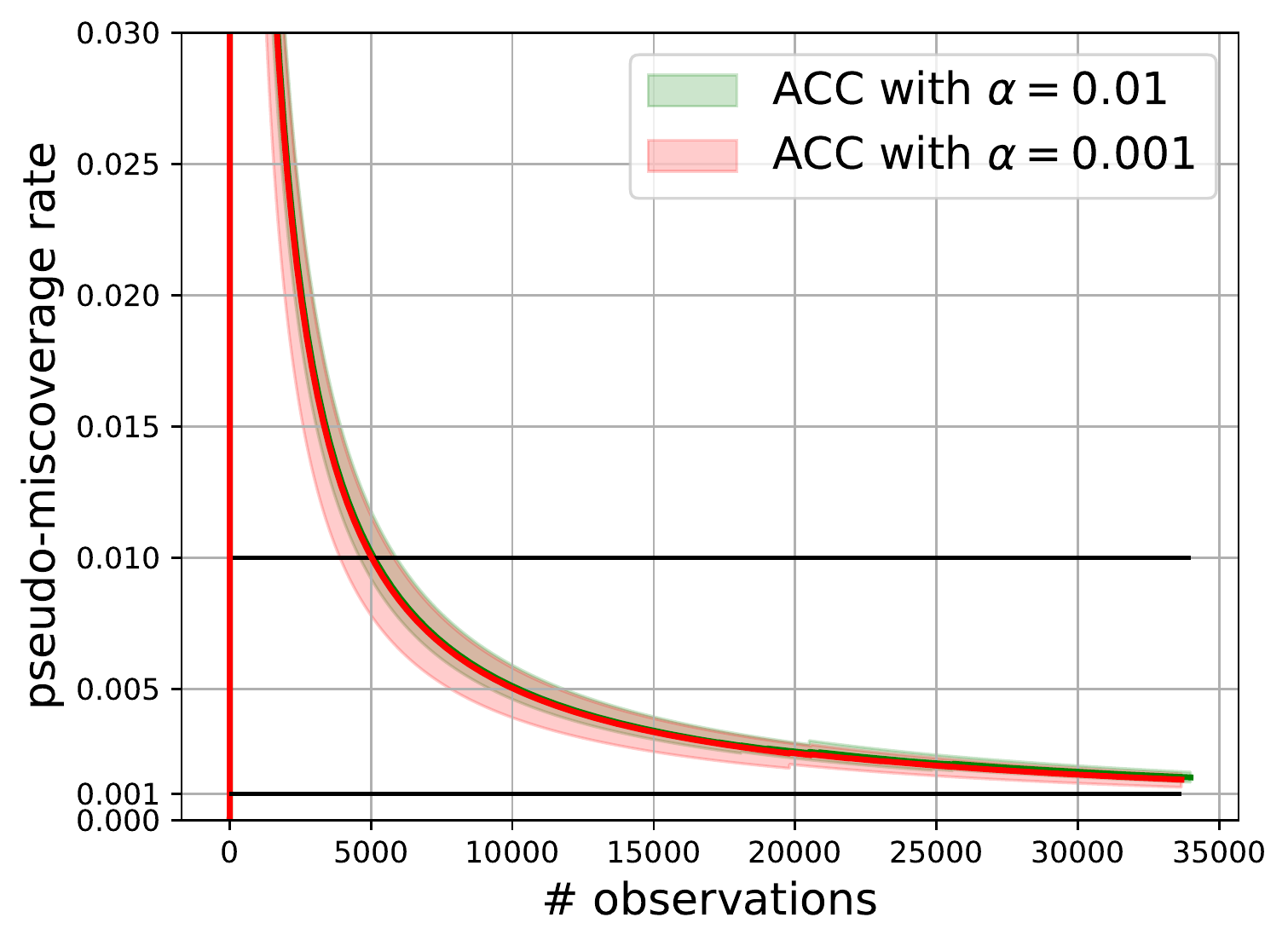}
    \label{fig:eth:K5}
  }
  \caption{
    \MR{
      Miscoverage rates on local Ethereum network data, where
      price manipulation occurs at around 24000. 
    Each $K$, we consider two different $\alpha \in \{0.01, 0.001\}$ as desired miscoverage rates
    with $\beta=1$.
    We conducted 5 random experiments from which we compute the mean pseudo-miscoverage rates
    in a solid red line for $\alpha=0.001$ and a solid green line for $\alpha=0.01$,
    where the variances of rates are depicted in transparent regions.
    For each $\alpha$, if an empirical pseudo-miscoverage rate is below or around of the $\alpha$ line,
    it empirically justifies the correctness of \ACC. 
    For various $K$ and $\alpha$, \ACC consistently satisfies a desired miscoverage rate.
    As $K$ gets larger, each base prediction set needs to satisfy a more conservative desired miscoverage $\alpha_k$ as $\alpha_k = \alpha / K$, leading to larger base prediction intervals.
    Moreover, consensus sets are composed of these larger, multiple base prediction intervals. 
    Thus, the consensus sets rarely mis-cover data, resulting in the conservative empirical pseudo-miscoverage rates.
    Moreover, 
    given $K$, \ACC with a smaller $\alpha$ tends to produce a more conservative empirical pseudo-miscoverage rate since $\alpha_k$ gets smaller, as before.
    Note that at the time of price manipulation,
    the pseudo-miscoverage rate is slightly increased due to the miscoverage rate increase of base prediction sets and a transient period of violating Assumption \ref{ass:mainassumption},
    as also mentioned in Figure \ref{fig:local:miscoverage}.
    }
  }
  \label{fig:ethnetwork:variousKalpha}
\end{figure*}

\begin{figure*}[tb!]
  \centering
  \small
  \newcommand\SCALE{0.6}
  \newcommand\MBW{2.8cm}
  \newcommand\MBH{1.4cm}
  \newcommand\GAP{0.5cm}
  \newcommand\GAPROW{0.8cm}
  \newcommand\GAPCOL{0.5cm}

  \subfigure[Prediction consensus] {
  \scalebox{\SCALE}{
    \begin{tikzpicture}

      \node [
        draw,
        fill=white,
        minimum width=\MBW,
        minimum height=\MBH,
      ] (S1) at (0, 0) {
	\makecell{
          prediction set
          \\
          $\Ch_{t, 1}$
        }
      };

      \node [
        above=\GAPROW of S1
      ] (CS1) {
	\makecell{
          example $\x_t$
          \\
          label $\y_{t-1, 1}$
        }
      };

      \node [
        draw=black,
        rounded corners,
        thick,
        fit={(S1) (CS1)},
        label={above:\makecell{source 1}}
      ] (SRC1) {};

      \node [
        draw,
        fill=white,
        very thick,
        minimum width=\MBW,
        minimum height=\MBH,
        below=\GAPROW of S1
      ] (C) {
	\makecell{
          Consensus
          \\
          $\Ch_t(\x_t)$
        }
      };

      \node [
        draw,
        fill=white,
        minimum width=\MBW,
        minimum height=\MBH,
        below left=\GAP of C
      ] (S2) {
	\makecell{
          prediction set
          \\
          $\Ch_{t, 2}$
        }
      };

      \node [
        above=\GAPROW of S2
      ] (CS2) {
	\makecell{
          example $\x_t$ \\
          label $\y_{t-1, 2}$
        }
      };

      \node [
        draw=black,
        rounded corners,
        thick,
        fit={(S2) (CS2)},
        label={above:\makecell{source 2}}
      ] (SRC2) {};

      \node [
        draw,
        fill=gray!20,
        minimum width=\MBW,
        minimum height=\MBH,
        below right=\GAP of C
      ] (S3) {
	\makecell{
          prediction set
          \\
          $\Ch_{t, 3}$
        }
      };

      \node [
        above=\GAPROW of S3 %
      ] (CS3) {
	\makecell{
          example $\x_{t}$
          \\
          label $\y_{t-1, 3}$
        }
      };

      \node [
        draw=black,
        rounded corners,
        thick,
        fit={(S3) (CS3)},
        label={above:\makecell{source 3}}
      ] (SRC3) {};

      
      \draw[-stealth] (S1.south) to [bend right=0] node[midway,left] {\footnotesize$\Ch_{t, 1}(\x_t)$} (C.north);
      \draw[-stealth] (S2.east) to [bend right=25] node[midway,below,xshift=1.5ex] {\footnotesize$\Ch_{t, 2}(\x_t)$} (C.south);
      \draw[-stealth] (S3.west) to [bend left=25] node[midway,below,xshift=-1.5ex] {\footnotesize$\Ch_{t, 3}(\x_t)$} (C.south);

      \draw[-stealth] (CS1) to [bend left=0] node[midway,right] {} (S1);
      \draw[-stealth] (CS2) to [bend left=0] node[midway,above] {} (S2);
      \draw[-stealth] (CS3) to [bend left=0] node[midway,above] {} (S3);

    \end{tikzpicture}
  }
  }
  \hspace{3ex}
  \subfigure[Price consensus] {
  \scalebox{\SCALE}{
    \begin{tikzpicture}

      \node [
        draw,
        fill=white,
        minimum width=\MBW,
        minimum height=\MBH,
      ] (S1) at (0, 0) {
	\makecell{
          prediction set
          \\
          $\Ch_{t, 1}$
        }
      };

      \node [
        above=\GAPROW of S1
      ] (CS1) {
	\makecell{
          observation $\x_t$
          \\
          price $\y_{t-1, 1}$
        }
      };

      \node [
        draw=black,
        rounded corners,
        thick,
        fit={(S1) (CS1)},
        label={above:\makecell{MM 1} }
      ] (SRC1) {};

      \node [
        draw,
        fill=white,
        very thick,
        minimum width=\MBW,
        minimum height=\MBH,
        below=\GAPROW of S1
      ] (C) {
	\makecell{
          Consensus
          \\
          $\Ch_t(\x_t)$
        }
      };

      \node [
        draw,
        fill=white,
        minimum width=\MBW,
        minimum height=\MBH,
        below left=\GAP of C
      ] (S2) {
	\makecell{
          prediction set
          \\
          $\Ch_{t, 2}$
        }
      };

      \node [
        above=\GAPROW of S2
      ] (CS2) {
	\makecell{
          observation $\x_t$ \\
          price $\y_{t-1, 2}$
        }
      };

      \node [
        draw=black,
        rounded corners,
        thick,
        fit={(S2) (CS2)},
        label={above:\makecell{AMM 2}}
      ] (SRC2) {};

      \node [
        draw,
        fill=gray!20,
        minimum width=\MBW,
        minimum height=\MBH,
        below right=\GAP of C
      ] (S3) {
	\makecell{
          prediction set
          \\
          $\Ch_{t, 3}$
        }
      };

      \node [
        above=\GAPROW of S3 %
      ] (CS3) {
	\makecell{
          observation $\x_{t}$
          \\
          price $\y_{t-1, 3}$
        }
      };

      \node [
        draw=black,
        rounded corners,
        thick,
        fit={(S3) (CS3)},
        label={above:\makecell{AMM 3}}
      ] (SRC3) {};

      \node [
        draw=black,
        rounded corners,
        dashed,
        thick,
        fit={(C) (S1) (SRC2) (SRC3)},
        inner sep=2ex,
        label={above left:\makecell{On-chain}}
      ] {};
      
      \draw[-stealth] (S1.south) to [bend right=0] node[midway,left] {\footnotesize$\Ch_{t, 1}(\x_t)$} (C.north);
      \draw[-stealth] (S2.east) to [bend right=25] node[midway,below,xshift=1.5ex] {\footnotesize$\Ch_{t, 2}(\x_t)$} (C.south);
      \draw[-stealth] (S3.west) to [bend left=25] node[midway,below,xshift=-1.5ex] {\footnotesize$\Ch_{t, 3}(\x_t)$} (C.south);

      \draw[-stealth] (CS1) to [bend left=0] node[midway,right] {} (S1);
      \draw[-stealth] (CS2) to [bend left=0] node[midway,above] {} (S2);
      \draw[-stealth] (CS3) to [bend left=0] node[midway,above] {} (S3);

    \end{tikzpicture}
  }
  }
  
  \caption{
    Prediction consensus and its application to price consensus.
    In prediction consensus, we aim to construct a consensus set $\Ch_t$
    that likely contains a consensus label even under
    distribution shift and 
    the Byzantine adversaries (as shown in gray), which
    undermine consensus. 
    The consensus set is constructed  
    based on votings from base prediction sets $\Ch_{t, 1}$, $\Ch_{t, 2}$, and $\Ch_{t, 3}$ from
    multiple data sources. Each base prediction sets are updated via online machine learning, thus
    the consensus set is also indirectly updated based on data from each source.
  }
  \label{fig:problem}
\end{figure*}
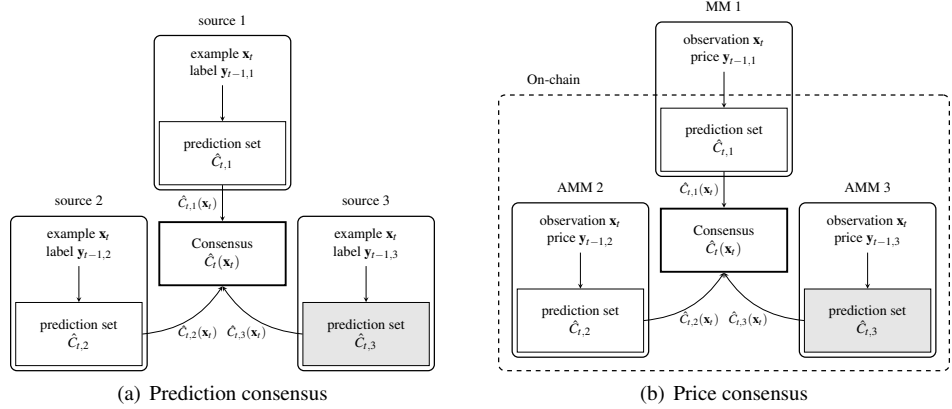

\section{MVP Algorithm}
\label{apdx:mvpalg}

\begin{algorithm}[thb!]
  \caption{MVP \cite{bastani2022practical} for the $k$-th source.
}
  \label{alg:smvp}
  \small
  \begin{algorithmic}[1]
    \Procedure{$\text{update}_{\text{MVP}}$}{$\Ch_{t, k}$, $\x_{t}$, $\y_{t, k}$; $\eta$}
    \State $n \gets n_{t, k}$
    \State $v \gets v_{t, k}$
    \State $err \gets \mathbbm{1} \( \y_{t, k} \notin \Ch_{t, k}(\x_{t}) \)$
    \State $n_{i} \gets n_{i} + 1$ if $\tau_{t, k}$ is in the $i$-th bin
    \State $v_{i} \gets v_{i} + \alpha - err$ if $\tau_{t, k}$ is in the $i$-th bin
    \State $w_0 \gets e^{\eta v_{0} / f(n_0)} - e^{-\eta v_0 / f(n_0)}$
    \If{$w_0 > 0$}
    \State $pos \gets True$
    \Else
    \State $pos \gets False$
    \EndIf
    \For{$i = 1, \dots, m-1$} \label{alg:loop}
    \State $w_i \gets e^{\eta v_i / f(n_i)} - e^{-\eta v_i / f(n_i)}$
    \If{$w_i > 0$}
    \State $pos \gets True$
    \EndIf
    \If{$w_i \cdot w_{i-1} \le 0$}
    \If{$|w_i| + |w_{i-1}| = 0$}
    \State $b_i \gets 1$
    \Else
    \State $b_i \gets \frac{|w_i|}{|w_i| + |w_{i-1}|}$
    \EndIf
    \If{$rand() \le b_i$}
    \State $\Comment{\text{$rand()$ randomly chooses a real number in $[0, 1)$}}$
    \State $\tau_{t+1, k} \gets 1 - \tau_{\text{max}} \(\frac{i}{m} - \frac{1}{rm} \)$
    \State \textbf{return} $\tau_{t+1, k}$ 
    \Else
    \State $\tau_{t+1, k} \gets 1 - \tau_{\text{max}} \frac{i}{m}$
    \State \textbf{return} $\tau_{t+1, k}$ 
    \EndIf
    \EndIf
    \EndFor
    \If{$pos$}
    \State $\tau_{t+1, k} \gets \tau_{\text{max}}$
    \Else
    \State $\tau_{t+1, k} \gets 0$
    \EndIf
    \State \textbf{return} $\tau_{t+1, k}$
    \EndProcedure
  \end{algorithmic}
\end{algorithm}

\input{impl.tex}

\onecolumn
\section{Lemmas and Proofs}

\subsection{A Special Case of Theorem \ref{thm:consensuscorrectness}}
\label{apdx:specialcaseanalysis}

We consider a simpler case of Theorem \ref{thm:consensuscorrectness}, assuming
$T \to \infty$ and a fixed but unknown $\beta$ Byzantine adversary $e \in \Es_{\beta}$.
This specialized asymptotic analysis highlights
the main proof ideas for the finite-sample guarantee of Theorem \ref{thm:consensuscorrectness}.
In particular, the same proof ideas are used in proving a general case in Lemma \ref{lem:local2global}.

\begin{lemma}
  \label{lem:specialcase}
  For any $k \in \{1, \dots, K\}$ and $\beta$ Byzantine adversary $e \in \Es_\beta$,
  if $\hat{C}_k$ satisfies
  \begin{align*}
    \Prob \left\{ \Y_{k} \notin \hat{C}_k(e(\X)) \right\} = \alpha_k,
  \end{align*}
  where the probability is taken over $\X$ and $\Y_k$,
  then the consensus set $\hat{C}$ satisfies
  \begin{align*}
    \Prob \left\{ Y \notin \hat{C}(e(\X)) \right\} \le \sum_{k=1}^K \alpha_k,
  \end{align*}
  where the probability is taken over $\X$ and $Y$.
\end{lemma}

\subsection{Proof of Lemma \ref{lem:specialcase}}

Let
$\Ks_{K-\beta} \coloneqq \{ \Ss \subseteq \{1, \dots, K\} \mid |\Ss| \ge K-\beta\}$
and
$\Ss^* \in \Ks_{K-\beta}$ be a set of indices of sources which are not manipulated by a $\beta$-Byzantine adversary. Then, we have

\begin{align}
  \Prob \left\{ Y \in \Ch(e(\X)) \right\} \nonumber
  &=
  \Prob\left\{ \sum_{k=1}^K \mathbbm{1}\( Y \in \Ch_{k}(e(\X))\) \ge K - \beta  \right\} \nonumber\\
  &=
  \Prob\left\{ \bigvee_{\Ss \in \Ks_{K-\beta}} \bigwedge_{k \in \Ss} \( Y \in \Ch_{k}(e(\X)) \)  \right\} \nonumber\\
  &\ge
  \Prob \left\{ \bigwedge_{k \in \Ss^*} \( Y \in \Ch_{k}(e(\X)) \)  \right\}
  \nonumber
  \\
  &=
  1 - 
  \Prob \left\{ \bigvee_{k \in \Ss^*} \( Y \notin \Ch_{k}(e(\X)) \)  \right\}
  \nonumber
  \\
  &\ge
  1 - 
  \sum_{k \in \Ss^*} \Prob \left\{ Y \notin \Ch_{k}(e(\X))  \right\}
  \nonumber
  \\
  &=
  1 - 
  \sum_{k \in \Ss^*} \Prob \left\{ \Y_{k} \notin \Ch_{k}(e(\X))  \right\}
  \label{eq:spelem:labeldistassumption}
  \\
  &\ge
  1 - 
  \sum_{k=1}^K \Prob \left\{ \Y_{k} \notin \Ch_{k}(e(\X))  \right\}
  \nonumber
  \\
  &=
  1 - 
  \sum_{k=1}^K \alpha_k.
  \nonumber
\end{align}
Here,
recall that $p$ is a probability density function of $\X$, $Y$, and $\Y_k$.
then
(\ref{eq:spelem:labeldistassumption}) holds as follows:
\begin{align}
  \Prob \left\{ Y \notin \Ch_{k}(e(\X)) \right\}
  &= \Exp \mathbbm{1} \left( Y \notin \Ch_{k}(e(\X)) \right)
  \nonumber
  \\
  &= \int \mathbbm{1} \left( y \notin \Ch_{k}(e(\x)) \right) p(\x) p( y \mid e(\x) ) ~\mathrm{d}\x \mathrm{d}y
  \nonumber
  \\
  &= \int \mathbbm{1} \left( \y_{k} \notin \Ch_{k}(e(\x)) \right) {p}(\x) {p}( \y_{k} \mid e(\x) ) ~\mathrm{d}\x \mathrm{d}\y_{k}
  \label{eq:spelem:ass1}
  \\
  &= \Prob \left\{ \Y_{k} \notin \Ch_{k}(e(\X)) \right\},
  \nonumber
\end{align}
where
(\ref{eq:spelem:ass1}) holds due to Assumption \ref{ass:mainassumption}.
This proves the claim.


\subsection{A Supporting Lemma for Theorem \ref{thm:consensuscorrectness}}

The Theorem \ref{thm:consensuscorrectness} proof exploits the following lemma.
Intuitively, the lemma connects the miscoverage rate of a consensus set
to the miscoverage rate of a base prediction set.

\begin{lemma}
  \label{lem:local2global}
  For any
  $T \in \naturalnum$,
  $i \in \{0, \dots, T\}$,
  $k \in \{1, \dots, K\}$,
  $\x_{1:i}$, $\y_{1:i}$, $y_{1:i}$, $\Ch_{1:i, k}$, and $\Ch_{1:i}$,
  if any $k$-th source learner $L_k$  satisfies
  \begin{align}
    \max_{\substack{p_{i+1} \in \Ps' \\ e_{i+1} \in \Es_\beta}} 
    \Exp
    \cdots
    \max_{\substack{p_T \in \Ps' \\ e_T \in \Es_\beta}}
    \Exp
    \sum_{t=i+1}^T \mathbbm{1}\( \Y_{t, k} \notin \Ch_{t, k}(e_t(\X_{t})) \)
    \le \delta_{i, k},
    \label{eq:lem:local2global:cond}
  \end{align}
  where 
  the $t$-th expectation for $i+1 \le t \le T$ is taken over
  $\X_{t} \sim p_t(\x)$, $\Y_{t} \sim p_t(\y\mid e_t(\X_{t}))$, and $\Ch_{t, k} \sim L_k (\x_{1:t-1}, \y_{1:t-1, k}, \Ch_{1:t-1, k})$,
  then
  a consensus learner $L$ satisfies
  \begin{align*}
    \max_{\substack{p_{i+1} \in \Ps \\ e_{i+1} \in \Es_\beta}} 
    \Exp
    \cdots
    \max_{\substack{p_T \in \Ps \\ e_T \in \Es_\beta}}
    \Exp
    \sum_{t=i+1}^T \mathbbm{1}\( Y_t \notin \Ch_t(e_t(\X_{t})) \)
    \le \sum_{k=1}^K \delta_{i, k},
  \end{align*}
  where
  the $t$-th expectation for $i+1 \le t \le T$ is taken over
  $\X_{t} \sim p_t(\x)$, $\Y_{t} \sim p_t(\y\mid e_t(\X_{t}))$, $Y_t \sim p_t(y \mid \X_t)$, and $\Ch_{t} \sim L(\x_{1:t-1}, \y_{1:t-1}, \Ch_{1:t-1})$.
\end{lemma}

\subsection{Proof of Lemma \ref{lem:local2global}}
\label{apdx:proof:lem:local2global}

Consider
$\bar{p}_t \in \Ps$
and
$\bar{e}_t \in \Es_\beta$
for all $t \in \{i+1, \dots, T\}$
that satisfy the following:
\begin{align}
  \max_{\substack{p_{i+1} \in \Ps \\ e_{i+1} \in \Es_\beta}}      
  \Exp                
  \cdots
  \max_{\substack{p_T \in \Ps \\ e_T \in \Es_\beta}}
  \Exp
  \sum_{t=i+1}^T \mathbbm{1}\( Y_t \notin \Ch_t(e_t(\X_{t})) \)
  =
  \sum_{t=i+1}^T \Prob \left\{ Y_t \notin \Ch_t(\bar{e}_t(\X_{t})) \right\} \nonumber,
\end{align}
where
the probability at time $t$ is take over $\X_{t} \sim \bar{p}_t(\x)$, $\Y_{t} \sim \bar{p}_t( \y \mid \bar{e}_t(\X_{t}))$, $Y_t \sim \bar{p}_t(y_t \mid \X_t)$, and $\Ch_{t} \sim L(\x_{1:t-1}, \y_{1:t-1}, \Ch_{1:t-1})$.
Then, for the same $\bar{p}_t$ and $\bar{e}_t$ we have
\begin{align}
  \sum_{t=i+1}^T \Prob \left\{ \Y_{t, k} \notin \Ch_t(\bar{e}_t(\X_{t})) \right\}
  \le
  \max_{\substack{p_{i+1} \in \Ps' \\ e_{i+1} \in \Es_\beta}} 
    \Exp
    \cdots
    \max_{\substack{p_T \in \Ps' \\ e_T \in \Es_\beta}}
    \Exp
    \sum_{t=i+1}^T \mathbbm{1}\( \Y_{t, k} \notin \Ch_{t, k}(e_t(\X_{t})) \)
    \le \delta_{i, k}  \label{eq:keylem:prop1},
\end{align}
where
the probability at time $t$ is take over $\X_{t} \sim \bar{p}_t(\x)$, $\Y_{t, k} \sim \bar{p}_t(\y_{t, k} \mid \bar{e}_t(\X_{t}))$, and $\Ch_{t, k} \sim L_k(\x_{1:t-1}, \y_{1:t-1, k}, \Ch_{1:t-1, k})$.

Then, we consider the upper bound of $\sum_{t=i+1}^T \Prob \left\{ Y_t \notin \Ch_t(\bar{e}_t(\X_{t})) \right\}$.
In particular,
letting
$\Ks_{K-\beta} \coloneqq \{ \Ss \subseteq \{1, \dots, K\} \mid |\Ss| \ge K-\beta\}$
and
$\Ss_t^* \in \Ks_{K-\beta}$ be a set of indices of sources which are not manipulated by a $\beta$-Byzantine adversary at time $t$,
we have
\begin{align}
  \sum_{t=i+1}^T \Prob \left\{ Y_t \in \Ch_t(\bar{e}_t(\X_{t})) \right\} \nonumber
  &=
  \sum_{t=i+1}^T \Prob\left\{ \sum_{k=1}^K \mathbbm{1}\( Y_t \in \Ch_{t, k}(\bar{e}_t(\X_{t}))\) \ge K - \beta  \right\} \nonumber\\
  &=
  \sum_{t=i+1}^T \Prob\left\{ \bigvee_{\Ss_t \in \Ks_{K-\beta}} \bigwedge_{k \in \Ss_t} \( Y_t \in \Ch_{t, k}(\bar{e}_t(\X_{t})) \)  \right\} \nonumber\\
  &\ge
  \sum_{t=i+1}^T \Prob \left\{ \bigwedge_{k \in \Ss_t^*} \( Y_t \in \Ch_{t, k}(\bar{e}_t(\X_{t})) \)  \right\} \nonumber
  \\
  &=
  T - i - 
  \sum_{t=i+1}^T \Prob \left\{ \bigvee_{k \in \Ss_t^*} \( Y_t \notin \Ch_{t, k}(\bar{e}_t(\X_{t})) \)  \right\}
  \nonumber
  \\
  &\ge
  T - i -
  \sum_{t=i+1}^T \sum_{k \in \Ss_t^*} \Prob \left\{ Y_t \notin \Ch_{t, k}(\bar{e}_t(\X_{t}))  \right\}
  \nonumber
  \\
  &=
  T - i - 
  \sum_{t=i+1}^T \sum_{k \in \Ss_t^*} \Prob \left\{ \Y_{t, k} \notin \Ch_{t, k}(\bar{e}_t(\X_{t}))  \right\}
  \label{eq:keylem:labeldistassumption}
  \\
  &\ge
  T - i - 
  \sum_{t=i+1}^T \sum_{k=1}^K \Prob \left\{ \Y_{t, k} \notin \Ch_{t, k}(\bar{e}_t(\X_{t}))  \right\}
  \nonumber
  \\
  &=
  T - i - 
  \sum_{k=1}^K \sum_{t=i+1}^T \Prob \left\{ \Y_{t, k} \notin \Ch_{t, k}(\bar{e}_t(\X_{t}))  \right\}
  \nonumber
  \\
  &\ge
  T - i - 
  \sum_{k=1}^K \delta_{i, k},
  \label{eq:keylem:prop1apply}
\end{align}
where (\ref{eq:keylem:prop1apply}) holds due to (\ref{eq:keylem:prop1}).
Here,
(\ref{eq:keylem:labeldistassumption}) holds as follows:
\begin{align}
  \Prob \left\{ Y_t \notin \Ch_{t, k}(\bar{e}_t(\X_{t})) \right\}
  &= \Exp \mathbbm{1} \left( Y_t \notin \Ch_{t, k}(\bar{e}_t(\X_{t})) \right)
  \nonumber
  \\
  &= \int \mathbbm{1} \left( y_t \notin \Ch_{t, k}(\bar{e}_t(\x_{t})) \right) \bar{p}_t(\x_t) \bar{p}_t( y_t \mid \bar{e}_t(\x_t) ) ~\mathrm{d}\x_t \mathrm{d}y_t
  \nonumber
  \\
  &= \int \mathbbm{1} \left( \y_{t, k} \notin \Ch_{t, k}(\bar{e}_t(\x_{t})) \right) \bar{p}_t(\x_t) \bar{p}_t( \y_{t, k} \mid \bar{e}_t(\x_t) ) ~\mathrm{d}\x_t \mathrm{d}\y_{t, k}
  \label{eq:keylem:ass1}
  \\
  &= \Prob \left\{ \Y_{t, k} \notin \Ch_{t, k}(\bar{e}_t(\X_{t})) \right\},
  \nonumber
\end{align}
where
(\ref{eq:keylem:ass1}) holds due to Assumption \ref{ass:mainassumption}.
This proves the claim.

\subsection{Proof of Theorem \ref{thm:consensuscorrectness}}
\label{apdx:proof:them:consensuscorrectness}

To avoid clutter, consider that the expectation at time $t$ is take over $\X_{t} \sim p_t(\x)$, $\Y_{t} \sim p_t(\y\mid e_t(\X_{t}))$, $Y_t \sim p_t(y \mid \X_t)$, and $\Ch_{t} \sim L(\z_{1:t-1})$.
We first prove a general statement; for any $i \in \{1, \dots, T\}$, we have
\begin{align*}
  T \ep_{T, k}
  \ge
  T \Vs_k(\Fs_k, T, \alpha_k, \beta, L_k)
  \ge
  \sum_{t=1}^{i} \mathbbm{1}\( \y_{t, k} \notin \Ch_{t, k}(e_t(\x_{t})) \)
  +
  \max_{\substack{p_{i+1} \in \Ps' \\ e_{i+1} \in \Es_\beta}}
  \Exp
  \cdots
  \max_{\substack{p_T \in \Ps' \\ e_T \in \Es_\beta}}
  \Exp    
  \sum_{t=i+1}^T \mathbbm{1}\( \Y_{t, k} \notin \Ch_{t, k}(e_t(\X_{t})) \)
  - T \alpha_k
\end{align*}
for some $(e_t(\x_1), \dots, e_t(\x_i))$, $(\y_{1,k}, \dots, \y_{i, k})$, and $(\Ch_{1, k}, \cdots, \Ch_{i, k})$.
Thus, we have
\begin{align*}
  \max_{\substack{p_{i+1} \in \Ps' \\ e_{i+1} \in \Es_\beta}}
  \Exp
  \cdots
  \max_{\substack{p_T \in \Ps' \\ e_T \in \Es_\beta}}
  \Exp    
  \sum_{t=i+1}^T \mathbbm{1}\( \Y_{t, k} \notin \Ch_{t, k}(e_t(\X_{t})) \)
  \le
  T\ep_{T, k} + T \alpha_k - \sum_{t=1}^{i} \mathbbm{1}\( \y_{t, k} \notin \Ch_{t, k}(e_t(\x_{t})) \).
\end{align*}

Due to Lemma \ref{lem:local2global},     
\begin{align*}
  \max_{\substack{p_{i+1} \in \Ps \\ e_{i+1} \in \Es_\beta}} 
  \Exp
  \cdots
  \max_{\substack{p_T \in \Ps \\ e_T \in \Es_\beta}}
  \Exp
  \sum_{t=i+1}^T \mathbbm{1}\( Y_t \notin \Ch_t(e_t(\X_{t})) \)
  &\le
  \sum_{k=1}^K\( T\ep_{T, k} + T \alpha_k - \sum_{t=1}^{i} \mathbbm{1}\( \y_{t, k} \notin \Ch_{t, k}(e_t(\x_{t})) \)  \)
  \\
  &= \sum_{k=1}^K T \ep_{T, k} + \sum_{k=1}^K T \alpha_k -
  \sum_{k=1}^K \sum_{t=1}^{i} \mathbbm{1}\( \y_{t, k} \notin \Ch_{t, k}(e_t(\x_{t})) \) 
\end{align*}
By rearranging terms, we have
\begin{align*}
  \sum_{k=1}^K \sum_{t=1}^{i} \mathbbm{1}\( \y_{t, k} \notin \Ch_{t, k}(e_t(\x_{t})) \) +
  \max_{\substack{p_{i+1} \in \Ps \\ e_{i+1} \in \Es_\beta}} 
  \Exp
  \cdots
  \max_{\substack{p_T \in \Ps \\ e_T \in \Es_\beta}}
  \Exp
  \sum_{t=i+1}^T \mathbbm{1}\( Y_t \notin \Ch_t(e_t(\X_{t})) \)
  -  \sum_{k=1}^K T \alpha_k
  &\le
  \sum_{k=1}^K  T \ep_{T, k}.
\end{align*}
By setting $i=0$, we have
\begin{align*}
  \frac{1}{T} \max_{\substack{p_{1} \in \Ps \\ e_{1} \in \Es_\beta}} 
  \Exp
  \cdots
  \max_{\substack{p_T \in \Ps \\ e_T \in \Es_\beta}}
  \Exp
  \sum_{t=1}^T \mathbbm{1}\( Y_t \notin \Ch_t(e_t(\X_{t})) \)
  -  \sum_{k=1}^K \alpha_k
  &\le
  \sum_{k=1}^K \ep_{T, k},
\end{align*}
as claimed.

\subsection{Proof of Lemma \ref{lem:valueconnection}}
\label{apdx:proof:lem:valueconnection}

We have 
\begin{align}
  \Vs_k(\Fs_k, T, \alpha_k, \beta, L_k)
  &=
  \max_{\substack{p_1 \in \Ps \\ e_1 \in \Es_\beta}}
  \Expop_{\substack{\X_{1} \sim p_1(\x) \\ \Y_{1, k} \sim p_1(\y_k \mid e_1(\X_{1})) \\ \Ch_{1, k} \sim L_k(\cdot)}}
  \dots
  \max_{\substack{p_T \in \Ps \\ e_T \in \Es_\beta}}
  \Expop_{\substack{\X_{T} \sim p_T(\x) \\ \Y_{T, k} \sim p_T(\y_k \mid e_T(\X_{T})) \\ \Ch_{T, k} \sim L_k(\cdot)}}
  \left\{
  \frac{1}{T} \sum_{t=1}^T \mathbbm{1}\( \Y_{t, k} \notin \Ch_{t, k}(e_t(\X_{t})) \) - \alpha
  \right\} \label{eq:lem:valueconnection:1}
  \\
  &\le
  \max_{\substack{p_1 \in \Ps'}}
  \Expop_{\substack{(\X_{1}, \Y_{1, k}) \sim p_1 \\ \Ch_{1, k} \sim L_k(\cdot)}}
  \dots
  \max_{\substack{p_T \in \Ps'}}
  \Expop_{\substack{(\X_{T}, \Y_{T, k}) \sim p_T \\ \Ch_{T, k} \sim L_k(\cdot)}}
  \left\{
  \frac{1}{T} \sum_{t=1}^T \mathbbm{1}\( \Y_{t, k} \notin \Ch_{t, k}(\X_{t}) \) - \alpha
  \right\}  \label{eq:lem:valueconnection:2}
  \\
  &\le
  \max_{\substack{p_1 \in \Ps'}}
  \Expop_{\substack{(\X_{1}, \Y_{1, k}) \sim p_1 \\ \Ch_{1, k} \sim L_k(\cdot)}}
  \dots
  \max_{\substack{p_T \in \Ps'}}
  \Expop_{\substack{(\X_{T}, \Y_{T, k}) \sim p_T \\ \Ch_{T, k} \sim L_k(\cdot)}}
  \left|
  \frac{1}{T} \sum_{t=1}^T \mathbbm{1}\( \Y_{t, k} \notin \Ch_{t, k}(\X_{t}) \) - \alpha
  \right| \nonumber
  \\
  &= \Vs'(\Fs_k, T, \alpha_k, L_k). \nonumber
\end{align}
Here, the first inequality holds as
the Byzantine adversaries $e_i$ for $i \in \{1, \dots, T\}$ may not choose to
manipulate examples for the $k$-th source in (\ref{eq:lem:valueconnection:1}),
but (\ref{eq:lem:valueconnection:2}) is equivalent to
always manipulating examples for the $k$-th source due to the maximum over $\Ps'$, thus forming an upper bound.

\subsection{Proof of Lemma \ref{lem:mvpbound}}
\label{apdx:proof:lem:mvpbound}

Let
$m = |\Fs|$,
$B_i = \big[ \frac{i-1}{m}, \frac{i}{m} \big)$,
$B_m = \[ \frac{m-1}{m}, 1 \]$,
$S_i = \{ t \in \{1, \dots, T\} \mid \tau_t \in B_i \}$,
$f(n) \coloneqq \sqrt{(n+1) \log_2^2 (n+2)}$, and
$K_1 = \sum_{n=0}^\infty \frac{1}{f(n)^2}$.  
Based on Theorem 3.1 of \cite{bastani2022practical},
the MVP learner is threshold-calibrated, multi-valid, which means that
for all $i \in \{1, \dots, m \}$
the value of the learner is bounded as follows:
\begin{align*}
  \max_{p_1 \in \Ps'} \Expop_{\substack{(X_1, Y_1) \sim p_1 \\ \Ch_1 \sim L_{\text{MVP}}(\cdot)}}
  \cdots
  \max_{p_T \in \Ps'} \Expop_{\substack{(X_T, Y_T) \sim p_T \\ \Ch_T \sim L_{\text{MVP}}(\cdot)}}
  \left| \frac{1}{S_i} \sum_{t \in S_i} \text{Miscover}(\Ch_t, X_t, Y_t) - \alpha \right|
\le \frac{f(|S_i|)}{|S_i|} \sqrt{4 K_1 m \ln m }
\end{align*}
if the distribution over scores $s_t(X_t, Y_t)$ for any $t \in \{1, \dots, T\}$ is smooth enough
and
$\eta = \sqrt{\frac{\ln m}{2 K_1 m}}$.
Thus, we have
\begin{align*}
  \Vs'(\Fs, T, \alpha, L_{\text{MVP}})
  &=
  \max_{\substack{p_1 \in \Ps'}}
  \Expop_{\substack{(X_{1}, Y_{1}) \sim p_1 \\ \Ch_{1} \sim L_{\text{MVP}}(\cdot)}}
  \dots
  \max_{\substack{p_T \in \Ps'}}
  \Expop_{\substack{(X_{T}, Y_{T}) \sim p_T \\ \Ch_{T} \sim L_{\text{MVP}}(\cdot)}}
  \left|
  \frac{1}{T} \sum_{t=1}^T \text{Miscover}(\Ch_t, X_t, Y_t) - \alpha
  \right|
  \\
  &=
  \max_{\substack{p_1 \in \Ps'}}
  \Expop_{\substack{(X_{1}, Y_{1}) \sim p_1 \\ \Ch_{1} \sim L_{\text{MVP}}(\cdot)}}
  \dots
  \max_{\substack{p_T \in \Ps'}}
  \Expop_{\substack{(X_{T}, Y_{T}) \sim p_T \\ \Ch_{T} \sim L_{\text{MVP}}(\cdot)}}
  \left|
  \frac{1}{T} \sum_{t=1}^T \( \text{Miscover}(\Ch_t, X_t, Y_t) - \alpha \)
  \right|
  \\
  &=
  \max_{\substack{p_1 \in \Ps'}}
  \Expop_{\substack{(X_{1}, Y_{1}) \sim p_1 \\ \Ch_{1} \sim L_{\text{MVP}}(\cdot)}}
  \dots
  \max_{\substack{p_T \in \Ps'}}
  \Expop_{\substack{(X_{T}, Y_{T}) \sim p_T \\ \Ch_{T} \sim L_{\text{MVP}}(\cdot)}}
  \left|
  \frac{1}{T} \sum_{i=1}^m S_i\frac{1}{S_i}\sum_{t \in S_i} \( \text{Miscover}(\Ch_t, X_t, Y_t) - \alpha \)
  \right|
  \\
  &\le
  \max_{\substack{p_1 \in \Ps'}}
  \Expop_{\substack{(X_{1}, Y_{1}) \sim p_1 \\ \Ch_{1} \sim L_{\text{MVP}}(\cdot)}}
  \dots
  \max_{\substack{p_T \in \Ps'}}
  \Expop_{\substack{(X_{T}, Y_{T}) \sim p_T \\ \Ch_{T} \sim L_{\text{MVP}}(\cdot)}}
  \frac{1}{T} \sum_{i=1}^m S_i
  \left|
  \frac{1}{S_i}\sum_{t \in S_i} \( \text{Miscover}(\Ch_t, X_t, Y_t) - \alpha \)
  \right|
  \\
  &\le
  \max_{\substack{p_1 \in \Ps'}}
  \Expop_{\substack{(X_{1}, Y_{1}) \sim p_1 \\ \Ch_{1} \sim L_{\text{MVP}}(\cdot)}}
  \dots
  \max_{\substack{p_T \in \Ps'}}
  \Expop_{\substack{(X_{T}, Y_{T}) \sim p_T \\ \Ch_{T} \sim L_{\text{MVP}}(\cdot)}}
  \frac{1}{T} \sum_{i=1}^m T
  \left|
  \frac{1}{S_i}\sum_{t \in S_i} \( \text{Miscover}(\Ch_t, X_t, Y_t) - \alpha \)
  \right|
  \\
  &\le
  \sum_{i=1}^m
  \max_{\substack{p_1 \in \Ps'}}
  \Expop_{\substack{(X_{1}, Y_{1}) \sim p_1 \\ \Ch_{1} \sim L_{\text{MVP}}(\cdot)}}
  \dots
  \max_{\substack{p_T \in \Ps'}}
  \Expop_{\substack{(X_{T}, Y_{T}) \sim p_T \\ \Ch_{T} \sim L_{\text{MVP}}(\cdot)}}
  \left|
  \frac{1}{S_i}\sum_{t \in S_i} \( \text{Miscover}(\Ch_t, X_t, Y_t) - \alpha \)
  \right|
  \\
  &\le
  \sum_{i=1}^m
  \frac{f(|S_i|)}{|S_i|} \sqrt{4 K_1 m \ln m},
\end{align*}
as claimed, considering that $3.3 \le K_1 \le 3.4$.

%% file: impl.tex
\section{Implementation Details}
\label{s:impl}

\para{Interval construction.}
Enumerating elements of a consensus set $\Ch_t$ in (\ref{eq:consensusset})
given $\Ch_{t, k}$, $K$, and $\beta$
is non-trivial if $\Ys$ is continuous.
We use the following heuristic for regression:
instead of enumerating all elements in $\Ys$,
we enumerate over the edge of intervals.
Let $\Ch_{t, k}(\x_t) = [l_{t, k}, u_{t, k}]$.
Then, we consider the finite set of candidates
$\Ys_{\text{candi}} \coloneqq \{ l_{t, k} \mid 1 \le k \le K \} \cup \{ u_{t, k} \mid 1 \le k \le K \}$.
Given this, we choose an element voted by at least $K - \beta$ intervals,
\ie $\Ys_{\text{maj}} = \{ y \in \Ys_{\text{candi}} \mid \sum_{k=1}^K \mathbbm{1}\( y \in\Ch_{t, k}(\x_t) \) \ge K - \beta\}$. Then, our final interval is $[min(\Ys_{\text{maj}}), max(\Ys_{\text{maj}})]$ if $\Ys_{\text{maj}}$ is not empty or  $\emptyset$ otherwise.
\MR{
  For example, consider three base prediction intervals, $[0, 2], [1, 4], [3, 5]$, where $K-\beta=2$.
  Then, $\Ys_{\text{maj}} = \{ 1, 2, 3, 4 \}$, so the final interval is $[1, 4]$.
}
The constructed internal is the over-approximation of $\Ch_{t}(\x_t)$
due to the convexity of intervals
(\eg if the two edge of an interval are voted by $K - \beta$,
so all elements between two edges are).  
For enumerating elements of a base prediction set $\Ch_{t, k}(\x_t)$,
we use a  closed form solution due to the Gaussian distribution, as in \cite{Park2020PAC}.
Specifically,
given $\tau_{t, k}$ and
letting $\mu$ and $\sigma$ be the mean and standard deviation of $\MR{\bar{s}_{t, k}}(\x_t, \y_{t, k})$, respectively,
we have $\Ch_{t, k}(\x_t) = [\mu - \sigma \sqrt{c}, \mu+ \sigma \sqrt{c}]$,
where
$c = -2 \ln (\tau_{t, k} \cdot \MR{s^{\text{max}}_{t,k}} ) -2 \ln \sigma - \ln (2 \pi)$.
If $c < 0$, $\Ch_{t, k}$ returns an empty set. 

\para{Practical mitigation to control the violation of Assumption \ref{ass:mainassumption}.}
\MR{
The consensus set (\ref{eq:consensusset}) is redefined to mitigate the violation of Assumption \ref{ass:mainassumption} as follows:
\begin{align}
  \Ch_{t}(\x_t) \coloneqq \left\{ y \in \Ys \vmid \sum_{k=1}^K \mathbbm{1}\( y \in \Ch_{t, k}^\nu(\x_{t}) \) \ge K - \beta \right\},
  \label{eq:consensussetnu}
\end{align}
where $\Ch^\nu$ increases the volume of the prediction set $\Ch$
by the factor of $\nu$ for $\nu \ge 0$.
If $\Ch(x) = [a, b]$, we use $\Ch^\nu(x)= [m - (d+\nu), m + (d+\nu)]$,
where $m = (a+b)/2$ and $d = m - a$.
As we always increase the volume, $\Ch^\nu$ covers at least $1 - \alpha$ data if $\Ch$ covers $1-\alpha$ data due to a provided correctness guarantee. 
We control $\nu$ to handle potential risk in violating Assumption \ref{ass:mainassumption}. 
}

\para{Noise variances update for Kalman filter.}
To update noise variances, we consider a
gradient descent method.
In particular, we conduct gradient descent update on noise variances $w_{t-1, k}$ and $v_{t-1, k}$,
which minimizes the standard negative log-likelihood of 
the score function on an observation,
considering reparameterization
$w_{t-1, k} \coloneqq e^{\bar{w}_{t-1,k}}$ and
$v_{t-1, k} \coloneqq e^{\bar{v}_{t-1,k}}$ to avoid non-positive variances, \ie
$\ell(\bar{w}_{t-1, k}, \bar{v}_{t-1, k}) \coloneqq - \ln s_{t, k}(\x_t, \y_{t, k})$.
Letting $\gamma_{\text{noise}}$ be the learning rate of noise parameters,
we update noise log-variances $\bar{w}_{t, k}$ and $\bar{v}_{t, k}$ as follows:
\begin{align}
  \text{update}_{\text{noise}}(\Ch, \x_t, \y_{t, k}) : \quad
  \bar{w}_{t, k} &\gets \bar{w}_{t-1, k} - \gamma_{\text{noise}} \nabla_{\bar{w}} \label{eq:noiseupdate}\\
  \bar{v}_{t, k} &\gets \bar{v}_{t-1, k} - \gamma_{\text{noise}} \nabla_{\bar{v}}, \nonumber
\end{align}
where
$\xi \coloneqq \sqrt{ \sigma_{t-1, k}^2 + w_{t-1, k}^2 + v_{t-1, k}^2}$ and
\begin{align*}
  \nabla_{\bar{w}} &= \( \frac{1}{\xi} - \frac{(\y_{t,k} - \mu_{t-1, k})^2}{\xi^3} \) \cdot \frac{w_{t-1,k}}{\xi} \cdot e^{\bar{w}}, \\
  \nabla_{\bar{v}} &= \( \frac{1}{\xi} - \frac{(\y_{t,k} - \mu_{t-1, k})^2}{\xi^3} \) \cdot \frac{v_{t-1,k}}{\xi} \cdot e^{\bar{v}}.
\end{align*}
Here, $\nabla_{\bar{w}}$ or $\nabla_{\bar{w}}$ is the gradient of $\ell(\bar{w}_{t-1, k}, \bar{v}_{t-1, k})$ with respect to
$\bar{w}_{t-1, k}$ or $\bar{v}_{t-1, k}$, respectively.


\para{Ethereum implementation.}
We implemented \ACC along with MVP and the Kalman filter for each base prediction set in \texttt{Solidity}
for the price market application.
To this end, we consider multi-thread implementation,
while Algorithm \ref{alg:acc} assumes a single-thread.
In particular,
we consider a swap pool based on UniswapV2.
Whenever, a swap operation is executed, \ie \texttt{UniswapV2Pair.sol::swap($\cdot$)}, at the $k$-th pool,
we call $\text{update}(\cdot)$ in Algorithm \ref{alg:acc} at the end of the swap operation,
where we use the spot price from the reserves of two tokens as the observation $\y_{t, k}$.
During the MVP update, we need a random number generator $rand()$;
we use block difficulty and timestamp for the source of randomness,
but this can be improved via random number generator oracles. 
Along with base prediction sets for each pool,
we implement a consensus set construction part via (\ref{eq:consensusset}) in a smart contract,
and whenever a user reads the consensus set,
$K$ base prediction sets from pre-specified $K$ pools are read.
For fixed point operations, we use the \texttt{PRBMath} math library \cite{prbmath}. 

We evaluate our Solidity implementation in forked Ethereum mainnet.
In particular, 
we use \texttt{anvil} in \texttt{foundry} \cite{foundry} as
a local Ethereum node, where it mines a block whenever a transaction arrives.
Then, we deploy three AMMs, based on UniswapV2 by initializing reserves of two tokens,
along with a customized swap function as mentioned above. 
Next, we also deploy the \ACC smart contract.
Once the markets by three AMMs are ready,
we execute a trader, interfacing with the local blockchain via a Web3 library,
which randomly swaps two tokens from a randomly chosen AMM.
Finally, we execute an arbitrageur that exploits
arbitrage opportunities across AMMs, which eventually contributes to balancing the prices
of three markets.
Optionally, we also execute an adversary that chooses an AMM and conducts a huge swap operation
to mimic price manipulation.




\para{Hyper-parameters.}
\MR{
  We provide our choice of hyper-parameters that potentially minimize consensus set size. 
  Here, to avoid data snooping, we only use non-manipulated data (from the first part of each dataset)
  for the hyper-parameter selection before algorithm execution and evaluation.
  However, choosing hyper-parameters during the operation of the algorithm is acceptable.
}

\MR{
  The consensus set in (\ref{eq:consensussetnu}) has a hyper-parameter $\nu$.
  We use $\nu=0$ for USD/ETH price data and Ethereum simulation
  as the arbitrageurs are actively involved in these cases.
  But, the INV/ETH market is minor, so the arbitrage is not active enough;
  thus, we use $\nu=\max(prices_1) - \min(prices_1)$, where
  $prices_1$ is a list of prices from all sources at time $t=1$.
Note that see Section \ref{sec:consensussets} for the choice of \ACC hyper-parameters on $\alpha$, $K$, and $\beta$.
\ACC can be used any base prediction sets
that satisfy the condition in Theorem \ref{thm:consensuscorrectness},
which can be hyper-parameter lighter.
This also implies that the choice of hyper-parameters does not affect the guarantee by Theorem \ref{thm:consensuscorrectness} but does affect the prediction set size, as discussed in Section \ref{sec:scorefunc}.
}

\MR{
  For the Kalman filter,
  we use a gradient descent method with $\gamma_{\text{noise}}=10^{-3}$ (non-divergent for all our cases)
  along with $\bar{w}_{0, k} = \bar{v}_{0,k} = 4.6$ for the USD/ETH dataset and
  $\bar{w}_{0, k} = \bar{v}_{0,k} = 0.1$ for the INV/ETH dataset,
  which make consensus set size on hold-out first data
  around $50$ and $0.5$, respectively.
  For after gradient update, we truncate $\bar{w}_{t, k}$ to have at least
  $4.6$ for the USD/ETH dataset
  and
  $0.1$ for the INV/ETH dataset
  to avoid
  too small state variances, which are chosen using the same criteria in choosing $\bar{w}_{0, k}$ and $\bar{v}_{0, k}$.
  If state variances are too small, they are not adaptive to large changes in observations.
}

\MR{
For MVP,
we chose hyper-parameters based on suggested in \cite{bastani2022practical}:
$\eta = 5$ (non-divergent for all our cases)
$m = 100$ (large enough granularity, while we use $20$ for Ethereum simulation due to a Solidity contract memory limit),
$r = 1000$ (which needs to be sufficiently large), 
and
$\tau_\text{max}=1$ (as scores lie between 0 and 1 due to (\ref{eq:kfnormscore})).
}
